\documentclass[aps,prd,amsmath,amssymb,eqsecnum,nofootinbib,10pt,notitlepage]{revtex4-1}
\usepackage{graphicx}       
\usepackage{graphics}
\usepackage[retainorgcmds]{IEEEtrantools}
\usepackage{array}
\usepackage{multirow} 

\allowdisplaybreaks[1]

\begin{document}

\title{Transport Equation Approach to Calculations of Hadamard Green functions\\
and non-coincident DeWitt coefficients}

\author{Adrian C. Ottewill}
\email{adrian.ottewill@ucd.ie}
\affiliation{Complex and Adaptive Systems Laboratory and School of Mathematical Sciences,\\ University College Dublin, Belfield, Dublin 4, Ireland.}
\author{Barry Wardell}
\email{barry.wardell@aei.mpg.de}
\affiliation{Max-Planck-Institut f\"ur Gravitationsphysik, Albert-Einstein-Institut, 14476 Potsdam, Germany}

\begin{abstract}
Building on an insight due to Avramidi, we provide a system of transport equations for determining
key fundamental bi-tensors, including derivatives of the world-function, $\sigma\left( x,x' \right)$, the square root of the Van Vleck determinant, $\Delta^{1/2}\left( x,x' \right)$, and the tail-term, $V(x,x')$, appearing in the Hadamard form of the Green function. These bi-tensors are central to a broad range of problems from radiation reaction to quantum field theory in curved spacetime
and quantum gravity. Their transport equations may be used either 
in a semi-recursive approach to determining their covariant Taylor series expansions, or as the basis of numerical calculations. To illustrate the power of the semi-recursive approach, we present an implementation in \textsl{Mathematica} which computes very high order covariant series expansions of these objects. Using this code, a moderate laptop can, for example, calculate the coincidence limit $[a_7(x,x)]$ and $V(x,x')$ to order $(\sigma^a)^{20}$ in a matter of minutes. Results may be output in either a compact notation or in \textsl{xTensor} form. In a second application of the approach, we present a scheme for numerically integrating the transport equations as a system of coupled ordinary differential equations. As an example application of the scheme, we integrate along null geodesics to solve for $V(x,x')$ in Nariai and Schwarzschild spacetimes. 
\end{abstract}


\maketitle

\section{Introduction}
In a recent paper~\cite{QL} we presented methods for obtaining coordinate expansions for
the (tail part of the) retarded Green function in spherically symmetric spacetimes.  By using computer algebra to obtain high
order Taylor series (of order $(\Delta x^\alpha)^{50}$), and applying the theory of Pad\'e approximants we 
were able to obtain accurate expressions in remarkably large regions.
Using these expressions, we were able to present the first complete matched expansion calculation of the self-force 
in a model `black hole' spacetime, the Nariai spacetime~\cite{Casals:Dolan:Ottewill:Wardell:2009}, and are currently applying the method to 
Schwarzschild spacetime.  Our ultimate goal in this programme is to work in more general spacetimes,
 especially Kerr spacetime. A key component of the matched expansion approach is knowledge of the 
Green function for points close together (i.e., in a \emph{quasilocal} region). As we move away from specific symmetry conditions, we can no longer 
 rely on methods based on a special choice of coordinates in the construction of our quasilocal
 solution and are led instead to consider other techniques such as transport equations and covariant expansion methods. 

Covariant methods for calculating the Green function of the wave operator and the corresponding heat kernel, briefly reviewed in Sec.~\ref{sec:review}
below, are central to a broad range of problems from radiation reaction to quantum field theory in curved spacetime
and quantum gravity. There is an extremely extensive literature on this topic; here we provide only a very brief overview referring the reader to the reviews by Vassilevich~\cite{Vassilevich:2003} and Poisson~\cite{Poisson:2003} and references therein for a more complete discussion. These methods have 
evolved from pioneering work by Hadamard~\cite{Hadamard} on the classical theory and
DeWitt~\cite{DeWitt:1960, DeWitt:1965} on the quantum theory.  The central objects in the Hadamard and DeWitt covariant expansions are 
geometrical bi-tensor coefficients $a_n^{AB'}(x,x')$ which are commonly called DeWitt\footnote{The Hadamard and DeWitt coefficients also appear in the literature under several other guises. They may be called DeWitt, Gilkey, Minakshisundaram, Schwinger or Seeley coefficients, or any combination thereof (yielding acronyms such as DWSC, DWSG and HDMS). In the coincidence limit, it has been proposed that they be called Hadamard-Minakshisundaram-DeWitt (HaMiDeW) \cite{Gibbons} coefficients. For the remainder of this paper, we will refer to them as either DeWitt (for the coefficients $a_k{}^A{}_{B'}$) or Hadamard (for the coefficients $V_r{}^A{}_{B'}$) coefficients.} coefficients in the
physics literature. These coefficients are closely related to the short proper-time
asymptotic expansion of the heat kernel of an elliptic operator in a Riemannian space 
and so are commonly called heat kernel coefficients in the mathematics literature. 
Traditionally most attention has focused on the \textit{diagonal value}  of the heat kernel $K^{A}{}_{A}(x,x;s)$,
since the coincidence limits $a_n^{A}{}_{A}(x,x)$ play a central role in the classical theory 
of spectral invariants~\cite{Gilkey} and in the quantum theory of the effective action and trace anomalies~\cite{Birrell:Davies}.  
By contrast, for the quasilocal part of the matched expansion approach to radiation reaction \cite{Ottewill:Wardell:2008, Ottewill:Wardell:2009} we seek
expansions valid for $x$ and $x'$ as far apart as geometrical methods permit.   

The classical approach to the calculation of these coefficients in the physics literature was to use a recursive 
approach developed by DeWitt \cite{DeWitt:1965} in the 1960s. Although these recursive methods work well
 for the first few terms in the expansion \cite{Christensen:1976vb,Christensen:1978yd}, 
 and may be implemented in a tensor software package \cite{Christensen:1995}, the amount of calculation required to compute subsequent terms quickly becomes prohibitively long, even when implemented as a computer program.
  An alternative approach, more common in the
mathematics literature, is to use pseudo-differential operators and invariance theory~\cite{Gilkey}, where a basis of 
curvature invariants of the appropriate
structure is constructed~\cite{Fulling:1992} and then their coefficients determined by explicit evaluation in simple spacetimes.
However, here too, the size of the basis grows rapidly and there seems little prospect of reaching orders 
comparable to those we obtained in the highly symmetric configurations previously studied. 

An extremely elegant, non-recursive approach to the calculation of DeWitt coefficients has been given by Avramidi~\cite{Avramidi:1986,Avramidi:2000}. 
As his motivation was to study the effective action in quantum gravity he was primarily interested in the coincidence limit of the DeWitt 
coefficients, while in the self-force problem, as noted above, we require point-separated expressions.  In addition, Avramidi introduced his method in the language 
of quantum mechanics, quite distinct from the language of transport equations, such as the Raychaudhuri equation, more familiar to 
discussions of geodesics among relativists.  In this paper we present 
Avramidi's approach in the language of transport equations and show that it is ideal for numerical and symbolic computation. 
In so doing we are building on the work of D\'ecanini and Folacci~\cite{Decanini:Folacci:2005a} who wrote many of the equations we present (we
indicate below where we deviate from their approach) and 
implemented them explicitly by hand. However, calculations by hand are long and inevitably prone to error, particularly for higher spin and 
for higher order terms in the series and are quite impractical for the very high order expansions we would like for radiation reaction calculations.
Instead, we use the transport equations as the basis for \textsl{Mathematica} code for algebraic calculations and \textsl{C} code for numerical calculations. 
Rather than presenting our higher order results in excessively long equations (our non-canonical expression for $a_7(x,x)$ for a scalar field contains $2\,987\,366$ terms!),
we have made these codes freely available online \cite{AvramidiCode,TransportCode}. 

In Sec.~\ref{sec:review}, we provide a brief review of Green functions, bi-tensors and covariant expansions,
outlining the relations between the classical and quantum theories.

In Sec.~\ref{sec:avramidi}, we detail the principles that we consider to encapsulate the key insights of the Avramidi approach
and use these to write down a set of transport equations for the key bi-tensors of the theory. These provide an
adaptation of the Avramidi approach which is ideally suited to implementation on a computer either numerically or symbolically.

In Sec.~\ref{sec:symbolic}, we describe a semi-recursive approach to solving for covariant expansions and briefly describe our \textsl{Mathematica}
implementation of it and its interface with the tensor software package \textsl{xTensor} \cite{xTensor}. 

In Sec.~\ref{sec:numerical}, we present a numerical implementation of the transport equation approach to the calculation of the bi-scalar $V(x,x')$ appearing in the Hadamard form of the Green function along null geodesics.

In the Appendix, we give canonical expressions for the coincidence limits of the first five terms in the Hadamard expansion of $V(x,x')$.

Given our motivation in studying the radiation reaction problem we shall phrase all the discussion of this paper in 4-dimensional
spacetime. The reader is referred to work by D\'ecanini and Folacci~\cite{Decanini:Folacci:2005a} for a discussion of the corresponding
situation in spacetimes of more general (integer) dimension. We do note, however, that the DeWitt coefficients are purely geometric bi-tensors, formally independent of the spacetime dimension.

Throughout this paper, we use units in which $G=c=1$ and adopt the sign conventions of \cite{Misner:Thorne:Wheeler:1974}. We denote symmetrization of indices using brackets (e.g. $(\alpha \beta)$) and exclude indices from symmetrization by surrounding them by vertical bars (e.g. $(\alpha | \beta | \gamma)$). Roman letters are used for free indices and Greek letters for indices summed over all spacetime dimensions. Capital letters are used to denote the spinorial/tensorial indices appropriate to the field being considered.

\section{A Brief Review of Green functions, Bi-tensors and Covariant Expansions}
\label{sec:review}

\subsection{Classical Green functions}

We take an arbitrary field $\varphi^{A}(x)$,
and consider wave operators 
which are second order partial differential operators of the form~\cite{Avramidi:2000}
\begin{equation}
\label{eq:Wave-operator}
\mathcal{D}^{A}{}_B  = \delta^{A}{}_B (\square - m^2) - P^{A}{}_B
\end{equation}
where $\square \equiv g^{\alpha\beta}\nabla_{\alpha}\nabla_{\beta}$, $g^{\alpha\beta}$ is the (contravariant) metric tensor, $\nabla_{\alpha}$ is the covariant derivative defined by a connection $\mathcal{A}^{A}{}_{B\alpha}$:
$\nabla_{\alpha}\varphi^{A}= \partial_{\alpha}\varphi^{A}+  \mathcal{A}^{A}{}_{B\alpha} \varphi^{B}$, $m$ is the mass of the field and $P^{A}{}_B(x)$ is a possible potential term.

In the classical theory of wave propagation in curved spacetime, a fundamental object is the retarded Green function, $G_{\mathrm{ret}}{}^{B}{}_{C'} \left( x,x' \right)$. It is a solution of the inhomogeneous wave equation,
\begin{equation}
\label{eq:Wave}
\mathcal{D}^{A}{}_B G_{\mathrm{ret}}{}^{B}{}_{C'} \left( x,x' \right) = - 4\pi \delta^{A}{}_{C'}\delta \left( x,x' \right) ,
\end{equation}
 with support on and within the past light-cone of the field point. (The factor of $4\pi$ is a matter of convention, our choice here is consistent with 
 Ref.~\cite{Poisson:2003}.) 
Finding the retarded Green function globally can be extremely hard. However, provided $x$ and $x'$ are sufficiently close (within a normal neighborhood\footnote{More precisely, the Hadamard form requires that $x$ and $x'$ lie within a \emph{causal domain} -- a \emph{convex normal neighborhood} with causality condition attached. This effectively requires that $x$ and $x'$ be connected by at most one non-spacelike geodesic which stays within the causal domain. However, as we expect the term \emph{normal neighborhood} to be more familiar to the reader, we will use it throughout this paper, with implied assumptions of convexity and a causality condition.\label{def:causal domain}}), we can use the Hadamard form for the retarded Green function solution \cite{Hadamard,Friedlander},
which in 4 spacetime dimensions takes the form
\begin{equation}
\label{eq:Hadamard}
G_{\mathrm{ret}}{}^{A}{}_{B'}\left( x,x' \right) = \theta_{-} \left( x,x' \right) \left\lbrace U^{A}{}_{B'} \left( x,x' \right) \delta \left( \sigma \left( x,x' \right) \right) - V^{A}{}_{B'} \left( x,x' \right) \theta \left( - \sigma \left( x,x' \right) \right) \right\rbrace ,
\end{equation}
where $\theta_{-} \left( x,x' \right)$ is analogous to the Heaviside step-function, being $1$ when $x'$ is in the causal past of $x$, and $0$ otherwise, $\delta\left( \sigma\left(x,x'\right)\right)$ is the covariant form of the Dirac delta function, $U^{AB'}\left( x,x' \right)$ and $V^{AB'}\left( x,x' \right)$ are symmetric bi-spinors/tensors and  are regular for $x' \rightarrow x$. The bi-scalar $\sigma \left( x,x' \right)$ is the Synge~\cite{Poisson:2003} world function, which  is equal to one half of the squared geodesic distance between $x$ and $x'$.
The first term, involving $U^{A}{}_{B'} \left( x,x' \right)$, in Eq.~(\ref{eq:Hadamard}) represents the \emph{direct} part of the Green function 
while the second term, involving $V^{A}{}_{B'} \left( x,x' \right)$,  is known as the \emph{tail} part of the Green function. This tail term represents back-scattering off the spacetime geometry and is, for example, responsible for the quasilocal contribution to the self-force.

Within the Hadamard approach, the symmetric bi-scalar $V^{AB'}\left( x,x' \right)$ is expressed in terms of a formal expansion in increasing powers of $\sigma$ \cite{Decanini:Folacci:2005a}:
\begin{equation}
\label{eq:V}
V^{AB'}\left( x,x' \right) = \sum_{r=0}^{\infty} V_{r}{}^{AB'}\left( x,x' \right) \sigma ^{r}\left( x,x' \right)
\end{equation}
The coefficients $U^{AB'}$ and  $V_{r}{}^{AB'}$ are determined by imposing the wave equation, using the identity $\sigma_{;\alpha} \sigma^{;\alpha}= 2 \sigma =\sigma_{;\alpha'} \sigma^{;\alpha'}$,
and setting the coefficient of each \textsl{manifest} power of $\sigma$ equal to zero. Since $V^{A}{}_{B'}$ is symmetric for self-adjoint wave operators we are free to apply the wave 
equation either at $x$ or at $x'$; here we choose to apply it at $x'$. We find that 
$U^{AB'} \left( x,x' \right) = \Delta^{1/2} \left( x,x' \right) g^{AB'}\left( x,x' \right)$,  where $\Delta \left( x,x' \right)$ is the Van Vleck-Morette determinant defined as~\cite{Poisson:2003}
\begin{equation}
\Delta \left( x,x' \right) = - \left[ -g \left( x \right) \right] ^{-1/2} \det \left( -\sigma _{;\alpha \beta '} \left( x,x' \right) \right) \left[ -g \left( x' \right) \right] ^{-1/2} 
 = \det \left( -g^{\alpha'}{}_\alpha \left( x,x' \right) \sigma ^{;\alpha}{}_{ \beta '} \left( x,x' \right) \right)
\end{equation}
with $g^{\alpha'}{}_\alpha\left( x,x' \right)$ being the bi-vector of parallel transport (defined fully below) and where $g^{AB'}$ is the bi-tensor of parallel transport appropriate to the tensorial nature of the field, eg.
\begin{equation}
 g^{A B'} = \begin{cases}
             1 & \text{(scalar)}\\
             g^{a b'} & \text{(electromagnetic)} \\
             g^{a' (a} g^{b) b'} & \text{(gravitational)},
            \end{cases}
\end{equation}
where the higher spin fields are taken in Lorentz gauge. 
In making this identification we have used the transport equation for the Van Vleck-Morette determinant:
\begin{equation}
\label{eq:VVtransport}
\sigma^{;\alpha} \nabla_{\alpha} \ln \Delta = (4 - \square \sigma) .
\end{equation}
The coefficients $V_{r}^{AB'}\left( x,x' \right)$ satisfy the recursion relations
\begin{subequations}
\label{eq:RecursionV}
\begin{align}
\label{eq:recursionVn}
  \sigma ^{;\alpha'} (\Delta ^{-1/2} V^{AB'}_{r})_{;\alpha'}  
 + \left( r+1 \right)  \Delta ^{-1/2}  V_{r}^{AB'} + {\frac{1}{2r}} \Delta ^{-1/2}  \mathcal{D}^{B'}{}_{C'} V^{AC'}_{r-1} = 0 
\end{align}
for $r \in \mathbb{N}$ along with the `initial condition'
\begin{eqnarray}
\label{eq:recursionV0}
\sigma ^{;\alpha'} (\Delta ^{-1/2} V_0^{AB'}){}_{;\alpha'} 
+ \Delta ^{-1/2} V^{AB'}_0 + {\frac{1}{2}}\Delta ^{-1/2} \mathcal{D}^{B'}{}_{C'} ( \Delta ^{1/2} g^{AC'}) &=& 0 .
\end{eqnarray}
\end{subequations}
These are transport equations which may be solved in principle within a normal neighborhood by direct integration along the geodesic from $x$ to $x'$. The complication is that the calculation of $ V^{AB'}_{r}$ requires the calculation of second derivatives of $ V^{AB'}_{r-1}$ in directions off the geodesic;
we address this issue below.

Finally we emphasize that the Hadamard expansion~\eqref{eq:V} is an ansatz not a Taylor series. For example, in deSitter spacetime for a conformally invariant scalar theory all  the $V_r$'s are non-zero while $V\equiv0$.

\subsection{The quantum theory}
In curved spacetime a fundamental object of interest is the Feynman Green 
function defined for a quantum field $\hat \varphi^A(x)$ in the state $| \Psi \rangle$ by
\begin{equation}
G_\mathrm{f}^{AB'}(x,x')= i  \langle \Psi | \mathrm{T}\left[ \hat \varphi^A(x) \hat \varphi^{B'}(x') \right] | \Psi \rangle .
\end{equation}
where $\mathrm{T}$ denotes time-ordering.
The Feynman Green function may be related to the advanced and retarded Green functions of the classical theory by
the covariant commutation relations~\cite{DeWitt:1965}
\begin{equation}
G_\mathrm{f}{}^{AB'}(x,x')= \frac{1}{8\pi} \left(G_\mathrm{adv}^{AB'}(x,x') + G_\mathrm{ret}^{AB'}(x,x')\right)
+  \frac{i}{2}  \langle \Psi | \hat \varphi^A(x) \hat \varphi^{B'}(x')+ \hat \varphi^{B'}(x')\hat \varphi^A(x)| \Psi \rangle .
\end{equation}
The anticommutator function $\langle \Psi | \hat \varphi^A(x) \hat \varphi^{B'}(x')+ \hat \varphi^{B'}(x')\hat \varphi^A(x)| \Psi \rangle$
clearly satisfies the homogeneous wave equation so that the Feynman Green function satisfies the equation
\begin{equation}
 \mathcal{D}^{A}{}_B G_{\mathrm{f}}{}^{B}{}_{C'} \left( x,x' \right) = - \delta^A{}_{C'} \delta(x,x') .
\end{equation}

Using the proper-time formalism~\cite{DeWitt:1965}, the identity 
\begin{equation}
i \int\limits_0^\infty \mathrm{d}s\> e^{- \epsilon s} \exp({ i s x}) = - \frac{1}{x + i \epsilon}, \qquad (\epsilon > 0),
\end{equation}
allows the causal properties of the Feynman function to be encapsulated in the formal expression
\begin{equation}
G_\mathrm{f}{}^{A}{}_{C'} \left( x,x' \right)
 = i \int\limits_0^\infty \mathrm{d}s\> e^{- \epsilon s} \exp(i s \mathcal{D})^{A}{}_{B} \delta^B{}_{C'} \delta(x,x')
\end{equation}
where the limit $\epsilon \rightarrow 0+$ is understood.
The integrand
\begin{align}
    K^{A}{}_{C'}(x,x';s) = \exp(i s \mathcal{D})^{A}{}_{B} \delta^B{}_{C'} \delta(x,x')
\end{align}
clearly satisfies the Schr\"odinger/heat equation
\begin{align}
    \frac {1}{i} \frac{\partial K^{A}{}_{C'}}{\partial s} (x,x';s)= \mathcal{D}^A{}_{B}K^{B}{}_{C'}(x,x';s)
\end{align}
together with the initial condition 
$
    K^{A}{}_{B'}(x,x';0) = \delta^A{}_{B'}(x,x') 
$.
The trivial way in which the mass $m$ enters these equations allows it to be eliminated through the
prescription
\begin{align}
\label{eq:Kmsq}
    K^{A}{}_{C'}(x,x';s) = e^{ - i m^2 s }K_0{}^{A}{}_{C'}(x,x';s),
\end{align}
with the massless heat kernel satisfying the equation
\begin{align}
\label{eq:K0heateqn}
    \frac {1}{i} \frac{\partial K_0{}^{A}{}_{C'}}{\partial s} (x,x';s)= (\delta^{A}{}_B \square  - P^{A}{}_B)K_0{}^{B}{}_{C'}(x,x';s)
\end{align}
together with the `initial condition' 
$  K_0{}^{A}{}_{B'}(x,x';0) = \delta^A{}_{B'}\delta(x,x') $.

In $4$-dimensional Minkowski spacetime without potential, the massless heat kernel is readily obtained as
\begin{align}
    K_0{}^{A}{}_{B'}(x,x';s) =  \frac{1}{(4 \pi s)^2} \exp \left(-\frac{\sigma}{2 i s}\right) \delta^{A}{}_{B'} \qquad\textrm{(flat spacetime)} .
\end{align}
This motivates the ansatz~\cite{DeWitt:1965} that in general the massless heat kernel allows the representation
\begin{align}
\label{eq:K0general}
    K_0{}^{A}{}_{B'}(x,x';s) \sim  \frac{1}{(4 \pi s)^2}\exp \left(-\frac{\sigma}{2 i s}\right)  \Delta^{1/2}\left( x,x' \right) \Omega_0{}^{A}{}_{B'}(x,x';s)\ ,
\end{align}
where $\Omega{}^{A}{}_{B'}(x,x';s)$ possesses the following asymptotic expansion as $s \rightarrow 0+$:
\begin{align}
\label{eq:dewittseries}
    \Omega{}^{A}{}_{B'}(x,x';s) \sim   \sum\limits_{r=0}^\infty a_r^{A}{}_{B'}(x,x') (i s)^r\ ,
\end{align}
where $a_0{}^{A}{}_{B'}(x,x)= \delta^{A}{}_{B'}$ and $a_r{}^{A}{}_{B'}(x,x')$ has dimension $(\textrm{length})^{-2 r}$.
The inclusion of the explicit factor of $\Delta^{1/2}$ is simply a matter of convention; by including it we are following DeWitt, but many authors, including D\'ecanini and Folacci, choose instead to include it in the
series coefficients 
\begin{align}
A_r{}^{A}{}_{B'}(x,x')=\Delta^{1/2}a_r{}^{A}{}_{B'}(x,x').
\end{align}
It is clearly trivial to convert between the two conventions and, in any case, the coincidence limits agree.

Now, requiring our expansion to satisfy Eq.~(\ref{eq:K0heateqn}) and using the symmetry of $\Omega{}^{A}{}_{B'}(x,x';s)$ to allow operators to act at $x'$, we find that
$\Omega{}^{A}{}_{B'}(x,x';s)$ must satisfy
\begin{equation}
\frac{1}{i}\frac{\partial \Omega{}^{AB'}}{\partial s} + \frac{1}{i s} \sigma^{;\alpha'} \Omega^{AB'}{}_{;\alpha'} 
 = \Delta^{-1/2} (\delta^{B'}{}_{C'} \square  - P^{B'}{}_{C'})\left(\Delta^{1/2}  \Omega{}^{AC'}(x,x';s)\right) .
\end{equation}
Inserting the expansion Eq.~(\ref{eq:dewittseries}), the coefficients $a_{n}^{AB'}\left( x,x' \right)$ satisfy the recursion relations
\begin{subequations}
\begin{align}
 \label{eq:Recursiona}
\sigma^{;\alpha'} a_{r+1}^{\phantom{n}AB'}{}_{;\alpha'} 
 + \left( r+1 \right)  a_{r+1}^{\phantom{n}AB'} -
 \Delta^{-1/2} (\delta^{B'}{}_{C'} \square  - P^{B'}{}_{C'})\left(\Delta^{1/2}  a_r{}^{AC'}\right) = 0 
\end{align}
for $r \in \mathbb{N}$ along with the `initial condition'
\begin{eqnarray}
\sigma ^{;\alpha'} a_0{}^{AB'}{}_{;\alpha'} &=& 0 ,
\end{eqnarray}
\end{subequations}
with the implicit requirement that they be regular as $x' \to x$.

To compare the DeWitt approach to the Hadamard approach we may start by rewriting the Hadamard recursion relations
\eqref{eq:RecursionV} as  
\begin{align}
\label{eq:RecursionVmult}
&  \sigma ^{;\alpha'} ((-2)^{r+1} r! \Delta ^{-1/2} V^{AB'}_{r})_{;\alpha'}  
 + \left( r+1 \right) ( (-2)^{r+1} r!  \Delta ^{-1/2}  V_{r}^{AB'}) \nonumber \\
&\qquad- \Delta ^{-1/2}  \bigl(\delta^{B'}{}_{C'}\square'  - P^{B'}{}_{C'}\bigr)\bigl ( \Delta ^{1/2} \,(-2)^{r}  (r-1)!  \Delta ^{-1/2}  V_{r-1}^{AC'}\bigr) 
    + m^2\bigl ( \,(-2)^{r}  (r-1)!  \Delta ^{-1/2}  V_{r-1}^{AB'}\bigr) = 0 
\end{align}
which can be taken to include $r=0$ with the formal identification $ (-1)!  \Delta ^{-1/2}  V_{-1}^{AB'}=g^{AB'}=a_0{}^{AB'}$.
Comparing \eqref{eq:RecursionVmult} and \eqref{eq:Recursiona}, one can see that the massless Hadamard and (mass-independent) DeWitt coefficients are related by
\begin{subequations}
\begin{align}
\label{eq:HadamardToDeWitt}
 a_{r+1}{}^{A}{}_{B'}(x,x') &=  
 {(-2)^{r+1} r!}  \Delta ^{-1/2}( x,x' )V_r^{(m^2=0)}{}^{A}{}_{B'}(x,x')  ,\\
\label{eq:DeWitttoHadamard}
V_r^{(m^2=0)}{}^{A}{}_{B'}(x,x') &=(-1)^{r+1} \frac{\Delta^{1/2}\left( x,x' \right)}{2^{r+1} r!} 
 a_{r+1}{}^{A}{}_{B'}(x,x')  .
\end{align}
\end{subequations}

We can also relate the Hadamard  coefficients for a theory of mass $m$ and the (mass-independent) DeWitt coefficients. We start by noting that from 
\eqref{eq:Kmsq}, \eqref{eq:K0general} and \eqref{eq:dewittseries} the massive heat kernel has the asymptotic expansion 
\begin{align}
    K{}^{A}{}_{B'}(x,x';s) \sim  \frac{1}{(4 \pi s)^2}\exp \left(-\frac{\sigma}{2 i s}\right)  \Delta^{1/2}\left( x,x' \right) \sum\limits_{r=0}^\infty \left(\sum\limits_{k=0}^r \frac{(-m^2)^{r-k}}{(r-k)!} a_k^{A}{}_{B'}(x,x') \right)(i s)^r .
\end{align}
It follows from linearity that the massive Hadamard coefficients may be obtained from \eqref{eq:DeWitttoHadamard} with the replacement
\begin{equation}
a_r{}^{A}{}_{B'}(x,x') \to \sum\limits_{k=0}^r \frac{(-m^2)^{r-k}}{(r-k)!} a_r{}^{A}{}_{B'}(x,x')
\end{equation}
yielding
\begin{equation}
\label{eq:v-a-relation}
V_r{}^{A}{}_{B'}(x,x') = (-1)^{r+1}\frac{\Delta^{1/2}\left( x,x' \right)}{2^{r+1} r!} \sum \limits_{k=0}^{r+1}
 \frac{(-m^2)^{r+1-k}}{(r+1-k)!} a_k{}^{A}{}_{B'}(x,x')
\end{equation}
with inverse
\begin{equation}
a_{r+1}{}^{A}{}_{B'}(x,x') = \Delta^{-1/2} \sum \limits_{k=0}^{r}
(-2)^{k+1} \frac{k!}{(r-k)!} (m^2)^{r-k} V_k{}^{A}{}_{B'}(x,x') +  \frac{(m^2)^{r+1}}{(r+1)!} .
\end{equation}

These relations enable us to relate the `tail term' of the
massive theory to that of the massless theory by
\begin{equation}
V(x,x'){}^{A}{}_{B'} = \sum\limits_{r=0}^\infty   V_r^{(m^2=0)}{}^{A}{}_{B'} (x,x')  
\frac{\left(2\sigma\right)^r r! J_r\left( (-2 m^2 \sigma)^{1/2} \right)}{ (-2 m^2 \sigma)^{r/2}} 
+ m^2 \Delta^{1/2} \frac{J_1\left( (-2 m^2 \sigma)^{1/2} \right)}{ (-2 m^2 \sigma)^{1/2}} \delta{}^{A}{}_{B'}.
\end{equation}
where $J_r(x)$ are Bessel functions of the first kind. This last expression is obtained by using \eqref{eq:HadamardToDeWitt} in \eqref{eq:v-a-relation}, substituting the result into \eqref{eq:V} and interchanging the order of summation (upon doing so, the sum over $k$ yields the Bessel functions).

\subsection{Classical Approach to Covariant Expansion Calculations}

The Synge world-function, $\sigma (x,x')$ is a bi-scalar (i.e., a scalar at $x$ and at $x'$) defined to be equal to half the square of the geodesic distance between $x$ and $x'$.
The world-function is defined through the fundamental identity
\begin{equation}
\label{eq:SigmaDefiningEq}
 \sigma_\alpha \sigma^\alpha = 2 \sigma = \sigma_{\alpha'} \sigma^{\alpha'},
\end{equation}
together with the `initial' conditions $\lim\limits_{x'\to x} \sigma (x,x')=0 $
and $\lim\limits_{x'\to x} \sigma_{a b} (x,x')= g_{ab}(x) $.
Here, we indicate derivatives at the (un-)primed point by (un-)primed indices:
\begin{align}
 \sigma^a &\equiv \nabla^a \sigma & \sigma_a &\equiv \nabla_a \sigma &
 \sigma^{a'} &\equiv \nabla^{a'} \sigma & \sigma_{a'} &\equiv \nabla_{a'} \sigma .
\end{align}
$\sigma^a $ is a vector at $x$ of length equal to the geodesic distance between $x$ and $x'$, tangent to the geodesic at $x$ and oriented in the direction $x'\to x$ while $\sigma^{a'} $ is a vector at $x'$ of length equal to the geodesic distance between $x$ and $x'$, tangent to the geodesic at $x'$ and oriented in the opposite direction.

The covariant derivatives of $\sigma$ may be written as
\begin{align}
 \sigma^{a}(x,x') &= (s - s') u^{a} & \sigma^{a'}(x,x') &= (s' - s) u^{a'}
\end{align}
where $s$ is an affine parameter and $u^{a}$ is tangent to the geodesic.
For time-like geodesics, $s$ may be taken as the proper time along the geodesic while
$u^{a}$ is the 4-velocity tangent to the geodesic and
\begin{equation}
\label{eq:sigma-def}
 \sigma (x,x') = -\frac{1}{2} (s - s')^2 .
\end{equation}
Similarly, for space-like geodesics, $s$ may be taken as the spatial geodesic distance along the geodesic and
\begin{equation}
 \sigma (x,x') = +\frac{1}{2} (s - s')^2 .
\end{equation}
For null geodesics, $u^{a}$ is null and  $\sigma (x,x') = 0$.

Another bi-tensor of frequent interest is the  bi-vector of parallel transport, $g_{a b'}$ defined by the transport equation
\begin{equation}
\sigma^{\alpha} g_{a b' ;\alpha}  = 0 = \sigma^{\alpha'} g_{a b' ;\alpha'} 
\end{equation}
with initial condition $\lim\limits_{x'\to x} g_{a b'} (x,x')= g_{ab}(x) $.
From the definition of a geodesic it follows that
\begin{equation}
 g_{a \alpha'} \sigma^{\alpha'} = - \sigma_{a} \qquad \mathrm{and} \qquad g_{\alpha a '} \sigma^{\alpha} = - \sigma_{a'}.
\end{equation}

Given a bi-tensor $T_{a}$ at $x$, the parallel transport bi-vector allows us define $\bar{T}_{a'}$, a bi-tensor at $x'$, obtained by parallel transporting $T_{a}$ along the geodesic from $x$ to $x'$ and vice-versa,
\begin{align}
 T_{\alpha} g^{\alpha}_{\phantom{a} a'} &= \bar{T}_{a'} &
 \bar{T}_{\alpha'} g^{\alpha'}_{\phantom{a'} a} &= T_{a}.
\end{align}
These are consistent as $g^{a}_{\phantom{a} \alpha'} g_{ b} {}^{\alpha'}= \delta^a{}_b$
and $g_{ \alpha} {}^{a'} g^{\alpha}{}_{b'} = \delta^{a'}{}_{b'}$.

Any sufficiently smooth bi-tensor $T_{a_1 \cdots a_m a_1' \cdots a_n'}$ may be expanded in a local covariant Taylor series about the point $x$.
To do so it is convenient and conventional to first use the bi-vector of parallel transport to transport all tensor indices to $x$, for example:
\begin{equation}
\label{eq:AxExpansion}
g_{b_1}{}^{b_1'} \cdots  g_{b_n}{}^{b_n'} T_{a_1 \cdots a_m b_1' \cdots b_n'} (x,x')= \sum_{k=0}^\infty \frac{(-1)^k}{k!} t_{a_1 \cdots a_m b_1 \cdots b_n \, \alpha_1 \cdots \alpha_k} (x) \sigma^{\alpha_1} \cdots \sigma^{\alpha_k} = \sum_{k=0}^\infty \frac{(-1)^k}{k!} T_{a_1 \cdots a_m b_1 \cdots b_n \, (k)}
\end{equation}
where the $t_{a_1 \cdots a_m b_1 \cdots b_n \, \alpha_1 \cdots \alpha_k}$ are the coefficients of the series and are local tensors at $x$
and $T_{a_1 \cdots a_m b_1 \cdots b_n \, (k)}$ is defined as this coefficient contracted with the corresponding $\sigma^{\alpha_i}$. Similarly, we can also expand about $x'$:
\begin{equation}
g^{a_1}{}_{a_1'} \cdots  g^{a_m}{}_{a_m'}  T_{a_1 \cdots a_m b_1' \cdots b_n'} (x,x')= \sum_{k=0}^\infty \frac{(-1)^k}{k!} t_{a_1' \cdots a_m' b_1' \cdots b_n' \, \alpha_1' \cdots \alpha_k'} (x') \sigma^{\alpha_1'} \cdots \sigma^{\alpha_k'}.
\end{equation}

For many fundamental bi-tensors, one would typically use the DeWitt approach \cite{DeWitt:1960} to determine the coefficients in these expansions as follows:
\begin{enumerate}
 \item Take covariant derivatives of the defining equation for the bi-tensor (the number of derivatives required depends on the order of the term to be found).
 \item Replace all known terms with their coincidence limit, $x \rightarrow x'$.
 \item Sort covariant derivatives, introducing Riemann tensor terms in the process.
 \item Take the coincidence limit $x' \rightarrow x$ of the result.
\end{enumerate}
This method allows all coefficients to be determined recursively in terms of lower order coefficients and Riemann tensor polynomials. Although this method proves effective for determining the lowest few order terms by hand and can be readily implemented in software, it does not scale well and it is not long before the computation time required to calculate the next term is prohibitively large. This issue can be understood from the fact that the calculation yields extremely large intermediate expressions which simplify tremendously in the end. It is therefore desirable to find an alternative approach which is more efficient and better suited to implementation in software. In the following sections, we will describe one such approach which proves to be highly efficient. 

\section{Avramidi Approach to Covariant Expansion Calculations}
\label{sec:avramidi}
The traditional DeWitt \cite{DeWitt:1960} approach to the calculation of covariant expansions of fundamental bi-tensors is to derive a set of recursion relations for the coefficients of the series. Avramidi \cite{Avramidi:1986} has proposed an alternative, extremely elegant non-recursive method for the calculation of these coefficients. Translated into the language of transport equations, this approach emphasizes two fundamental principles when doing calculations:
\begin{enumerate}
 \item When expanding about $x$, always try to take derivatives at $x'$. The result is that derivatives only act on the $\sigma^a$'s and not on the coefficients.
 \item Where possible, whenever taking a covariant derivative, $\nabla_{a'}$, contract the derivative with $\sigma^{a'}$.
\end{enumerate}
Applying these two principles, Avramidi has derived non-recursive\footnote{Avramidi retains the recurrence relations for the DeWitt coefficients, $a_k$ (and hence the Hadamard coefficients, $V_r$). However, all other relations are non-recursive.} expressions for the coefficients of covariant expansions of several bi-tensors. As Avramidi's derivations use a rather abstract notation, we will now briefly review his technique in a more explicit notation. We will also extend the derivation to include several other bi-tensors and note that Eqs. \eqref{eq:xi-transport}, \eqref{eq:eta-transport}, \eqref{eq:gamma-transport}, \eqref{eq:lambda-eq}, \eqref{eq:vanVleck-transport2}, \eqref{eq:sqrtdelta-zeta}, \eqref{eq:V0-transport} and \eqref{eq:Vr-transport} were previously written down and used by D\'ecanini and Folacci \cite{Decanini:Folacci:2005a}.

Throughout this section, we fix the base point $x$ and allow it to be connected to any other point $x'$ by a geodesic. In all cases, we expand about the fixed point, $x$.

Defining the transport operators $D$ and $D'$ as
\begin{align}
 D & \equiv \sigma^\alpha \nabla_\alpha & D' & \equiv \sigma^{\alpha'} \nabla_{\alpha'},
\end{align}
we can rewrite Eq.~(\ref{eq:SigmaDefiningEq}) as
\begin{align}
 (D-2)\sigma &= 0 & (D'-2)\sigma &= 0.
\end{align}
Differentiating  these equations at $x$ and at $x'$, we get
\begin{align}
\left( D - 1 \right) \sigma^{a} &= 0 & \left( D- 1 \right) \sigma^{a'} &= 0 &\left( D' - 1 \right) \sigma^{a} &= 0 & \left( D' - 1 \right) \sigma^{a'} &= 0.
\end{align}
Defining
\begin{align}
 \eta^{a}_{\phantom{a} b'} &\equiv \sigma^{a}_{\phantom{b} b'} &
 \xi^{a'}_{\phantom{a'} b'} &\equiv  \sigma^{a'}_{\phantom{a'} b'},
\end{align}
the second pair of these equations can be rewritten as
\begin{align}
\label{eq:sigma-eta-xi}
 \sigma^{a} &= \eta^{a}_{\phantom{a} \alpha'} \sigma^{\alpha'} &
 \sigma^{a'} &= \xi^{a'}_{\phantom{a'} \alpha'} \sigma^{\alpha'}.
\end{align}
Finally, we define $\gamma^{a'}_{\phantom{a'} b}$, the inverse of $\eta^{a}_{\phantom{a} b'}$,
\begin{equation}
\label{eq:gamma-def}
 \gamma^{a'}_{\phantom{a'} b} \equiv (\eta^{b}_{\phantom{b} a'})^{-1}.
\end{equation}
and also introduce the definition
\begin{equation}
 \lambda^{a}_{\phantom{a} b} \equiv \sigma^{a}_{\phantom{a} b}.
\end{equation}

We will now derive transport equations for each of these newly introduced quantities along with some others which will be defined as required. Many of these derivations involve considerable index manipulations and are most easily (and accurately) done using a tensor software package such as \textsl{xTensor}~\cite{xTensor}.

The transport equations of this section may be derived in a recursive manner, making use of the identities
\begin{align}
\label{eq:transport-deriving-primed}
D'(\sigma_{a_1'\dots a_n' a_{n+1}'}) &= \nabla_{a_{n+1}'} (D' \sigma_{a_1'\dots a_n'}) - \xi^{\alpha'}{}_{a_{n+1}'}  \nabla_{\alpha'}\sigma_{a_1'\dots a_n'}            
 + \sigma^{\alpha'} R^{c'}{}_{a_1'a_{n+1}'\alpha'}\sigma_{c'\dots a_n'}   + \dots + \sigma^{\alpha'} R^{c'}{}_{a_n'a_{n+1}'\alpha'}\sigma_{a_1'\dots c'}\\
\label{eq:transport-deriving-unprimed}
D'(\sigma^{b}{}_{a_1'\dots a_n'}) &= \nabla^{b} (D' \sigma_{a_1'\dots a_n'}) - \eta^{b}{}_{\alpha'}  \nabla^{\alpha'}\sigma_{a_1'\dots a_n'} .
\end{align}
and its generalisation, given below.
This method is naturally algorithmic and well suited to implementation on a computer, thus allowing for the automated derivation of a transport equation for an arbitrary number of derivatives of a bi-tensor.

\subsection{Transport equation for $\xi^{a'}_{~b'}$}
Taking a primed derivative of the second equation in (\ref{eq:sigma-eta-xi}), we obtain
\begin{equation}
 \xi^{a'}_{\phantom{a'} b'} = \xi^{a'}_{\phantom{a'} \alpha' b'} \sigma^{\alpha'} + \xi^{a'}_{\phantom{a'} \alpha'}\xi^{\alpha'}_{\phantom{\alpha'} b'}.
\end{equation}

We now commute the last two covariant derivatives in the first term on the right hand side of this equation and rearrange to obtain:
\begin{equation}
\label{eq:xi-transport}
D' \xi^{a'}_{\phantom{a'} b'} + \xi^{a'}_{\phantom{a'} \alpha'} \xi^{\alpha'}_{\phantom{\alpha'} b'}  - \xi^{a'}_{\phantom{a'} b'} + R^{a'}_{\phantom{a'} \alpha' b' \beta'} \sigma^{\alpha'} \sigma^{\beta'} = 0
\end{equation}

\subsection{Transport equation for $\eta^{a}_{~b'}$}
Taking a primed derivative of the first equation in (\ref{eq:sigma-eta-xi}), we obtain
\begin{equation}
 \eta^{a}_{\phantom{a} b'} = \eta^{a}_{\phantom{a} \alpha' b'} \sigma^{\alpha'} + \eta^{a}_{\phantom{a} \alpha'}\xi^{\alpha'}_{\phantom{\alpha'} b'}
\end{equation}

In this case, since $\sigma^a$ is a scalar at $x'$, we can commute the two primed covariant derivatives in the first term on the right hand side of this equation without introducing a Riemann term. Rearranging, we obtain:
\begin{equation}
\label{eq:eta-transport}
D' \eta^{a}_{\phantom{a} b'} + \eta^{a}_{\phantom{a} \alpha'} \xi^{\alpha'}_{\phantom{\alpha'} b'}  - \eta^{a}_{\phantom{a} b'} = 0
\end{equation}

\subsection{Transport equation for $\gamma^{a'}_{~b}$}
Solving Eq.~(\ref{eq:eta-transport}) for $\xi^{a'}_{\phantom{a'} b'}$ and using \eqref{eq:gamma-def}, we obtain
\begin{align}
\label{eq:xi-solve-eta-gamma}
 \xi^{a'}_{\phantom{a'} b'} &= \delta^{a'}_{\phantom{a'} b'} - \gamma^{a'}_{\phantom{a'} \alpha} \left(D' \eta^{\alpha}_{\phantom{\alpha'} b'}\right)\nonumber \\
	&= \delta^{a'}_{\phantom{a'} b'} + \left( D' \gamma^{a'}_{\phantom{a'} \alpha} \right) \eta^{\alpha}_{\phantom{\alpha'} b'},
\end{align}
Next, substituting Eq.~(\ref{eq:xi-solve-eta-gamma}) into Eq.~(\ref{eq:xi-transport}) and rearranging, we obtain a transport equation for $\gamma^{a'}_{\phantom{a'} b}$:
\begin{equation}
\label{eq:gamma-transport}
 (D')^2 \gamma^{a'}_{\phantom{a'} b} + D' \gamma^{a'}_{\phantom{a'} b} + R^{a'}_{\phantom{a'} \alpha' \gamma' \beta'} \gamma^{\gamma'}_{\phantom{a'} b} \sigma^{\alpha'} \sigma^{\beta'}
 =0.
\end{equation}

\subsection{Equation for  $\lambda^{a}_{~b}$}
Differentiating Eq.~(\ref{eq:SigmaDefiningEq}) at $x$ and $x'$, we obtain
\begin{equation}
 \eta^{a}_{\phantom{a} b'} = \lambda^{a}_{\phantom{a} \alpha} \eta^{\alpha}_{\phantom{\alpha} b'} + D \eta^{a}_{\phantom{a} b'}
\end{equation}
which is easily rearranged to give an equation for $\lambda^{a}_{\phantom{a} b}$:
\begin{equation}
\label{eq:lambda-eq}
 \lambda^{a}_{\phantom{a} b} = \delta^{a}_{\phantom{a} b} - (D \eta^{a}_{\phantom{a} \alpha'}) \gamma^{\alpha'}_{\phantom{\alpha'} b}.
\end{equation}

\subsection{Transport equation for $\sigma^{a'}_{~~ b' c'}$}
Applying the identity \eqref{eq:transport-deriving-primed} to \eqref{eq:xi-transport} and simplifying the resulting expression, we obtain
\begin{align}
\label{eq:transport-sigma-ppp}
(D'-1) \sigma^{a'}_{\phantom{a'} b' c'}
+ \sigma^{\alpha'}_{\phantom{\alpha'} c'} \sigma^{a}_{\phantom{a} \alpha' b'}
+ \sigma^{\alpha'}_{\phantom{\alpha'} b'} \sigma^{a'}_{\phantom{a'} \alpha' c'}
+ \sigma^{a'}_{\phantom{a'} \alpha'} \sigma^{\alpha'}_{\phantom{\alpha'} b' c'}
+ R^{a'}_{\phantom{a'} \alpha' b' \beta' ; c'} \sigma^{\alpha'} \sigma^{\beta'} \nonumber \\
- R^{a'}_{\phantom{a'} \alpha' \beta' b'} \sigma^{\beta'} \sigma^{\alpha'}_{\phantom{\alpha'} c'}
- R^{a'}_{\phantom{a'} \alpha' \beta' c'} \sigma^{\beta'} \sigma^{\alpha'}_{\phantom{\alpha'} b'}
+ R^{\alpha'}_{\phantom{\alpha'} b' \beta' c'} \sigma^{\beta'} \sigma^{a'}_{\phantom{a'} \alpha'}
= 0
\end{align}

\subsection{Transport equation for $\sigma^{a}_{~b' c'}$}
Applying the identity \eqref{eq:transport-deriving-unprimed} to \eqref{eq:xi-transport} and simplifying the resulting expression, we obtain
\begin{equation}
\label{eq:transport-sigma-upp}
(D'-1) \sigma^{a}_{\phantom{a} b' c'}
+ \sigma^{\alpha'}_{\phantom{\alpha'} b'} \sigma^{a}_{\phantom{a} \alpha' c'}
+ \sigma^{\alpha'}_{\phantom{\alpha'} c'} \sigma^{a}_{\phantom{a} \alpha' b'}
+ \sigma^{a}_{\phantom{a} \alpha'} \sigma^{\alpha'}_{\phantom{\alpha'} b' c'}
+ R^{\alpha'}_{\phantom{\alpha'} b' \beta' c'} \sigma^{a}_{\phantom{a} \alpha'} \sigma^{\beta'}
= 0
\end{equation}

\subsection{Transport equation for $\sigma^{a'}_{~~b' c' d'}$}
Applying the identity \eqref{eq:transport-deriving-primed} to \eqref{eq:transport-sigma-ppp} and simplifying the resulting expression, we obtain
\begin{multline}
\label{eq:sigma-pppp-transport}
(D'-1)\sigma^{a'}_{\phantom{a'} b' c' d'} 
+ \sigma^{a'}_{\phantom{a'} \alpha' b' c'} \sigma^{\alpha'}_{\phantom{\alpha'} d'} 
+ \sigma^{a'}_{\phantom{a'} \alpha' b' d'} \sigma^{\alpha'}_{\phantom{\alpha'} c'} 
+ \sigma^{a'}_{\phantom{a'} \alpha' c' d'} \sigma^{\alpha'}_{\phantom{\alpha'} b'}
+ \sigma^{a'}_{\phantom{a'} \alpha' b'} \sigma^{\alpha'}_{\phantom{\alpha'} c' d'} 
+ \sigma^{a'}_{\phantom{a'} \alpha' c'} \sigma^{\alpha'}_{\phantom{\alpha'} b' d'} 
+ \sigma^{a'}_{\phantom{a'} \alpha' d'} \sigma^{\alpha'}_{\phantom{\alpha'} b' c'} \\
+ \sigma^{a'}_{\phantom{a'} \alpha'} \sigma^{\alpha'}_{\phantom{\alpha'} b' c' d'}
+ R^{a'}_{\phantom{a'} \alpha' \beta' c'} R^{\alpha'}_{\phantom{\alpha'} d' \gamma' b'} \sigma^{\beta'} \sigma^{\gamma'}
+ R^{a'}_{\phantom{a'} \alpha' \beta' b'} R^{\alpha'}_{\phantom{\alpha'} d' \gamma' c'} \sigma^{\beta'} \sigma^{\gamma'}
+ R^{a'}_{\phantom{a'} \alpha' \beta' d'} R^{\alpha'}_{\phantom{\alpha'} c' \gamma' b'} \sigma^{\beta'} \sigma^{\gamma'}\\
- R^{a'}_{\phantom{a'} \beta' \alpha' d'} R^{\alpha'}_{\phantom{\alpha'} b' \gamma' c'} \sigma^{\beta'} \sigma^{\gamma'}
- R^{a'}_{\phantom{a'} \beta' \alpha' c'} R^{\alpha'}_{\phantom{\alpha'} b' \gamma' d'} \sigma^{\beta'} \sigma^{\gamma'}
+ R^{a'}_{\phantom{a'} \beta' b' \gamma' ; c' d'} \sigma^{\beta'} \sigma^{\gamma'}
+ R^{\alpha'}_{\phantom{\alpha'} b' \beta' c' ; d'} \sigma^{\beta'} \sigma^{a'}_{\phantom{a'} \alpha'}\\
- R^{a'}_{\phantom{a'} \alpha' \beta' c' ; d'} \sigma^{\beta'} \sigma^{\alpha'}_{\phantom{\alpha'} b'}
- R^{a'}_{\phantom{a'} \alpha' \beta' b' ; d'} \sigma^{\beta'} \sigma^{\alpha'}_{\phantom{\alpha'} c'}
- R^{a'}_{\phantom{a'} \alpha' \beta' b' ; c'} \sigma^{\beta'} \sigma^{\alpha'}_{\phantom{\alpha'} d'}
+ R^{\alpha'}_{\phantom{\alpha'} c' \beta' d'} \sigma^{\beta'} \sigma^{a'}_{\phantom{a'} b' \alpha'}
+ R^{\alpha'}_{\phantom{\alpha'} b' \beta' d'} \sigma^{\beta'} \sigma^{a'}_{\phantom{a'} c' \alpha'}\\
+ R^{\alpha'}_{\phantom{\alpha'} b' \beta' c'} \sigma^{\beta'} \sigma^{a'}_{\phantom{a'} d' \alpha'}
- R^{a'}_{\phantom{a'} \alpha' \beta' d'} \sigma^{\beta'} \sigma^{\alpha'}_{\phantom{\alpha'} b' c'}
- R^{a'}_{\phantom{a'} \alpha' \beta' c'} \sigma^{\beta'} \sigma^{\alpha'}_{\phantom{\alpha'} b' d'}
- R^{a'}_{\phantom{a'} \alpha' \beta' b'} \sigma^{\beta'} \sigma^{\alpha'}_{\phantom{\alpha'} c' d'} = 0
\end{multline}

\subsection{Transport equation for $\sigma^{a}_{~b' c' d'}$}
Applying the identity \eqref{eq:transport-deriving-unprimed} to \eqref{eq:transport-sigma-ppp} and simplifying the resulting expression, we obtain
\begin{multline}
\label{eq:sigma-uppp-transport}
 (D'-1)\sigma^{a}_{\phantom{a} b' c' d'}  + \sigma^{a}_{\phantom{a} \alpha' b' c'} \sigma^{\alpha'}_{\phantom{\alpha'} d'} + \sigma^{a}_{\phantom{a} \alpha' b' d'} \sigma^{\alpha'}_{\phantom{\alpha'} c'} + \sigma^{a}_{\phantom{a} \alpha' c' d'} \sigma^{\alpha'}_{\phantom{\alpha'} b'} + \sigma^{a}_{\phantom{a} \alpha' b'} \sigma^{\alpha'}_{\phantom{\alpha'} c' d'} + \sigma^{a}_{\phantom{a} \alpha' c'} \sigma^{\alpha'}_{\phantom{\alpha'} b' d'} + \sigma^{a}_{\phantom{a} \alpha' d'} \sigma^{\alpha'}_{\phantom{\alpha'} b' c'} \\
+ \sigma^{a}_{\phantom{a} \alpha'} \sigma^{\alpha'}_{\phantom{\alpha'} b' c' d'} 
+ R^{\alpha'}_{\phantom{\alpha'} b' \beta' c'; d'} \sigma^{\beta'} \sigma^{a}_{\phantom{a} \alpha'}
+ R^{\alpha'}_{\phantom{\alpha'} b' \beta' c'} \sigma^{\beta'} \sigma^{a}_{\phantom{a} d' \alpha'}
+ R^{\alpha'}_{\phantom{\alpha'} b' \beta' d'} \sigma^{\beta'} \sigma^{a}_{\phantom{a} c' \alpha'}
+ R^{\alpha'}_{\phantom{\alpha'} c' \beta' d'} \sigma^{\beta'} \sigma^{a}_{\phantom{a} b' \alpha'} = 0
\end{multline}

\subsection{Transport equation for $g_{a'}^{~~ b}$}
The bi-vector of parallel transport is defined by the transport equation
\begin{equation}
\label{eq:bi-transport}
 D'g_{a'}^{\phantom{a'} b} = \sigma^{\alpha'}  g_{a' \phantom{b} ;\alpha'}^{\phantom{a'} b} \equiv 0.
\end{equation}

\subsection{Transport equation for $g_{a b' ; c'}$}
Let
\begin{equation}
 A_{a b c} = g_{b}^{\phantom{b} \alpha'} g_{c}^{\phantom{c} \beta'} g_{a \alpha' ; \beta'}
\end{equation}
Applying $D'$ and commuting covariant derivatives, we obtain a transport equation for $A_{a b c}$:
\begin{equation}
\label{eq:A-transport}
 D' A_{a b c} + A_{a b \alpha} \xi^{\beta'}_{\phantom{\beta'} \gamma'} g_{\beta'}^{\phantom{\beta'} \alpha} g_{c}^{\phantom{c} \gamma'} - g_{a}^{\phantom{a} \alpha'} g_{b}^{\phantom{b} \beta'} g_{c}^{\phantom{c} \gamma'} R_{\alpha' \beta' \gamma' \delta'} \sigma^{\delta'} = 0
\end{equation}

\subsection{Transport equation for $g_{a b' ; c}$}
Let
\begin{equation}
 B_{a b c} = g_{b}^{\phantom{b} \beta'} g_{a \beta' ; c}
\end{equation}
Applying $D'$ and rearranging, we obtain a transport equation for $B_{\alpha \beta \gamma}$:
\begin{equation}
\label{eq:B-transport}
 D' B_{a b c} = - A_{a b \alpha} \eta^{\alpha}_{\phantom{\alpha} \beta'} g_{c}^{\phantom{c} \beta'}
\end{equation}

\subsection{Transport equation for $g_{a ~~ ; c' d'}^{~ b'}$}
Applying $D'$ to $g_{a \phantom{b'} ; c' d'}^{\phantom{a} b'}$, we obtain
\begin{equation}
 D' g_{a \phantom{b'} ; c' d'}^{\phantom{a} b'} = \sigma^{\alpha'} g_{a \phantom{b'} ; c' d' \alpha'}^{\phantom{a} b'}.
\end{equation}
Commuting covariant derivatives on the right hand side, this becomes
\begin{equation}
 D' g_{a \phantom{b'} ; c' d'}^{\phantom{a} b'} = \sigma^{\beta'} \left(g_{a \phantom{b'} ;\beta' c' d'}^{\phantom{a} b'}  
+ R^{b'}_{\phantom{b'} \alpha' \beta' d'} g^{\phantom{a} \alpha'}_{a \phantom{\alpha'} ;c'}
+ R^{b'}_{\phantom{b'} \alpha' \beta' c'} g^{\phantom{a} \alpha'}_{a \phantom{\alpha'} ;d'}
- R^{\alpha'}_{\phantom{\alpha'} c' \beta' d'} g^{\phantom{a} b'}_{a \phantom{b'} ;\alpha'}
+ R^{b'}_{\phantom{b'} \alpha' \beta' c' ; d'} g^{\phantom{a} \alpha'}_{a \phantom{\alpha'}}
\right).
\end{equation}
Bringing $\sigma^{\beta'}$ inside the derivative in the first time on the right hand side, and noting that $\sigma^{\beta'}  g_{a \phantom{b'} ;\beta'}^{\phantom{a} b'}= 0$, this then yields a transport equation for $g_{a \phantom{b'} ;c' d'}^{\phantom{a} b'}$:
\begin{multline}
\label{eq:d2Iinv-transport}
 D' g_{a \phantom{b'} ; c' d'}^{\phantom{a} b'} = 
  -\sigma^{\beta'}_{\phantom{\beta'} c'} g_{a \phantom{b'} ;\beta' d'}^{\phantom{a} b'}  
  -\sigma^{\beta'}_{\phantom{\beta'} d'} g_{a \phantom{b'} ;\beta' c'}^{\phantom{a} b'}  
  -\sigma^{\beta'}_{\phantom{\beta'} c' d'} g_{a \phantom{b'} ;\beta'}^{\phantom{a} b'}  \\
+ R^{b'}_{\phantom{b'} \alpha' \beta' d'} \sigma^{\beta'} g^{\phantom{a} \alpha'}_{a \phantom{\alpha'} ;c'}
+ R^{b'}_{\phantom{b'} \alpha' \beta' c'} \sigma^{\beta'} g^{\phantom{a} \alpha'}_{a \phantom{\alpha'} ;d'}
- R^{\alpha'}_{\phantom{\alpha'} c' \beta' d'} \sigma^{\beta'} g^{\phantom{a} b'}_{a \phantom{b'} ;\alpha'}
+ R^{b'}_{\phantom{b'} \alpha' \beta' c' ; d'} \sigma^{\beta'} g^{\phantom{a} \alpha'}_{a \phantom{\alpha'}}.
\end{multline}

\subsection{Transport equation for $\zeta = \ln \Delta^{1/2}$}
The Van Vleck-Morette determinant, $\Delta$ is a bi-scalar defined by
\begin{equation}
 \Delta \left(x,x'\right) \equiv \det \left[ \Delta^{\alpha'}_{\phantom{\alpha'} \beta'} \right], ~~~ \Delta^{\alpha'}_{\phantom{\alpha'} \beta'} \equiv -g^{\alpha'}_{\phantom{\alpha'} \alpha} \sigma^{\alpha}_{\phantom{\alpha} \beta'} = -g^{\alpha'}_{\phantom{\alpha'} \alpha} \eta^{\alpha}_{\phantom{\alpha} \beta'}
\end{equation}

By Eq.~(\ref{eq:eta-transport}), we can write the second equation here as:

\begin{equation}
 \Delta^{\alpha'}_{\phantom{\alpha'} \beta'} = -g^{\alpha'}_{\phantom{\alpha'} \alpha} \left( D' \eta^{\alpha}_{\phantom{\alpha} \beta'} + \eta^{\alpha}_{\phantom{\alpha} \gamma'} \xi^{\gamma'}_{\phantom{\gamma'} \beta'} \right)
\end{equation}

Since $ D' g^{\alpha'}_{\phantom{\alpha'} \alpha} = g^{\alpha'}_{\phantom{\alpha'} \alpha ; \beta'} \sigma^{\beta'} = 0$, we can rewrite this as

\begin{equation}
 \Delta^{\alpha'}_{\phantom{\alpha'} \beta'} = D' \Delta^{\alpha'}_{\phantom{\alpha'} \beta'} + \Delta^{\alpha'}_{\phantom{\alpha'} \gamma'} \xi^{\gamma'}_{\phantom{\gamma'} \beta'}
\end{equation}

Introducing the inverse $(\Delta^{-1})^{\alpha'}_{\phantom{\alpha'} \beta'}$ and multiplying it by the above, we obtain
\begin{equation}
\label{eq:vanVleck-transport}
 4 = \xi^{\alpha'}_{\phantom{\alpha'} \alpha'} + D'(\ln \Delta)
\end{equation}
where we have used the matrix identity $\delta \ln \det \mathbf{M} = \rm{Tr} \mathbf{M}^{-1} \delta \mathbf{M}$ to convert the trace to a determinant. This can also be written in terms of $\Delta^{1/2}$:
\begin{equation}
\label{eq:vanVleck-transport2}
 D'\zeta=  \frac{1}{2} \left(4 - \xi^{\alpha'}_{\phantom{\alpha'} \alpha'} \right)
\end{equation}

\subsection{Transport equation for the Van Vleck-Morette determinant, $\Delta^{1/2}$}
By the definition of $\zeta$, the Van Vleck-Morette determinant is given by
\begin{equation}
\label{eq:sqrtdelta-zeta}
 \Delta^{1/2} = e^{\zeta},
\end{equation}
and so satisfies the transport equation
\begin{equation}
 D' \Delta^{1/2} = \frac12 \Delta^{1/2} \left( 4 - \xi^{\alpha'}_{\phantom{\alpha'} \alpha'} \right).
\end{equation}

\subsection{Equation for $\Delta^{-1/2} D (\Delta^{1/2})$}
Defining $\tau = \Delta^{-1/2} D (\Delta^{1/2})$, it is immediately clear that
\begin{align}
\label{eq:tau-eq}
 \tau = \Delta^{-1/2} D (\Delta^{1/2}) = D \zeta .
\end{align}

\subsection{Equation for $\Delta^{-1/2} D' (\Delta^{1/2})$}
Defining $\tau' = \Delta^{-1/2} D' (\Delta^{1/2})$, it is immediately clear that
\begin{align}
\label{eq:tau-p-eq}
 \tau' = \Delta^{-1/2} D' (\Delta^{1/2}) = D' \zeta .
\end{align}

\subsection{Equation for $\nabla_{a'} \Delta$}
To derive an equation for $\nabla_{a'} \Delta$, we note that 
\begin{align}
\label{eq:VV-def}
 \Delta &\equiv \det \left[-g^{a'}_{\phantom{a'} \alpha} \eta^{\alpha}_{\phantom{\alpha} b'}\right] = - \det \left[ \eta^{a}_{\phantom{a} b'}\right] \det \left[g_{a}^{\phantom{a} a'} \right],
\end{align}
and make use of Jacobi's matrix identity
\begin{align}
\label{eq:d-det-A}
 \mathrm{d} \left(\det \mathbf{A}\right) &= \mathrm{tr} \left( \mathrm{adj} \left( \mathbf{A} \right) \mathrm{d} \mathbf{A} \right) \nonumber \\
 &= \left(\det \mathbf{A} \right) \mathrm{tr} \left( \mathbf{A}^{-1} \mathrm{d} \mathbf{A} \right)
\end{align}
where the operator $\mathrm{d}$ indicates a derivative. Applying \eqref{eq:d-det-A} to \eqref{eq:VV-def}, we obtain an equation for $\nabla_{a'}\Delta$:
\begin{equation}
\label{eq:nabla-delta}
 \nabla_{a'} \Delta = - \Delta \left[g_{\alpha'}^{\phantom{\alpha'} \alpha} g_{\alpha \phantom{\alpha'} ;a'}^{\phantom{\alpha} \alpha'} + \gamma^{\alpha'}_{\phantom{\alpha'} \alpha} \sigma^{\alpha}_{\phantom{\alpha} \alpha' a'} \right] .
\end{equation}
As a consistency check we note that contracting with $\sigma^{a'}$ and using Eq.~(\ref{eq:eta-transport}) we recover Eq.~(\ref{eq:vanVleck-transport}).

\subsection{Equation for $\Box' \Delta$}
Applying Jacobi's identity twice, together with $ \mathrm{d} (\mathbf{A}^{-1})=  -\mathbf{A}^{-1} (\mathrm{d} \mathbf{A})\mathbf{A}^{-1}$, we find an identity for the second derivative of the determinant of a matrix:
\begin{equation}
 \mathrm{d}^2\left(\det \mathbf{A}\right) = \left(\det \mathbf{A}\right) \left(\mathrm{tr}\left(\mathbf{A}^{-1} \mathrm{d}\mathbf{A}\right)\mathrm{tr}\left(\mathbf{A}^{-1} \mathrm{d}\mathbf{A}\right) - \mathrm{tr}\left(\mathbf{A}^{-1} \mathrm{d}\mathbf{A} \mathbf{A}^{-1} \mathrm{d}\mathbf{A}\right) + \mathrm{tr}\left(\mathbf{A}^{-1} \mathrm{d}^2 \mathbf{A}\right)\right).
\end{equation}
Using this identity in Eq.~\eqref{eq:VV-def}, we obtain an equation for $\Box'\Delta$,
\begin{multline}
\label{eq:box-delta}
 \Box'\Delta = \Delta \left[ \left(g_{\alpha'}^{\phantom{\alpha'} \alpha} g_{\alpha \phantom{\alpha'} ;\mu'}^{\phantom{\alpha} \alpha'} + \gamma^{\alpha'}_{\phantom{\alpha'} \alpha} \sigma^{\alpha}_{\phantom{\alpha} \alpha' \mu'} \right) \left(g_{\beta'}^{\phantom{\beta'} \beta} g_{\beta \phantom{\beta'} }^{\phantom{\beta} \beta'; \mu'} + \gamma^{\beta'}_{\phantom{\beta'} \beta} \sigma^{\beta \phantom{\beta'} \mu'}_{\phantom{\beta} \beta'} \right)
- \left( g_{\alpha'}^{\phantom{\alpha'} \alpha} g_{\alpha \phantom{\beta'} ;\mu'}^{\phantom{\alpha} \beta'} g_{\beta'}^{\phantom{\beta'} \beta} g_{\beta}^{\phantom{\beta} \alpha' ; \mu'} \right) 
\right. \\
 ~\left.
 - \left( \gamma^{\alpha'}_{\phantom{\alpha'} \alpha} \sigma^{\alpha}_{\phantom{\alpha} \beta' \mu'} \gamma^{\beta'}_{\phantom{\beta'} \beta} \sigma^{\beta \phantom{\alpha'} \mu'}_{\phantom{\beta} \alpha'}\right) 
 + \left( g_{\alpha'}^{\phantom{\alpha'} \alpha} g_{\alpha \phantom{\alpha'} ;\mu'}^{\phantom{\alpha} \alpha' \phantom{;\mu'} \mu'}\right) 
 + \left( \gamma^{\alpha'}_{\phantom{\alpha'} \alpha} \sigma^{\alpha \phantom{\alpha'} \mu'}_{\phantom{\alpha} \alpha' \phantom{\mu'} \mu'}\right) 
\right] .
\end{multline}

\subsection{Equation for $\Box' \Delta^{1/2}$}
Noting that
\begin{align}
 \Box'\Delta^{1/2} &= \left(\frac12 \Delta^{-1/2}\Delta_{;\mu'}\right)^{;\mu'} = \frac12 \Delta^{-1/2}\Box'\Delta - \frac14 \Delta^{-3/2}\Delta^{;\mu'}\Delta_{;\mu'},
\end{align}
it is straightforward to use Eqs.~\eqref{eq:nabla-delta} and \eqref{eq:box-delta} to find an equation for $\Box'\Delta^{1/2}$:
\begin{multline}
\label{eq:box-sqrt-delta}
\Box' \Delta^{1/2} = \frac12 \Delta^{1/2} \left[\frac12
 \left( g_{\alpha'}^{\phantom{\alpha'} \alpha} g_{\alpha \phantom{\alpha'} ;\mu'}^{\phantom{\alpha} \alpha'} + \gamma^{\alpha'}_{\phantom{\alpha'} \alpha} \sigma^{\alpha}_{\phantom{\alpha} \alpha' \mu'} \right)
 \left( g_{\alpha'}^{\phantom{\alpha'} \alpha} g_{\alpha}^{\phantom{\alpha} \alpha'  ;\mu'} + \gamma^{\alpha'}_{\phantom{\alpha'} \alpha} \sigma^{\alpha \phantom{\alpha'}  ;\mu'}_{\phantom{\alpha} \alpha'}\right)
 - \left( g_{\alpha'}^{\phantom{\alpha'} \alpha} g_{\alpha \phantom{\beta'} ;\mu'}^{\phantom{\alpha} \beta'} g_{\beta'}^{\phantom{\beta'} \beta} g_{\beta}^{\phantom{\beta} \alpha' ; \mu'} \right) 
\right. \\
 ~\left.
 - \left( \gamma^{\alpha'}_{\phantom{\alpha'} \alpha} \sigma^{\alpha}_{\phantom{\alpha} \beta' \mu'} \gamma^{\beta'}_{\phantom{\beta'} \beta} \sigma^{\beta \phantom{\alpha'} \mu'}_{\phantom{\beta} \alpha'}\right) 
 + \left( g_{\alpha'}^{\phantom{\alpha'} \alpha} g_{\alpha \phantom{\alpha'} ;\mu'}^{\phantom{\alpha} \alpha' \phantom{;\mu'} \mu'}\right) 
 + \left( \gamma^{\alpha'}_{\phantom{\alpha'} \alpha} \sigma^{\alpha \phantom{\alpha'} \mu'}_{\phantom{\alpha} \alpha' \phantom{\mu'} \mu'}\right) 
\right] .
\end{multline}

\subsection{Transport equation for $V_0$}
As is given in Eq.~\eqref{eq:recursionV0}, $V_0$ satisfies the transport equation
\begin{equation}
\label{eq:V0-transport}
 \left(D'+1\right) V^{AB'}_0 + \frac{1}{2} V^{AB'}_0 \left( \xi^{\mu'}_{\phantom{\mu'} \mu'} - 4 \right) + \frac{1}{2} \mathcal{D}^{B'}{}_{C'} (\Delta^{1/2} g^{AC'}) = 0 ,
\end{equation}
or equivalently
\begin{equation}
\label{eq:V0-transport-equiv}
\left(D'+1\right) \left(   \Delta^{-1/2} V^{AB'}_0\right) + \frac{1}{2}  \Delta^{-1/2} \mathcal{D}^{B'}{}_{C'} (\Delta^{1/2} g^{AC'}) = 0.
\end{equation}

In particular, for a scalar field:
\begin{equation}
\label{eq:V0-transport-scalar}
 \left(D'+1\right) V_0 + \frac{1}{2} V_0 \left( \xi^{\mu'}_{\phantom{\mu'} \mu'} - 4 \right) + \frac{1}{2} (\Box'-m^2-P') \Delta^{1/2} = 0,
\end{equation}
where $P'\equiv P(x')$ is frequently taken to be proportional to the Ricci scalar: $P= \xi R$.

\subsection{Transport equations for $V_r$}
As is given in Eq.~\eqref{eq:recursionVn}, $V_{r}$ satisfies the transport equation
\begin{equation}
\label{eq:Vr-transport}
 \left(D'+r+1\right) V^{AB'}_r + \frac{1}{2} V^{AB'}_r \left( \xi^{\mu'}_{\phantom{\mu'} \mu'} - 4 \right) + \frac{1}{2r} \mathcal{D}^{B'}{}_{C'} V^{AC'}_{r-1} = 0,
\end{equation}
or equivalently
\begin{equation}
\label{eq:Vr-transport-equiv}
\left(D'+r+1\right) \left(  \Delta^{-1/2} V^{AB'}_r\right) + \frac{1}{2r}  \Delta^{-1/2} \mathcal{D}^{B'}{}_{C'} V^{AC'}_{r-1} = 0.
\end{equation}
Comparing with Eq.~(\ref{eq:V0-transport-equiv}), it is clear that Eq.~(\ref{eq:Vr-transport-equiv})  may be taken to include $r=0$ if we replace 
$V^{AC'}_{r-1}/r$ by $\Delta^{1/2} g^{AC'}$.

In particular, for a scalar field:
\begin{equation}
\label{eq:Vr-transport-scalar}
 \left(D'+r+1\right) V_r + \frac{1}{2} V_r \left( \xi^{\mu'}_{\phantom{\mu'} \mu'} - 4 \right) + \frac{1}{2r} (\Box'-m^2-P')  V_{r-1} = 0.
\end{equation}

Together with the earlier equations, the transport equation Eq.~\eqref{eq:V0-transport}  allows us to immediately solve for $V^{AB'}_0$ along a geodesic. 
To obtain the higher order $V_r$ we also need to determine $\square' V_{r-1}^{AB'}$. At first sight this appears to require integrating along a
family of neighbouring geodesics but, in fact, again we can write transport equations for it.
First we note the identity
\begin{align}
\label{eq:recursive-transport}
\nabla_{a'} (D' T^{AB'}{}_{a_1'\dots a_n'}) &=D'( \nabla_{a'} T^{AB'}{}_{a_1'\dots a_n'})  + \xi^{\alpha'}{}_{a'}  \nabla_{\alpha'}T^{AB'}{}_{a_1'\dots a_n'}
 + \sigma^{\alpha'} \mathcal{R}^{B'}{}_{C'a'\alpha'}T^{AC'}{}_{a_1'\dots a_n'}               
\nonumber \\
&\qquad - \sigma^{\alpha'} R^{c'}{}_{a_1'a'\alpha'}T^{AC'}{}_{c'\dots a_n'}   - \dots - \sigma^{\alpha'} R^{c'}{}_{a_n'a'\alpha'}T^{AC'}{}_{a_1'\dots c'}       
\end{align}
where $\mathcal{R}^{A}{}_{Bcd} = \partial_c \mathcal{A}^A{}_{Bd} -  \partial_c \mathcal{A}^A{}_{Bd} + \mathcal{A}^A{}_{Cd}\mathcal{A}^C{}_{Bc}
-\mathcal{A}^A{}_{Cd}\mathcal{A}^C{}_{Bc}$.
Working with $\tilde V^{AB'}_r=\Delta^{-1/2} V^{AB'}_r$, on differentiating Eq.~(\ref{eq:Vr-transport-equiv}) we obtain
a transport equation for the first derivative of $V^{AB'}_r$
\begin{align}
(D'+r+1) (\tilde V_r^{AB'}{}_{;a'}) + \xi^{\alpha'}{}_{a'} \tilde V_r^{AB'}{}_{;\alpha'} + \sigma^{\alpha'} \mathcal{R}^{B'}{}_{C'a'\alpha'}\tilde V^{AC'}_r + \frac{1}{2r} \left ( \Delta^{-1/2} \mathcal{D}^{B'}{}_{C'} \left(  \Delta^{1/2}  \tilde V^{AC'}_{r-1}\right)\right)_{;a'} = 0 .
\end{align}
As noted above, this equation also includes $r=0$ if we replace $\tilde V^{AC'}_{r-1}/r$ in this case by $g^{AC'}$:
\begin{align}
(D'+1) (\tilde V_0^{AB'}{}_{;a'}) + \xi^{\alpha'}{}_{a'} \tilde V_0^{AB'}{}_{;\alpha'} + \sigma^{\alpha'} \mathcal{R}^{B'}{}_{C'a'\alpha'}\tilde V^{AC'}_0 + 
\frac{1}{2} \left (\Delta^{-1/2} \mathcal{D}^{B'}{}_{C'} \left( \Delta^{1/2}  g^{AC'}\right)\right)_{;a'} = 0 .
\end{align}
Repeating the process
\begin{align}
&(D'+r+1) (\tilde V_r^{AB'}{}_{;a'b'}) + \xi^{\alpha'}{}_{b'} \tilde V_r^{AB'}{}_{;a'\alpha'} + \xi^{\alpha'}{}_{a'} \tilde V_r^{AB'}{}_{;\alpha'b'}
\nonumber\\&\qquad
+ \sigma^{\alpha'} \mathcal{R}^{B'}{}_{C'b'\alpha'}V_r^{AC'}{}_{;a'}  
+ \sigma^{\alpha'} \mathcal{R}^{B'}{}_{C'a'\alpha'}\tilde V_r^{AC'}{}_{;b'} + \xi^{\alpha'}{}_{a';b'} \tilde V_r^{AB'}{}_{;\alpha'}
- \sigma^{\alpha'} R^{\beta'}{}_{a'b'\alpha'}\tilde V_r^{AC'}{}_{;\beta'}
\nonumber\\&\qquad
+ \xi^{\alpha'}{}_{b'} \mathcal{R}^{B'}{}_{C'a'\alpha'}\tilde V^{AC'}_r+ \sigma^{\alpha'} \mathcal{R}^{B'}{}_{C'a'\alpha';b'}\tilde V^{AC'}_r + \frac{1}{2r} \left (\Delta^{-1/2} \mathcal{D}^{B'}{}_{C'} \left(\Delta^{1/2}  \tilde V^{AC'}_{r-1}\right)\right)_{;a'b'} = 0 ,
\end{align}
with the $\tilde V_0^{AB'}{}_{;a'b'}$ equation given by the same replacement as above.

Clearly, this process may be repeated as many times as necessary.  At each stage, we require two more derivatives on $\tilde V^{AC'}_{r-1}$ than on 
$\tilde V^{AC'}_{r}$, but this may be obtained by a bootstrap process grounded by the $\tilde V^{AC'}_0$ equation which involves only 
the fundamental objects $\Delta^{1/2}$  and $ g^{AC'}$, which we have explored above.
As with our previous equations, while this process quickly becomes too tedious to follow by hand, it is straightforward to 
programme.

For example, to determine $V_1$ for a scalar field we first need to solve the two transport equations
\begin{align}
(D'+1) (\tilde V_{0;a'}) + \xi^{c'}{}_{a'} \tilde V_{0;c'}  + 
\frac{1}{2} \left(\Delta^{-1/2} \left(\Box'-m^2-P'\right) \Delta^{1/2} \right)_{;a'} = 0 ,
\end{align}
and 
\begin{align}
&(D'+1) (\tilde V_{0;a'b'}) + \xi^{c'}{}_{b'} \tilde V_{0;a'c'} + \xi^{c'}{}_{a'} \tilde V_{0;c'b'}+
\nonumber\\&\qquad
+ \xi^{c'}{}_{a';b'} \tilde V_{0;c'}
- \sigma^{c'} R^{d'}{}_{a'b'c'}\tilde V_{0;d'}
+ \frac{1}{2} \left (\Delta^{-1/2} \left(\Box'-m^2-P'\right) \Delta^{1/2} \right)_{;a'b'} = 0 .
\end{align}

In the next two sections we show how the above system of transport equations can be solved either as a series expansion or numerically. 
For sufficiently simple spacetimes, it is also possible to find closed form solutions which provide a 
useful check on our results.

\section{Semi-recursive Approach to Covariant Expansions} \label{sec:symbolic}
In this section, we will investigate solutions to the transport equations of Sec.~\ref{sec:avramidi} in the form of covariant series expansions. The goal is to find covariant series expressions for the Hadamard and DeWitt coefficients. Several methods have been previously applied for doing such calculations, both by hand and using computer algebra \cite{Gilkey:1979,Gilkey:1980,Barvinsky:1985,Fulling:1990,Fulling:HeatKernel,Belkov:1993,Gusynin:1994,Belkov:1996,
Avramidi:Schimming:1996,Booth:1998,Fliegner:1998,Ven:1998,Dowker:1999,Gusynin:1999,Salcedo:2001,Salcedo:2004,Gayral:2006,
Anselmi:2007,Salcedo:2007,Matyjasek:Tryniecki:Zwierzchowska:2010}. However, this effort has been focused primarily on the calculation of the \emph{diagonal} coefficients. To our knowledge, only the work of D\'ecanini and Folacci \cite{Decanini:Folacci:2005a}, upon which our method is based, has been concerned with the off-diagonal coefficients.

Before proceeding further, it is helpful to see how covariant expansions behave under certain operations. First, applying the operator $D'$ to the covariant expansion of any bi-tensor $T_{a_1 \cdots a_m b_1' \cdots b_n'}$ about the point $x$, we obtain
\begin{align}
 g_{b_1}{}^{b_1'} \cdots  &g_{b_n}{}^{b_n'}  D' T_{a_1 \cdots a_m b_1' \cdots b_n'} (x,x') = D' g_{b_1}{}^{b_1'} \cdots  g_{b_n}{}^{b_n'} T_{a_1 \cdots a_m b_1' \cdots b_n'} (x,x') \nonumber\\
	&= \sum_{k=0}^\infty \frac{(-1)^k}{k!} k\,t_{a_1 \cdots a_m b_1 \cdots b_n \, \beta_1 \cdots \beta_k} (x) \sigma^{\beta_1} \cdots \sigma^{\beta_k}_{\phantom{\beta_k} \alpha'} \sigma^{\alpha'} = \sum_{k=0}^\infty \frac{(-1)^k}{k!} k\,t_{a_1 \cdots a_m b_1 \cdots b_n \, \beta_1 \cdots \beta_k} (x) \sigma^{\beta_1} \cdots \sigma^{\beta_k}
\end{align}
where the last equality is obtained by applying Eq.~(\ref{eq:sigma-eta-xi}). In other words, applying $D'$ to the $k$-th term in the series is equivalent to multiplying that term by $k$:
\begin{align*}
(D'T)_{(k)} = k T_{(k)} .
\end{align*}

Next we consider applying the operator $D$ to the covariant expansion of any bi-tensor $T_{a_1 \cdots a_m a_1' \cdots a_n'}$ about the point $x$. In this case, there will also be a term involving the derivative of the series coefficient, giving 
\begin{align}
 g_{b_1}{}^{b_1'} \cdots  &g_{b_n}{}^{b_n'} D  T_{a_1 \cdots a_m b_1' \cdots b_n'} (x,x') =D g_{b_1}{}^{b_1'} \cdots  g_{b_n}{}^{b_n'} T_{a_1 \cdots a_m b_1' \cdots b_n'} (x,x') \nonumber\\
	&= \sum_{k=0}^\infty \frac{(-1)^k}{k!} \left[ k\,t_{a_1 \cdots a_m b_1 \cdots b_n \, \beta_1 \cdots \beta_k} (x) \sigma^{\beta_1} \cdots \sigma^{\beta_k}_{\phantom{\beta_k} \alpha} \sigma^{\alpha} + t_{a_1 \cdots a_m b_1 \cdots b_n \, \beta_1 \cdots \beta_k ; \alpha} (x) \sigma^{\beta_1} \cdots \sigma^{\beta_k} \sigma^{\alpha}\right] \nonumber \\
	&= \sum_{k=0}^\infty \frac{(-1)^k}{k!} \left[ k\,t_{a_1 \cdots a_m b_1 \cdots b_n \, \beta_1 \cdots \beta_k} (x) \sigma^{\beta_1} \cdots \sigma^{\beta_k}+ t_{a_1 \cdots a_m b_1 \cdots b_n \, \beta_1 \cdots \beta_k ; \alpha} (x) \sigma^{\beta_1} \cdots \sigma^{\beta_k} \sigma^{\alpha}\right] .
\end{align}

We can also consider multiplication of covariant expansions. For example, for any two tensors, $S^{a}_{\phantom{a} b}$ and $T^{a}_{\phantom{a} b}$, with product $U^{a}_{\phantom{a} b} \equiv S^{a}_{\phantom{a}\alpha}T^{\alpha}_{\phantom{\alpha} b}$, say, we can relate their covariant expansions by
\begin{align}
 \sum_{n=0}^{\infty} \frac{(-1)^n}{n!}  u^{a}_{\phantom{a} b \, \beta_1 \cdots \beta_n} \sigma^{\beta_1} \cdots \sigma^{\beta_n} &=\left(\sum_{k=0}^{\infty} \frac{(-1)^k}{k!} s^{a}_{\phantom{a} \alpha \, \beta_1 \cdots \beta_k} \sigma^{\beta_1} \cdots \sigma^{\beta_k}\right) \left(\sum_{l=0}^{\infty} \frac{(-1)^l}{l!} t^{\alpha}_{\phantom{\alpha} b \, \beta_1 \cdots \beta_l} \sigma^{\beta_1} \cdots \sigma^{\beta_l}\right)\nonumber\\
	&= \sum_{n=0}^{\infty} \frac{(-1)^n}{n!} \sum_{k=0}^{n} \binom{n}{k} s^{a}_{\phantom{a} \alpha \, \beta_1 \cdots \beta_k} t^{\alpha}_{\phantom{\alpha} b \, \beta_{k+1} \cdots \beta_n} \sigma^{\beta_1} \cdots \sigma^{\beta_n},
\end{align}
or equivalently
\begin{align*}
U_{(n)} = \sum_{k=0}^{n} \binom{n}{k} S_{(k)} T_{(n-k)}.
\end{align*}

Finally, many of the equations derived in the previous section contain terms involving the Riemann tensor at $x'$, $R^{a'}_{\phantom{a'} b' c' d'}$. As all other quantities are expanded about $x$ rather than $x'$, we will also need to rewrite these Riemann terms in terms of their expansion about $x$:
\begin{align}
g^{a}{}_{a'}g_{b}{}^{b'} R^{a'}_{\phantom{a'} \alpha' b' \beta'} \sigma^{\alpha'} \sigma^{\beta'} &=  \sum_{k=0}^{\infty} \frac{(-1)^k}{k!} R^{a}_{\phantom{a} (\alpha |b| \beta ; \gamma_1 \cdots \gamma_k)} \sigma^{\alpha} \sigma^{\beta} \sigma^{\gamma_1} \cdots \sigma^{\gamma_k}\nonumber \\
	&=  \sum_{k=2}^{\infty} \frac{(-1)^k}{(k-2)!} \mathcal{K}^{a}_{\phantom{a} b \, (k)},
\end{align}
where we follow Avramidi \cite{Avramidi:1986} in introducing the definition
\begin{align}
\label{eq:Kdef}
 \mathcal{K}^{a}_{\phantom{a} b \,(n)} &\equiv R^{a}_{\phantom{a} (\alpha_1 |b| \alpha_2 ; \alpha_3 \cdots \alpha_n)} \sigma^{\alpha_1} \cdots \sigma^{\alpha_n}\nonumber \\
  &\equiv \bar{\mathcal{K}}^{a}_{\phantom{a} b \,  (n)}  \sigma^{\alpha_1} \cdots \sigma^{\alpha_n}.
\end{align}

These four considerations will now allow us to rewrite the transport equations of Sec.~\ref{sec:avramidi} as recursion relations for the coefficients of the covariant expansions of the tensors involved.

\subsection{Recursion relation for coefficients of the covariant expansion of  $\gamma^{a'}{}_{b}$}
Rewriting Eq.~(\ref{eq:gamma-transport}) in terms of covariant expansions, we find
\begin{align}
  \sum_{n=0}^\infty \frac{(-1)^n}{n!} &(n^2+n)\, g^a{}_{a'} \gamma^{a'}_{\phantom{a'} b \, \alpha_1 \cdots \alpha_n} (x) \sigma^{\alpha_1} \cdots \sigma^{\alpha_n} \nonumber \\
&  +   \left( \sum_{k=2}^{\infty} \frac{(-1)^k}{(k-2)!} \mathcal{K}^{a}_{\phantom{\alpha} \beta \, (k)} \right) 
  \left(\sum_{l=0}^\infty \frac{(-1)^l}{l!} g^{\beta}{}_{\beta'} \gamma^{\beta'}_{\phantom{\beta'} b \, \alpha_1 \cdots \alpha_l} (x) \sigma^{\alpha_1} \cdots \sigma^{\alpha_l}\right) = 0.
\end{align}
From this, the $n$-th term in the covariant series expansion of $g^a{}_{a'}\gamma^{a'}_{\phantom{a} b}$,
\begin{align*}
g^a{}_{a'}\gamma^{a'}_{\phantom{a} b} = \sum_{n=0}^{\infty} \frac{(-1)^n}{n!} \gamma^{a}_{\phantom{a} b \, (n)} ,
\end{align*}
 can be written recursively in terms of products of lower order terms in the series with $\mathcal{K}$:
\begin{align}
\label{eq:gamma_coefs}
   \gamma^{a}_{\phantom{a} b \,(0)} =& -\delta^{a}_{\phantom{a} b} ,&\qquad  \gamma^{a}_{\phantom{a} b \,(1)} =& \, 0 ,&
\gamma^{a}_{\phantom{a} b \, (n)} =& -\left(\frac{n-1}{n+1}\right) \sum_{k=0}^{n-2} \binom{n-2}{k}  \mathcal{K}^{a}_{\phantom{\alpha} \alpha \, (n-k)}  \gamma^{\alpha}_{\phantom{\beta'} b \, (k)}.
\end{align}
Many of the following recursion relations will make use of these coefficients; to illustrate their structure we write the next five explicitly,
\begin{gather*}
\gamma^{a}_{\phantom{a} b \,(2)} = \frac{1}{3} \mathcal{K}^{a}_{\phantom{a} b\,(2)} ,\qquad \gamma^{a}_{\phantom{a} b \,(3)} = \frac{1}{2} \mathcal{K}^{a}_{\phantom{a} b\,(3)}, \qquad \gamma^{a}_{\phantom{a} b \,(4)} = \frac{3}{5} \mathcal{K}^{a}_{\phantom{a} b\,(4)} -\frac{1}{5} \mathcal{K}^{a}_{\phantom{a} \alpha\,(2)} \mathcal{K}^{\alpha}_{\phantom{\alpha} b\,(2)}, \nonumber \\
\gamma^{a}_{\phantom{a} b \,(5)} = \frac{2}{3} \mathcal{K}^{a}_{\phantom{a} b\,(5)} -\frac{2}{3} \mathcal{K}^{a}_{\phantom{a} \alpha\,(3)} \mathcal{K}^{\alpha}_{\phantom{\alpha} b\,(2)}-\frac{1}{3} \mathcal{K}^{a}_{\phantom{a} \alpha\,(2)} \mathcal{K}^{\alpha}_{\phantom{\alpha} b\,(3)}, \nonumber \\
\gamma^{a}_{\phantom{a} b \,(6)} = \frac{5}{7} \mathcal{K}^{a}_{\phantom{a} b\,(6)} -\frac{10}{7} \mathcal{K}^{a}_{\phantom{a} \alpha\,(4)}
\mathcal{K}^{\alpha}_{\phantom{\alpha} b\,(2)}-\frac{10}{7} \mathcal{K}^{a}_{\phantom{a} \alpha\,(3)} \mathcal{K}^{\alpha}_{\phantom{\alpha} b\,(3)}-\frac{3}{7} \mathcal{K}^{a}_{\phantom{a} \alpha\,(2)} \mathcal{K}^{\alpha}_{\phantom{\alpha} b\,(4)}
+\frac{1}{7} \mathcal{K}^{a}_{\phantom{a} \alpha\,(2)} \mathcal{K}^{\alpha}_{\phantom{\alpha} \beta\,(2)}\mathcal{K}^{\beta}_{\phantom{\beta} b\,(2)} .
\end{gather*}
While one can give a closed form combinatoric expression, the recursive formula (\ref{eq:gamma_coefs}) is best suited to our needs.

\subsection{Recursion relation for coefficients of the covariant expansion of  $\eta^{a}_{~b'}$}
Since $\gamma^{a'}_{\phantom{a'} b}$ is the inverse of $\eta^{a}_{\phantom{a} b'}$, we have 
\begin{equation}
(g^a{}_{a'} \gamma^{a'}_{\phantom{a'} \alpha})( g_b{}^{b'}  \eta^{\alpha}_{\phantom{\alpha} b'}) = \delta^{a}_{\phantom{a} b}.
\end{equation}
Substituting in covariant expansion expressions for $g^a{}_{a'} \gamma^{a'}_{\phantom{a'} \alpha}$ and $g_b{}^{b'} \eta^{\alpha}_{\phantom{\alpha} b'}$, we find, using our standard notation,  that the $n$-th term in the covariant series expansion of $g_b{}^{b'} \eta^{a}_{\phantom{a} b'}$ is
\begin{align}
\label{eq:eta_coefs}
\eta^{a}_{\phantom{a} b \,(0)} &= -\delta^{a}_{\phantom{a} b} , &
 \eta^{a}_{\phantom{a} b \,(1)} &= 0, &\eta^{a}_{\phantom{a} b \,(n)} &= \sum_{k=2}^n \binom{n}{k}  \gamma^{\alpha}_{\phantom{\alpha} \beta\, (k)} \eta^{\beta}_{\phantom{\beta} b\,(n-k)}.
\end{align}
Again, to illustrate their structure we write the next five explicitly,
\begin{gather*}
\eta^{a}_{\phantom{a} b \,(2)} = -\frac{1}{3} \mathcal{K}^{a}_{\phantom{a} b\,(2)} ,\qquad \eta^{a}_{\phantom{a} b \,(3)} = -\frac{1}{2} \mathcal{K}^{a}_{\phantom{a} b\,(3)}, \qquad \eta^{a}_{\phantom{a} b \,(4)} = -\frac{3}{5} \mathcal{K}^{a}_{\phantom{a} b\,(4)} -\frac{7}{15} \mathcal{K}^{a}_{\phantom{a} \alpha\,(2)} \mathcal{K}^{\alpha}_{\phantom{\alpha} b\,(2)}, \nonumber \\
\eta^{a}_{\phantom{a} b \,(5)} = -\frac{2}{3} \mathcal{K}^{a}_{\phantom{a} b\,(5)} -\mathcal{K}^{a}_{\phantom{a} \alpha\,(3)} \mathcal{K}^{\alpha}_{\phantom{\alpha} b\,(2)}-\frac{4}{3} \mathcal{K}^{a}_{\phantom{a} \alpha\,(2)} \mathcal{K}^{\alpha}_{\phantom{\alpha} b\,(3)}, \nonumber \\
\eta^{a}_{\phantom{a} b \,(6)} = -\frac{5}{7} \mathcal{K}^{a}_{\phantom{a} b\,(6)} -\frac{11}{7} \mathcal{K}^{a}_{\phantom{a} \alpha\,(4)}
\mathcal{K}^{\alpha}_{\phantom{\alpha} b\,(2)}-\frac{25}{7} \mathcal{K}^{a}_{\phantom{a} \alpha\,(3)} \mathcal{K}^{\alpha}_{\phantom{\alpha} b\,(3)}-\frac{18}{7} \mathcal{K}^{a}_{\phantom{a} \alpha\,(2)} \mathcal{K}^{\alpha}_{\phantom{\alpha} b\,(4)}
-\frac{31}{21} \mathcal{K}^{a}_{\phantom{a} \alpha\,(2)} \mathcal{K}^{\alpha}_{\phantom{\alpha} \beta\,(2)}\mathcal{K}^{\beta}_{\phantom{\beta} b\,(2)} .
\end{gather*}

\subsection{Recursion relation for coefficients of the covariant expansion of  $\xi^{a'}_{~b'}$}
Writing Eq.~(\ref{eq:xi-solve-eta-gamma}) in terms of covariant series, it is immediately apparent that the $n$-th term in the covariant expansion of $ g^{a}{}_{a'}  g_{b}{}^{b'} \xi^{a'}_{\phantom{a'} b'}$ is 
\begin{gather}
\label{eq:xi_coefs}
 \xi^{a}_{\phantom{a} b \,(0)} = \delta^{a}_{\phantom{a} b}, \qquad 
\xi^{a}_{\phantom{a} b \,(1)} = 0\nonumber \\
\xi^{a}_{\phantom{a} b \,(n)} = n\,\eta^{a}_{\phantom{\alpha} b \,(n)} - \sum_{k=2}^{n-2} \binom{n}{k} k\, \gamma^{a}_{\phantom{a} \alpha\, (n-k)}   \eta^{\alpha}_{\phantom{\alpha} b \,(k)}.
\end{gather}
Once more, to illustrate their structure we write the next five explicitly,
\begin{gather*}
\xi^{a}_{\phantom{a} b \,(2)} = -\frac{2}{3} \mathcal{K}^{a}_{\phantom{a} b\,(2)} ,\qquad \xi^{a}_{\phantom{a} b \,(3)} = -\frac{3}{2} \mathcal{K}^{a}_{\phantom{a} b\,(3)}, \qquad \xi^{a}_{\phantom{a} b \,(4)} = -\frac{12}{5} \mathcal{K}^{a}_{\phantom{a} b\,(4)} -\frac{8}{15} \mathcal{K}^{a}_{\phantom{a} \alpha\,(2)} \mathcal{K}^{\alpha}_{\phantom{\alpha} b\,(2)}, \nonumber \\
\xi^{a}_{\phantom{a} b \,(5)} = -\frac{10}{3} \mathcal{K}^{a}_{\phantom{a} b\,(5)} -\frac{5}{3} \mathcal{K}^{a}_{\phantom{a} \alpha\,(3)} \mathcal{K}^{\alpha}_{\phantom{\alpha} b\,(2)}-\frac{5}{3} \mathcal{K}^{a}_{\phantom{a} \alpha\,(2)} \mathcal{K}^{\alpha}_{\phantom{\alpha} b\,(3)}, \nonumber \\
\xi^{a}_{\phantom{a} b \,(6)} = -\frac{30}{7} \mathcal{K}^{a}_{\phantom{a} b\,(6)} -\frac{24}{7} \mathcal{K}^{a}_{\phantom{a} \alpha\,(4)}
\mathcal{K}^{\alpha}_{\phantom{\alpha} b\,(2)}-\frac{45}{7} \mathcal{K}^{a}_{\phantom{a} \alpha\,(3)} \mathcal{K}^{\alpha}_{\phantom{\alpha} b\,(3)}-\frac{24}{7} \mathcal{K}^{a}_{\phantom{a} \alpha\,(2)} \mathcal{K}^{\alpha}_{\phantom{\alpha} b\,(4)}
-\frac{32}{21} \mathcal{K}^{a}_{\phantom{a} \alpha\,(2)} \mathcal{K}^{\alpha}_{\phantom{\alpha} \beta\,(2)}\mathcal{K}^{\beta}_{\phantom{\beta} b\,(2)} .
\end{gather*}

\subsection{Recursion relation for coefficients of the covariant expansion of  $\lambda^{a}_{~b}$}
Using Eq.~(\ref{eq:lambda-eq}), $ \lambda^{a}_{\phantom{a} b} = \delta^{a}_{\phantom{a} b} - (D \eta^{a}_{\phantom{a} \alpha'}) \gamma^{\alpha'}_{\phantom{\alpha'} b}$, we can write an equation for the $n$-th order coefficient of the covariant expansion of $\lambda^{a}_{\phantom{a} b}$. However, the expression involves the operator $D$ acting on the covariant series expansion of $g_b{}^{b'}\eta^{a}_{\phantom{a} b'}$, so we will first need to find an expression for that. As discussed in the beginning of this section, the derivative in $D$ will affect both the coefficient and the $\sigma^{a}$'s. When acting on the $\sigma^{a}$'s, it has the effect of multiplying the term by $n$ as was previously the case with $D'$. When acting on the coefficient, it will add a derivative to it and increase the order of the term (since we will then be adding a $\sigma^{a}$). 
In particular, given our definition \eqref{eq:Kdef}
\begin{align*}
 \mathcal{K}^{a}_{\phantom{a} b \, (n)} \equiv R^{a}_{\phantom{a} (\alpha_1 |b| \alpha_2 ; \alpha_3 \cdots \alpha_n)} \sigma^{\alpha_1} \cdots \sigma^{\alpha_n}
\end{align*}
we have 
\begin{align*}
D \mathcal{K}^{a}_{\phantom{a} b \, (n)} &= \sigma^{\alpha_{n+1}} \nabla_{\alpha_{n+1}} ( R^{a}_{\phantom{a} (\alpha_1 |b| \alpha_2 ; \alpha_3 \cdots \alpha_n)} \sigma^{\alpha_1} \cdots \sigma^{\alpha_n}  )\\
 &= R^{a}_{\phantom{a} (\alpha_1 |b| \alpha_2 ; \alpha_3 \cdots \alpha_n)\alpha_{n+1} } \sigma^{\alpha_1} \cdots \sigma^{\alpha_n}\sigma^{\alpha_{n+1}}+ n R^{a}_{\phantom{a} (\alpha_1 |b| \alpha_2 ; \alpha_3 \cdots \alpha_n)} \sigma^{\alpha_1} \cdots \sigma^{\alpha_n} \\
&= \mathcal{K}^{a}_{\phantom{a} b \, (n+1)} + n \mathcal{K}^{a}_{\phantom{a} b \, (n)}  .
\end{align*}
Here the first term is one order higher while the second keeps the order the same.

We now appeal to the fact that, from Eqs.~(\ref{eq:gamma_coefs}) and (\ref{eq:xi_coefs}), the terms in the expansion of $g_b{}^{b'}\eta^{a}_{\phantom{a} b'}$ consist solely of products of $\mathcal{K}^{a}_{\phantom{a} b \, (n)}$. This means that we can apply the preceding rules when $D$ acts on $\mathcal{K}^{a}_{\phantom{a} b \, (n)}$, and when encountering compound expressions (i.e., consisting of more than a single $\mathcal{K}^{a}_{\phantom{a} b \, (n)}$), use the normal rules for differentiation (product rule, distributivity, etc.).
To illustrate this explicitly
\begin{align*}
g_b{}^{b'}D\eta^{a}_{\phantom{a} b'}&=D(g_b{}^{b'}\eta^{a}_{\phantom{a} b'}) = D\left( \eta^{a}_{\phantom{a} b \,(0)} 
- \eta^{a}_{\phantom{a} b \,(1)}+\frac{1}{2!}\eta^{a}_{\phantom{a} b \,(2)}
-\frac{1}{3!}\eta^{a}_{\phantom{a} b \,(3)}+\frac{1}{4!}\eta^{a}_{\phantom{a} b \,(4)}+\cdots \right)\\
&= D\left(-\delta^{a}_{\phantom{a} b} +\frac{1}{2!}(-\frac{1}{3} \mathcal{K}^{a}_{\phantom{a} b\,(2)} )-\frac{1}{3!}(-\frac{1}{2} \mathcal{K}^{a}_{\phantom{a} b\,(3)})+\frac{1}{4!}(-\frac{3}{5} \mathcal{K}^{a}_{\phantom{a} b\,(4)} -\frac{7}{15} \mathcal{K}^{a}_{\phantom{a} \alpha\,(2)} \mathcal{K}^{\alpha}_{\phantom{\alpha} b\,(2)})+\cdots \right)\\
&= \frac{1}{2!}\left(-\frac{1}{3} \right)(\mathcal{K}^{a}_{\phantom{a} b\,(3)} +2\mathcal{K}^{a}_{\phantom{a} b\,(2)} )-\frac{1}{3!}\left(-\frac{1}{2}\right)( \mathcal{K}^{a}_{\phantom{a} b\,(4)}+3\mathcal{K}^{a}_{\phantom{a} b\,(3)})+\frac{1}{4!}\left(-\frac{3}{5}\right)( \mathcal{K}^{a}_{\phantom{a} b\,(5)} +4\mathcal{K}^{a}_{\phantom{a} b\,(4)})\\
&\qquad +\frac{1}{4!}\left(-\frac{7}{15} \right)\left((\mathcal{K}^{a}_{\phantom{a} \alpha\,(3)}+2\mathcal{K}^{a}_{\phantom{a} \alpha\,(2)} )\mathcal{K}^{\alpha}_{\phantom{\alpha} b\,(2)}+\mathcal{K}^{a}_{\phantom{a} \alpha\,(2)}(\mathcal{K}^{\alpha}_{\phantom{\alpha} b\,(3)}+2\mathcal{K}^{\alpha}_{\phantom{\alpha} b\,(2)})\right)+\cdots \\
&= \frac{1}{2!}\left(-\frac{2}{3} \mathcal{K}^{a}_{\phantom{a} b\,(2)}\right)  
- \frac{1}{3!}\left(- \frac{1}{2} \mathcal{K}^{a}_{\phantom{a} b\,(3)} \right)
+ \frac{1}{4!} \left(-\frac{2}{5} \mathcal{K}^{a}_{\phantom{a} b\,(4)}-\frac{28}{15}\mathcal{K}^{a}_{\phantom{a} \alpha\,(2)}\mathcal{K}^{\alpha}_{\phantom{\alpha} b\,(2)}\right)+\cdots
\end{align*}
We can then  write the general $n$-th term in the covariant series expansion of $g_b{}^{b'}D\eta^{a}_{\phantom{a} b'}$ symbolically as
\begin{equation}
 (D\eta^{a}_{\phantom{a} b} )_{(n)} = D^0  \eta^{a}_{\phantom{a} b \,(n)} - n D^+  \eta^{a}_{\phantom{a} b \,(n-1)}
\end{equation}
where $D^+$ signifies the contribution that raises the order by one and $D^0$ signifies the contribution that keeps the order the same. For example,
\begin{IEEEeqnarray*}{rCl?rCl}
D^+ \eta^{a}_{\phantom{a} b \,(2)} &=& -\frac{1}{3} \mathcal{K}^{\alpha}_{\phantom{\alpha} b\,(3)}, &
D^0 \eta^{a}_{\phantom{a} b \,(3)} &=& -\frac{3}{2} \mathcal{K}^{\alpha}_{\phantom{\alpha} b\,(3)}
\end{IEEEeqnarray*}
and so
\begin{equation*}
(D\eta^{a}_{\phantom{a} b} )_{(3)} = -\frac{3}{2} \mathcal{K}^{\alpha}_{\phantom{\alpha} b\,(3)} - 3\left(-\frac{1}{3} \mathcal{K}^{\alpha}_{\phantom{\alpha} b\,(3)}\right) = -\frac{1}{2} \mathcal{K}^{\alpha}_{\phantom{\alpha} b\,(3)}.
\end{equation*}
It is then straightforward to write an expression for the $n$-th term in the covariant series expansion of $\lambda^{a}_{\phantom{a} b}$:
\begin{gather}
  \lambda^{a}_{\phantom{a} b \,(0)} = \delta^{a}_{\phantom{a} b}, \qquad 
\lambda^{a}_{\phantom{a} b \,(1)} = 0\nonumber \\
\lambda^{a}_{\phantom{a} b \, (n)} = - \sum_{k=0}^{n-2} \binom{n}{k} \left( D^0 \eta^{a}_{\phantom{a} \alpha \,(n-k)} - (n-k) D^+ \eta^{a}_{\phantom{a} \alpha \,(n-k-1)} \right) \gamma^{\alpha}_{\phantom{\alpha'} b \, (k)} .
\end{gather}
The next five terms are given explicitly by
\begin{gather*}
\lambda^{a}_{\phantom{a} b \,(2)} = -\frac{2}{3} \mathcal{K}^{a}_{\phantom{a} b\,(2)} ,\qquad \lambda^{a}_{\phantom{a} b \,(3)} = -\frac{1}{2} \mathcal{K}^{a}_{\phantom{a} b\,(3)}, \qquad \lambda^{a}_{\phantom{a} b \,(4)} = -\frac{2}{5} \mathcal{K}^{a}_{\phantom{a} b\,(4)} -\frac{8}{15} \mathcal{K}^{a}_{\phantom{a} \alpha\,(2)} \mathcal{K}^{\alpha}_{\phantom{\alpha} b\,(2)}, \nonumber \\
\lambda^{a}_{\phantom{a} b \,(5)} = -\frac{1}{3} \mathcal{K}^{a}_{\phantom{a} b\,(5)} - \mathcal{K}^{a}_{\phantom{a} \alpha\,(3)} \mathcal{K}^{\alpha}_{\phantom{\alpha} b\,(2)}- \mathcal{K}^{a}_{\phantom{a} \alpha\,(2)} \mathcal{K}^{\alpha}_{\phantom{\alpha} b\,(3)}, \nonumber \\
\lambda^{a}_{\phantom{a} b \,(6)} = -\frac{2}{7} \mathcal{K}^{a}_{\phantom{a} b\,(6)} -\frac{10}{7} \mathcal{K}^{a}_{\phantom{a} \alpha\,(4)}
\mathcal{K}^{\alpha}_{\phantom{\alpha} b\,(2)}-\frac{17}{7} \mathcal{K}^{a}_{\phantom{a} \alpha\,(3)} \mathcal{K}^{\alpha}_{\phantom{\alpha} b\,(3)}-\frac{10}{7} \mathcal{K}^{a}_{\phantom{a} \alpha\,(2)} \mathcal{K}^{\alpha}_{\phantom{\alpha} b\,(4)}
-\frac{32}{21} \mathcal{K}^{a}_{\phantom{a} \alpha\,(2)} \mathcal{K}^{\alpha}_{\phantom{\alpha} \beta\,(2)}\mathcal{K}^{\beta}_{\phantom{\beta} b\,(2)} .
\end{gather*}

\subsection{Recursion relation for coefficients of the covariant expansion of $A_{a b c}$}
We can rewrite Eq.~(\ref{eq:A-transport}) as
\begin{equation}
 (D'+1)(A_{a b \alpha} g^\alpha{}_{\alpha'}\gamma^{\alpha'}_{\phantom{\alpha} c}) + g_a{}^{\alpha'}g_b{}^{\beta'}R_{\alpha' \beta' \alpha \beta} \sigma^{\alpha} \gamma^{\beta}_{\phantom{\beta} c} = 0,
\end{equation}
which when rewritten in terms of covariant series becomes
\begin{equation}
 A_{a b c \, (k)} = -\frac{1}{n+1} \sum_{k=0}^{n} \binom{n}{k} k \,\mathcal{R}_{a b \alpha \, (k)}  \gamma^{\alpha}_{\phantom{\alpha} c \, (n-k)} + \sum_{k=0}^{n-2} \binom{n}{k} A_{a b \alpha \, (k)}\gamma^{\alpha}_{\phantom{\alpha} c \, (n-k)}
\end{equation}
where we follow Avramidi \cite{Avramidi:1986,Avramidi:2000} in defining
\begin{equation}
 \mathcal{R}_{a b c \, (n)} \equiv R_{a b (\alpha_1 |c| ; \alpha_2 \cdots \alpha_n)} \sigma^{\alpha_1} \cdots \sigma^{\alpha_n}.
\end{equation}

Alternatively, writing Eq.~(\ref{eq:A-transport}) directly in terms of covariant series, we obtain
\begin{equation}
 A_{a b c \, (k)} = \frac{n}{n+1} \mathcal{R}_{a b c \, (n)} - \frac{1}{n+1} \sum_{k=0}^{n-2} \binom{n}{k} A_{a b \alpha \, (k)} \xi^{\alpha}_{\phantom{\alpha} c \, (n-k)},
\end{equation}
which has the benefit of requiring half as much computation as the previous expression.

\subsection{Recursion relation for coefficients of the covariant expansion of $B_{a b c}$}
By Eq.~(\ref{eq:B-transport}), we can immediately write an equation for the coefficients of the covariant expansion of $B_{a b c}$:
\begin{equation}
 B_{a b c \, (n)} = \frac{1}{n} \sum_{k=0}^{n} \binom{n}{k} A_{a b \alpha \, (k)} \eta^{\alpha}_{\phantom{\alpha} c \, (n-k)}.
\end{equation}

\subsection{Covariant expansion of $\zeta$}
From Eq.~(\ref{eq:vanVleck-transport2}) we immediately obtain expressions for the coefficients of the covariant series of $\zeta$:
\begin{align}
 \zeta_{(0)} &=0 & \zeta_{(1)} &=0 & \zeta_{(n)} = -\frac{1}{2n} \xi^{\rho}_{\phantom{\rho} \rho\,(n)}
\end{align}

\subsection{Recursion relation for $\Delta^{1/2}$, $\Delta^{-1/2}$}
Since $\zeta = \ln{\Delta^{1/2}}$, we can write
\begin{equation}
 \Delta^{1/2} D' \zeta = D' \Delta^{1/2}.
\end{equation}
This allows us to write down a recursive equation for the coefficients of the covariant series expansion of $\Delta^{1/2}$,
\begin{equation}
 \Delta^{1/2}_{(n)} = \frac{1}{n} \sum_{k=2}^{n} \binom{n}{k} k\, \zeta_{(k)} \Delta^{1/2}_{(n-k)}.
\end{equation}
Similarly, the equation
\begin{equation}
 - \Delta^{-1/2} D' \zeta = D' \Delta^{-1/2}.
\end{equation}
allows us to write down a recursive equation for the coefficients of the covariant series expansion of $\Delta^{-1/2}$,
\begin{equation}
 \Delta^{-1/2}_{(n)} = - \frac{1}{n} \sum_{k=2}^{n} \binom{n}{k} k\, \zeta_{(k)} \Delta^{-1/2}_{(n-k)}.
\end{equation}

\subsection{Covariant expansion of $\tau$ and $\tau'$}
Equations~\eqref{eq:tau-eq} and \eqref{eq:tau-p-eq} may be immediately written as covariant series equations,
\begin{equation}
 \tau_{(n)} = D^0 \zeta_{(n)} - n D^+ \zeta_{(n-1)}, \qquad  \tau'_{(n)} = n \zeta_{(n)}.
\end{equation}

\subsection{Covariant expansion of covariant derivative at $x'$ of a bi-scalar}
Let $T(x,x')$ be a general bi-scalar. Writing $T$ as a covariant series,
\begin{equation}
 T(x,x') = \sum_{n=0}^{\infty} \frac{(-1)^n}{n!} T_{(n)} = \sum_{n=0}^{\infty} \frac{(-1)^n}{n!} t_{\alpha_1 \cdots \alpha_n} (x)\sigma^{\alpha_1} \cdots \sigma^{\alpha_n},
\end{equation}
and applying a covariant derivative at $x'$, we obtain
\begin{align}
\label{eq:cd-covex}
 g_a{}^{\alpha'} T_{;\alpha'} &= \sum_{n=0}^{\infty} \frac{(-1)^n}{n!}  g_a{}^{\alpha'}  T_{(n) ;\alpha'} \nonumber \\
	 &= \sum_{n=0}^{\infty} \frac{(-1)^n}{n!} n\, t_{(\alpha_1 \cdots \alpha_n)} \sigma^{\alpha_1} \cdots \sigma^{\alpha_{n-1}}  g_a{}^{\alpha'} \sigma^{\alpha_n}_{\phantom{\alpha_n} \alpha'}\nonumber \\
         &= \sum_{n=0}^{\infty} \frac{(-1)^n}{n!} n\, t_{(\alpha_1 \cdots \alpha_{n-1} \rho)} \sigma^{\alpha_1} \cdots \sigma^{\alpha_{n-1}}  g_a{}^{\alpha'}\eta^{\rho}_{\phantom{\rho} \alpha'} \nonumber \\
	 &= - \sum_{n=0}^{\infty} \frac{(-1)^n}{n!}  \sum_{k=0}^{n} \binom{n}{k} (T_{(k+1)})_{(-1) \,\rho} \eta^\rho{}_{a\,(n-k)}
\end{align}
where we have introduced the notation
\begin{equation}
 (T_{(n)})_{(-k)\, \alpha_{(n-k+1)} \cdots \alpha_{n}} \equiv t_{(\alpha_1 \cdots \alpha_{(n-k)} \alpha_{(n-k+1)} \cdots \alpha_{n})} \sigma^{\alpha_1} \cdots \sigma^{\alpha_{n-k}}.
\end{equation}

\subsection{Covariant expansion of d'Alembertian at $x'$ of a bi-scalar}
Let $T(x,x')$ be a general bi-scalar as in the previous section. Applying \eqref{eq:cd-covex} twice and taking care to include the term involving $g_{a}{}^{b'}$, we can then write the d'Alembertian, $\Box' T(x,x')$ at $x'$ in terms of covariant series,
\begin{equation}
\label{eq:box-covex}
 (\Box' T)_{(n)} =  -\sum_{k=0}^{n} \binom{n}{k} ((g_\alpha{}^{\alpha'} T_{;\alpha'})_{ (k+1)})_{(-1)\,\rho} \eta^{\rho \alpha}{}_{(n-k)} -\sum_{k=1}^{n} \binom{n}{k} A^{\alpha \rho}{}_{\rho\,(k)} (g_\alpha{}^{\alpha'} T_{;\alpha'})_{ (n-k)},
\end{equation}
where $A_{abc \, (n)}$ is the $n$-th term in the covariant series of the tensor defined in \eqref{eq:A-transport}.

\subsection{Covariant expansion of $\nabla_{a'} \Delta^{1/2}$}
Applying Eq.~(\ref{eq:cd-covex}) to the case $T=\Delta^{1/2}$, we obtain
\begin{equation}
 (g_a{}^{a'} \Delta^{1/2}_{;a'})_{(n)}= -\sum_{k=0}^{n} \binom{n}{k} (\Delta^{1/2}_{(k+1)})_{(-1) \,\rho} \eta^\rho{}_{a\,(n-k)}.
\end{equation}

\subsection{Covariant expansion of $\Box' \Delta^{1/2}$}
Applying Eq.~(\ref{eq:box-covex}) to the case $T=\Delta^{1/2}$, we obtain
\begin{equation}
(\Box'  \Delta^{1/2})_{(n)} =  -\sum_{k=0}^{n} \binom{n}{k} ((g_\alpha{}^{\alpha'}  \Delta^{1/2}_{;\alpha'})_{ (k+1)})_{(-1)} \eta^{\rho \alpha}{}_{(n-k)} -\sum_{k=1}^{n} \binom{n}{k}  A^{\alpha \rho}{}_{\rho\,(k)} (g_\alpha{}^{\alpha'}  \Delta^{1/2}_{;\alpha'})_{ (n-k)}
\end{equation}

\subsection{Covariant expansion of $V_0$}
The transport equation for $V_0$, Eq.~(\ref{eq:V0-transport}), can be written in the alternative form
\begin{equation}
 \left(D'+1\right) V_0 - V_0 \tau' + \frac{1}{2} (\Box'-m^2 - P') \Delta^{1/2} = 0.
\end{equation}
This equation is then easily written in terms of covariant expansion coefficients,
\begin{equation}
 V_{0\,(n)} = \frac{1}{n+1}\left( \sum_{k=0}^{n-2} \binom{n}{k} V_{0 (k)} \tau'_{(n-k)} - \frac{1}{2}\left((\Box '\Delta^{1/2})_{(n)} - m^2  \Delta^{1/2}_{(n)}  - \sum_{k=0}^{n} \binom{n}{k} P_{(k)}  \Delta^{1/2}_{(n-k)} \right)\right)
\end{equation}

\subsection{Covariant expansion of $V_r$}
The transport equation for $V_r$, Eq.~(\ref{eq:Vr-transport}) can also be written in the alternative form
\begin{equation}
 \left(D'+r+1\right) V_r - V_r \tau' + \frac{1}{2r} (\Box'-m^2 - P') V_{r-1} = 0.
\end{equation}
Again, this is easily written in terms of covariant expansion coefficients,
\begin{equation}
 V_{r\,(n)} = \frac{1}{r+n+1}\left( \sum_{k=0}^{n-2} \binom{n}{k} V_{r\, (k)} \tau'_{(n-k)} - \frac{1}{2r}\left((\Box' V_{r-1})_{(n)} - m^2  V_{r-1\,(n)} - \sum_{k=0}^{n} \binom{n}{k} P_{(k)}  V_{r-1\,(n-k)} \right)\right).
\end{equation}

\subsection{Results}
We have implemented the semi-recursive algorithm as a \emph{Mathematica} package which we are making freely available online \cite{AvramidiCode}. It serves as an efficient tool for easily computing high order covariant expansions. The high level of efficiency is illustrated in Tables \ref{table:a-times} and \ref{table:v0-times}, where we show the performance of our implementation when run on a desktop computer (2.4GHz processor). For each coefficient, we list the time, number of terms and memory consumed in the calculation of that term. We also list the number of terms after reduction to canonical form by the Invar \cite{Invar1,Invar2} package. Note that Invar currently only supports canonicalization of scalar invariants up to order $6$. We have therefore not canonicalized our expressions for $a_7$, $a_8$ and $a_9$, nor our expressions for the non-diagonal coefficients given in Table~\ref{table:v0-times}. We have also not canonicalized our expression for $a_6$, primarily due to memory constraints.

The expressions for the DeWitt coefficients produced by our code are valid for any spacetime of any dimension.  These may in turn be used to construct the Hadamard coefficients in any dimension, although we have limited ourselves to the $4$-dimensional case here. Given our motivation to study massless fields in vacuum spacetimes such as Schwarzschild and Kerr, it is possible to make further assumptions in order to reduce the number of terms which appear. It is straightforward to impose the fact that the field is massless and the Ricci tensor vanishes with the requirements
\begin{IEEEeqnarray}{rClrClrCl}
\mathcal{K}^\alpha{}_{\alpha\,(n)} &=& 0, \qquad & \mathcal{R}_{a}{}^{\beta}{}_{\beta\,(n)} = 0, \qquad & m &=& 0.
\end{IEEEeqnarray}
This is a conservative requirement: terms such as $tr\left(\left(\mathcal{K}^a{}_{b\,(3)}\right)_{(-2)}\right)$ will yield some terms involving a Ricci tensor after the symmetrization is explicitly expanded. However, as is shown in Tables~\ref{table:a-times} and \ref{table:v0-times}, it is sufficient to significantly reduce the number of terms in the expansions.

\begin{table}[htb]
 \begin{tabular}{|c||c|c|c|c|c||c|c|c|c|}
\hline
  \textbf{DeWitt} & \multicolumn{5}{c||}{\textbf{General}} & \multicolumn{4}{c|}{\textbf{Vacuum}}  \\ \cline{2-10}
  \textbf{Coefficient} & \textbf{Time} & \textbf{Terms} & \textbf{Memory} & \textbf{Canonical} & \textbf{Canonical (P=0)} & \textbf{Time} & \textbf{Terms} & \textbf{Memory} & \textbf{Canonical}\\
\hline
\hline
 $a_0$ & $0$ & $1$ & $16 \,\rm B$ & $1$ & $1$ & $0$ & $1$ & $16 \,\rm B$ & $1$\\ \hline
 $a_1$ & $0$ & $2$ & $432 \,\rm B$ & $2$ & $1$ & $0$ & $0$ & $16 \,\rm B$ & $0$\\ \hline
 $a_2$ & $0.003$ & $10$ & $5 \,\rm kB$ & $7$ & $4$ & $0$ & $2$ & $536 \,\rm B$ & $1$\\ \hline
 $a_3$ & $0.02$ & $91$ & $63 \,\rm kB$ & $26$ & $15$ & $0.003$ & $7$ & $5 \,\rm kB$ & $2$\\ \hline
 $a_4$ & $0.2$ & $1\,058$ & $949 \,\rm kB$ & $113$ & $68$ & $0.015$ & $56$ & $51 \,\rm kB$ & $5$\\ \hline
 $a_5$ & $3.6$ & $13\,972$ & $15 \,\rm MB$ & $611$ & $380$ & $0.1$ & $507$ & $559 \,\rm kB$ & $14$\\ \hline
 $a_6$ & $76$ & $199\,264$ & $254 \,\rm MB$ & - & - & $1.1$ & $4\,988$ & $6.3 \,\rm MB$ & $-$\\ \hline
 $a_7$ & $1489$ & $2\,987\,366$ & $4.4 \,\rm GB$ & - & - & $17$ & $51\,700$ & $75 \,\rm MB$ & $-$\\ \hline
 $a_8$ & - & - & - & - & - & $254$ & $554\,715$ & $910 \,\rm MB$ & $-$\\ \hline
 $a_9$ & - & - & - & - & - & $3373$ & $6\,098\,069$ & $10.9 \,\rm GB$ & $-$\\ \hline
 \end{tabular}
\caption{Calculation performance of our semi-recursive implementation of the Avramidi method for computing the coincident (diagonal) DeWitt coefficients, $a_{r\,(0)}$ for both general and vacuum spacetimes. In each case, we list the computation time (in seconds), number of terms, memory consumed and number of terms after canonicalization. In the general case, we also list the number of canonical terms when the potential $P$ is set to $0$. This would be the case, for example, for a minimally coupled scalar field.}
\label{table:a-times}
\end{table}

\begin{table}[htb]
 \begin{tabular}{|c||c|c|c||c|c|c|}
\hline
  \multirow{2}{*}{\textbf{Order}} & \multicolumn{3}{c||}{\textbf{General}} & \multicolumn{3}{c|}{\textbf{Vacuum, massless}} \\ \cline{2-7}
   & \textbf{Time} & \textbf{Terms} & \textbf{Memory} & \textbf{Time} & \textbf{Terms} & \textbf{Memory} \\
\hline
\hline
 $0$ & $0.001$ & $3$ & $760\,\rm B$ & $0.001$ & $0$ & $16\,\rm B$\\ \hline
 $1$ & $0.001$ & $2$ & $288\,\rm B$ & $0.001$ & $0$ & $16\,\rm B$\\ \hline
 $2$ & $0.002$ & $10$ & $3.8\,\rm kB$ & $0.001$ & $2$ & $432\,\rm B$\\ \hline
 $3$ & $0.003$ & $15$ & $6.1\,\rm kB$ & $0.002$ & $2$ & $432\,\rm B$\\ \hline
 $4$ & $0.005$ & $47$ & $22.0\,\rm kB$ & $0.003$ & $5$ & $2.5\,\rm kB$\\ \hline
 $5$ & $0.007$ & $81$ & $40.7\,\rm kB$ & $0.004$ & $7$ & $3.6\,\rm kB$\\ \hline
 $6$ & $0.014$ & $206$ & $112\,\rm kB$ & $0.009$ & $22$ & $12\,\rm kB$ \\ \hline
 $7$ & $0.024$ & $383$ & $221\,\rm kB$ & $0.015$ & $39$ & $23\,\rm kB$ \\ \hline
 $8$ & $0.047$ & $856$ & $526\,\rm kB$ & $0.019$ & $94$ & $59\,\rm kB$\\ \hline
 $9$ & $0.084$ & $1\,641$ & $1.03\,\rm MB$ & $0.03$ & $177$ & $115\,\rm kB$\\ \hline
 $10$ & $0.16$ & $3\,414$ & $2.25\,\rm MB$ & $0.05$ & $384$ & $260\,\rm kB$\\ \hline
 $11$ & $0.30$ & $6\,547$ & $4.51\,\rm MB$ & $0.1$ & $729$ & $515\,\rm kB$\\ \hline
 $12$ & $0.58$ & $13\,064$ & $9.34\,\rm MB$ & $0.19$ & $1480$ & $1.1\,\rm MB$ \\ \hline
 $13$ & $1.1$ & $24\,870$ & $18.5\,\rm MB$ & $0.33$ & $2811$ & $2.1\,\rm MB$ \\ \hline
 $14$ & $2.1$ & $48\,167$ & $37.1\,\rm MB$ & $0.61$ & $5485$ & $4.2\,\rm MB$ \\ \hline
 $15$ & $4.1$ & $90\,808$ & $72.3\,\rm MB$ & $1.1$ & $10\,320$ & $8.3\,\rm MB$ \\ \hline
 $16$ & $7.8$ & $172\,214$ & $141\,\rm MB$ & $2.1$ & $19\,637$ & $16\,\rm MB$ \\ \hline
 $17$ & $15$ & $321\,145$ & $271\,\rm MB$ & $3.7$ & $36\,556$ & $30\,\rm MB$\\ \hline
 $18$ & $28$ & $599\,460$ & $522\,\rm MB$ & $6.8$ & $68\,295$ & $58\,\rm MB$ \\ \hline
 $19$ & $53$ & $1\,106\,459$ & $987\,\rm MB$  & $12$ & $125\,852$ & $110\,\rm MB$\\ \hline
 $20$ & $99$ & $2\,039\,285$ & $1.81\,\rm GB$ & $23$ & $231\,837$ & $208\,\rm MB$\\ \hline
 \end{tabular}
\caption{Calculation performance of the Avramidi method for computing the terms $V_{0\,(n)}$ of order $n$ in the covariant series expansion of $V_0$ for both general ($4$ dimensional) and vacuum spacetimes. In each case, we list the computation time (in seconds), number of terms and memory consumed.}
\label{table:v0-times}
\end{table}

The relative compactness of our expressions after canonicalization means that they may be readily computed for a given choice of spacetime. For example, evaluating the expressions for the coincidence limits, $V_r(x,x)=V_{r(0)}(x)$, of the first five Hadamard coefficients given in the Appendix for Schwarzschild spacetime gives:
\begin{gather}
V_{0 (0)} = 0, \quad
V_{1 (0)} = \frac{M^2}{15r^6}, \quad
V_{2 (0)} = \frac{M^2}{1008r^9} (194 M - 81 r), \quad
V_{3 (0)} = \frac{M^2}{3150r^{12}} ( 210 r^2 - 1125 r M + 1454 M^2), \nonumber \\
V_{4 (0)} = \frac{-M^2}{1663200 r^{15} } (78750 r^3-1746182 M^3+1932801 r M^2-689775 M r^2).
\end{gather}
A similar calculation can be done for spacetimes with less symmetry (such as Kerr) without any additional difficulty other than the fact that the results are somewhat less compact.

\section{Numerical Solution of Transport Equations} \label{sec:numerical}
 In this section, we describe the implementation of the numerical solution 
of the transport equations of Sec.~\ref{sec:avramidi}.
We use the analytic results for $\sigma$, $\Delta^{1/2}$, $g_a{}^{b'}$ and $V_0$ for a scalar field in Nariai spacetime from Refs.~\cite{Nolan:2009} and \cite{Wardell:PhD} as a check on our numerical code.

For the purposes of numerical calculations, the operator $\mathcal{D}'$ acting on a general bi-tensor $T^{a' \ldots}_{\phantom{a' \ldots} b' \ldots}$ can be written as
\begin{equation}
 \mathcal{D}' T^{a' \ldots}_{\phantom{a' \ldots} b' \ldots} = (s'-s) \left( \frac{d}{ds'}T^{a' \ldots}_{\phantom{a' \ldots} b' \ldots} + T^{\alpha' \ldots}_{\phantom{\alpha' \ldots} b' \ldots} \Gamma^{a'}_{\alpha' \beta'} u^{\beta'} + \cdots - T^{a' \ldots}_{\phantom{a' \ldots} \alpha' \ldots} \Gamma^{\alpha'}_{b' \beta'} u^{\beta'} - \cdots\right),
\end{equation}
where $s'$ is the affine parameter, $\Gamma^{a'}_{b' c'}$ are the Christoffel symbols at $x'$ and $u^{a'}$ is the four velocity at $x'$. Additionally, we make use of the fact that
\begin{equation}
 \sigma^{a'} = (s'- s) u^{a'}.
\end{equation}
which allows us to write Eqs.~\eqref{eq:xi-transport}, \eqref{eq:eta-transport}, \eqref{eq:transport-sigma-ppp}, \eqref{eq:transport-sigma-upp}, \eqref{eq:sigma-pppp-transport}, \eqref{eq:sigma-uppp-transport}, \eqref{eq:bi-transport}, \eqref{eq:A-transport}, \eqref{eq:d2Iinv-transport}, \eqref{eq:vanVleck-transport2}, \eqref{eq:box-sqrt-delta} and \eqref{eq:V0-transport} as a system of coupled, tensor ordinary differential equations. These equations all have the general form:
\begin{equation}
\label{eq:numeric-transport}
 \frac{d}{ds'}T^{a' \ldots}_{\phantom{a' \ldots} b' \ldots} = (s')^{-1} A^{a' \ldots}_{\phantom{\alpha'} b' \ldots} + B^{a' \ldots}_{\phantom{\alpha'} b' \ldots} + s' C^{a' \ldots}_{\phantom{\alpha'} b' \ldots} - T^{\alpha' \ldots}_{\phantom{\alpha'} b' \ldots} \Gamma^{a'}_{\alpha' \beta'} u^{\beta'} - \cdots + T^{a' \ldots}_{\phantom{a' \ldots} \alpha' \ldots} \Gamma^{\alpha'}_{b' \beta'} u^{\beta'} + \cdots,
\end{equation}
where we have set $s=0$ without loss of generality and where $A^{a' \ldots}_{\phantom{\alpha'} b' \ldots} = 0 $ initially (i.e., at $s'=0$). It is not necessarily true, however, that the derivative of $A^{\alpha' \ldots}_{\phantom{\alpha'} b' \ldots}$ is zero initially. This fact is important when considering initial data for the numerical scheme.

Solving this system of equations along with the geodesic equations for the spacetime of interest will then yield a numerical value for $V_0$. Moreover, since $V=V_0$ along a null geodesic, the transport equation for $V_0$ will effectively give the full value of $V$ on the light-cone. We have implemented this numerical integration scheme for geodesics in Nariai and Schwarzschild spacetimes using the Runge-Kutta-Fehlberg method (with adaptive time stepping) provided by the GNU Scientific Library \cite{GSL}. The source code of our implementation is available online \cite{TransportCode}.

\subsection{Initial Conditions}
Numerical integration of the transport equations requires initial conditions for each of the bi-tensors involved. These initial conditions are readily obtained by considering the covariant series expansions of $V_0$, $\Delta^{1/2}$, $\xi^{a'}_{\phantom{a'} b'}$, $\eta^{a}_{\phantom{a} b'}$ and $g^{\phantom{a} b'}_{a}$ and their covariant derivatives at $x'$. Initial conditions for all bi-tensors used for calculating $V_0$ are given in Table~\ref{table:init}, where we list the transport equation for the bi-tensor, the bi-tensor itself and its initial value.

Additionally, as is indicated in Eq.~\eqref{eq:numeric-transport}, many of the transport equations will contain terms involving $(s')^{-1}$. These terms must obviously be treated with care in any numerical implementation. Then, for the initial time step ($s'=0$), we require analytic expressions for
\begin{equation}
\label{eq:numerical-limit}
 \lim_{s'\to 0} (s')^{-1} A^{\alpha' \ldots}_{\phantom{\alpha'} b' \ldots}
\end{equation}
which may then be used to numerically compute an accurate initial value for the derivative. This limit can be computed from the first order term in the covariant series of $A^{\alpha' \ldots}_{\phantom{\alpha'} b' \ldots}$, which is found most easily by considering the covariant series of its constituent bi-tensors. For this reason, we also list in Table~\ref{table:init} the limit as $s'\to 0$ of all required constituent bi-tensors multiplied by $(s')^{-1}$. In Table~\ref{table:init-transport} we list the terms $(s')^{-1} A^{\alpha' \ldots}_{\phantom{\alpha'} b' \ldots}$ for each transport equation involving $(s')^{-1}$, along with their limit as $s'\to 0$.

\begin{table}[htb]
\renewcommand{\arraystretch}{2}
 \begin{tabular}{|c||c|c|c|}
\hline
  \textbf{Equation} & \textbf{Bi-tensor} & \textbf{Initial Condition} & \textbf{$(s')^{-1}$ Initial Condition}\\
\hline
\hline
 \eqref{eq:xi-transport} & $\xi^{a'}_{\phantom{a'} b'}$ & $\delta^{a}_{\phantom{a} b}$ & $0$\\ \hline
 \eqref{eq:eta-transport} & $\eta^{a}_{\phantom{a} b'}$ & $-\delta_{\phantom{a} b}^{a}$ & $0$\\ \hline
 \eqref{eq:transport-sigma-ppp} & $\sigma^{a'}_{\phantom{a'} b' c'}$ & $0$ & $-\frac23 R^{a}_{\phantom{a} (\alpha |b| c)} u^{\alpha}$ \\ \hline 
 \eqref{eq:transport-sigma-upp} & $\sigma^{a}_{\phantom{a} b' c'}$ & $0$ & $\frac12 R^{a}_{\phantom{a} b \alpha c} u^{\alpha} - \frac13 R^{a}_{\phantom{a} (\alpha |b| c)} u^{\alpha}$\\ \hline 
 \eqref{eq:sigma-pppp-transport} & $\sigma^{a'}_{\phantom{a'} b' c' d'}$ & $-\frac23 R^{a}_{\phantom{a}  (c | b | d)}$ & $\frac12 R^{a}_{\phantom{a} (c |b| d ; \alpha)}u^{\alpha} - \frac23 R^{a}_{\phantom{a} (\alpha |b| d) ;c} u^{\alpha} - \frac23 R^{a}_{\phantom{a} (\alpha |b| c) ;d} u^{\alpha}$\\ \hline
 \eqref{eq:sigma-uppp-transport} & $\sigma^{a}_{\phantom{a} b' c' d'}$ & $-\frac13 R^{a}_{\phantom{a}  (c | b | d)} -\frac12  R^{a}_{\phantom{a}  b  c  d}$ & $-\frac12 R^{a}_{\phantom{a} (c |b| d ; \alpha)}u^{\alpha} + \frac23 R^{a}_{\phantom{a} (\alpha |b| d) ;c} u^{\alpha}$ \\ \hline
 \eqref{eq:bi-transport} & $g_{a'}^{\phantom{a'} b}$ & $\delta_{a}^{\phantom{a} b}$ & 0\\ \hline
 \eqref{eq:A-transport} & $g_{a \phantom{b'} ; c'}^{\phantom{a} b'}$ & $0$ & $\frac12 R^{b}_{\phantom{b} a \alpha c} u^{\alpha}$ \\ \hline
 \eqref{eq:d2Iinv-transport} & $g_{a \phantom{b'} ; c' d'}^{\phantom{a} b'}$ & $-\frac12 R^{b}_{\phantom{b} a c d}$ & $-\frac23 R^{b}_{\phantom{b} a c (d ; \alpha)} u^{\alpha}$ \\ \hline
 \eqref{eq:vanVleck-transport2} & $\Delta^{1/2}$ & $1$ & 0 \\ \hline
 \eqref{eq:V0-transport} & $V_0$ & $\frac12 m^2 + \frac12 \left(\xi-\frac16\right) R$ & $-\frac14 \left(\xi-\frac16\right) R_{;\alpha} u^{\alpha}$\\
\hline
 \end{tabular}
\caption{Initial conditions for tensors used in the numerical calculation of $V_0$ with $P=\xi R$.}
\label{table:init}
\end{table}

\begin{table}[htb]
\renewcommand{\arraystretch}{2}
 \begin{tabular}{|c||c|c|}
\hline
  \textbf{Equation} & \textbf{Terms involving $(s')^{-1}$} & \textbf{Initial condition for $(s')^{-1}$ terms} \\
\hline
\hline
 \eqref{eq:xi-transport} & $-(s')^{-1}\left(\xi^{a'}_{\phantom{a'} \alpha'}\xi^{\alpha'}_{\phantom{\alpha'} b'} - \xi^{a'}_{\phantom{a'} b'}\right)$ & $0$ \\ \hline
 \eqref{eq:eta-transport} & $-(s')^{-1}\left(\eta^{a}_{\phantom{a} \alpha'}\xi^{\alpha'}_{\phantom{\alpha'} b'} - \eta^{a}_{\phantom{a} b'}\right)$ & $0$ \\ \hline
 \eqref{eq:transport-sigma-ppp} & $(s')^{-1} \left( \sigma^{a'}_{\phantom{a'} b' c'} - \xi^{a'}_{\phantom{a'} \alpha'} \sigma^{\alpha'}_{\phantom{\alpha'} b' c'} - \xi^{\alpha'}_{\phantom{\alpha'} b'} \sigma^{a'}_{\phantom{a'} \alpha' c'} - \xi^{\alpha'}_{\phantom{\alpha'} c'} \sigma^{a'}_{\phantom{a'} \alpha' b'}\right)$ & $-\frac23 R^{a}_{\phantom{a} (b |\alpha| c)} u^{\alpha}$ \\\hline 
 \eqref{eq:transport-sigma-upp} & $(s')^{-1} \left( \sigma^{a}_{\phantom{a} b' c'} - \eta^{a}_{\phantom{a} \alpha'} \sigma^{\alpha'}_{\phantom{\alpha'} b' c'} - \xi^{\alpha'}_{\phantom{\alpha'} b'} \sigma^{a}_{\phantom{a} \alpha' c'} - \xi^{\alpha'}_{\phantom{\alpha'} c'} \sigma^{a}_{\phantom{a} \alpha' b'}\right)$ & $-\frac12 R^{a}_{\phantom{a} \alpha b c} u^{\alpha} -\frac13 R^{a}_{\phantom{a} (b |\alpha| c)} u^{\alpha}$ \\ \hline 
 \eqref{eq:sigma-pppp-transport} & \multicolumn{1}{|l|}{$(s')^{-1}\left( \sigma^{a'}_{\phantom{a'} b' c' d'} - \sigma^{a'}_{\phantom{a'} \alpha' b'}\sigma^{\alpha'}_{\phantom{\alpha'} c' d'}- \sigma^{a'}_{\phantom{a'} \alpha' c'}\sigma^{\alpha'}_{\phantom{\alpha'} b' d'} - \sigma^{a'}_{\phantom{a'} \alpha' d'}\sigma^{\alpha'}_{\phantom{\alpha'} b' c'}\right.$} & $-\frac32 R^{a}_{\phantom{a} (b |\alpha| c; d)} u^{\alpha}$\\
 & \multicolumn{1}{|r|}{$\left. - \sigma^{a'}_{\phantom{a'} \alpha' b' c'} \xi^{\alpha'}_{\phantom{\alpha'} d'} - \sigma^{a'}_{\phantom{a'} \alpha' b' d'} \xi^{\alpha'}_{\phantom{\alpha'} c'} - \sigma^{a'}_{\phantom{a'} \alpha' c' d'} \xi^{\alpha'}_{\phantom{\alpha'} b'} - \sigma^{\alpha'}_{\phantom{\alpha'} b' c' d'} \xi^{a'}_{\phantom{a'} \alpha'} \right)$} & \\ \hline
 \eqref{eq:sigma-uppp-transport} & \multicolumn{1}{|l|}{$(s')^{-1}\left( \sigma^{a}_{\phantom{a} b' c' d'} - \sigma^{a}_{\phantom{a} \alpha' b'}\sigma^{\alpha'}_{\phantom{\alpha'} c' d'}- \sigma^{a}_{\phantom{a} \alpha' c'}\sigma^{\alpha'}_{\phantom{\alpha'} b' d'} - \sigma^{a}_{\phantom{a} \alpha' d'}\sigma^{\alpha'}_{\phantom{\alpha'} b' c'} \right.$} & $\frac12 R^{a}_{\phantom{a} (c |\alpha| d) ;b}u^{\alpha} - \frac56 R^{a}_{\phantom{a} b \alpha (c ; d)}u^{\alpha} + \frac76 R^{a}_{\phantom{a} (d |\alpha b| ;c)}u^{\alpha}$ \\
 & \multicolumn{1}{|r|}{$\left. - \sigma^{a}_{\phantom{a} \alpha' b' c'} \xi^{\alpha'}_{\phantom{\alpha'} d'} - \sigma^{a}_{\phantom{a} \alpha' b' d'} \xi^{\alpha'}_{\phantom{\alpha'} c'} - \sigma^{a}_{\phantom{a} \alpha' c' d'} \xi^{\alpha'}_{\phantom{\alpha'} b'} - \sigma^{\alpha'}_{\phantom{\alpha'} b' c' d'} \eta^{a}_{\phantom{a} \alpha'} \right)$} &  \\ \hline
 \eqref{eq:bi-transport} & $0$ & $0$ \\ \hline
 \eqref{eq:A-transport} & $-(s')^{-1} g_{a \phantom{b'} ; \alpha'}^{\phantom{a} b'} \xi^{\alpha'}_{\phantom{\alpha'} c'}$ & $-\frac12 R^{b}_{\phantom{b} a \alpha c} u^{\alpha}$\\ \hline
 \eqref{eq:d2Iinv-transport} & $-(s')^{-1} \left(g_{a \phantom{b'} ; \alpha' d'}^{\phantom{a} b'} \xi^{\alpha'}_{\phantom{\alpha'} c'} + g_{a \phantom{b'} ; \alpha' c'}^{\phantom{a} b'} \xi^{\alpha'}_{\phantom{\alpha'} d'} + g_{a \phantom{b'} ; \alpha'}^{\phantom{a} b'} \sigma^{\alpha'}_{\phantom{\alpha'} c' d'}\right)$ & $-\frac23 R^{b}_{\phantom{b} a \alpha (c ; d)} u^{\alpha}$ \\ \hline
 \eqref{eq:vanVleck-transport2} & $-(s')^{-1} \Delta^{1/2} \left(\xi^{a'}_{\phantom{a'} a'} - \delta^{a'}_{\phantom{a'} a'}\right)$ & $0$ \\ \hline
 \eqref{eq:V0-transport} & $-(s')^{-1} \left[\left(\xi^{a'}_{\phantom{a'} a'} - \delta^{a'}_{\phantom{a'} a'}\right) V_0 + 2 V_0 + (\Box' -m^2 - \xi R)\Delta^{1/2} \right]$ & $\frac14 \left(\xi-\frac16\right) R_{;\alpha} u^{\alpha}$ \\
\hline
 \end{tabular}
\caption{Initial conditions for transport equations required for the numerical calculation of $V_0$  with $P=\xi R$.}
\label{table:init-transport}
\end{table}

\subsection{Results}
\begin{figure}
\begin{center}
 \includegraphics[width=7.2cm]{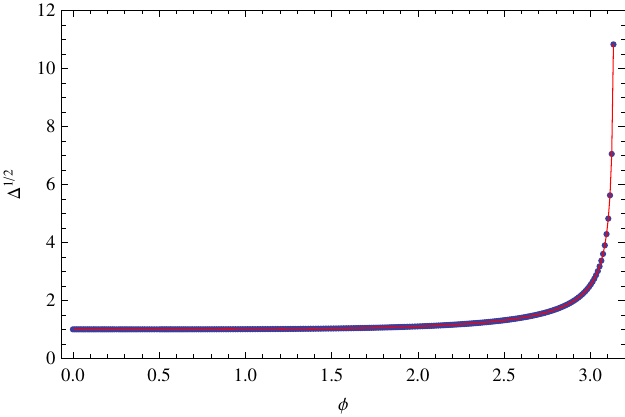}
 \includegraphics[width=8cm]{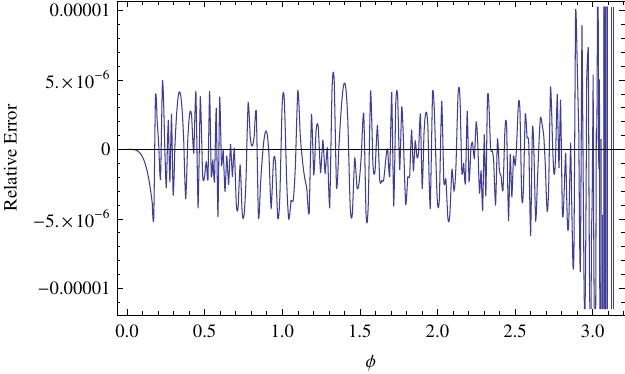}
\end{center}
\caption{\emph{Comparison of numerical and exact analytic calculations of $\Delta^{1/2}$ as a function of the angle, $\phi$, along an orbiting null geodesic in Nariai spacetime.} Left: The numerical calculation (blue dots) is a close match with the analytic expression (red line). Right: With parameters so that the code completes in $1$ minute, the relative error is within $0.0001\%$ up to the boundary of the normal neighborhood (at $\phi=\pi$).}
\label{fig:nariai-transport-VV-comparison}
\end{figure}

\begin{figure}
\begin{center}
 \includegraphics[width=7.9cm]{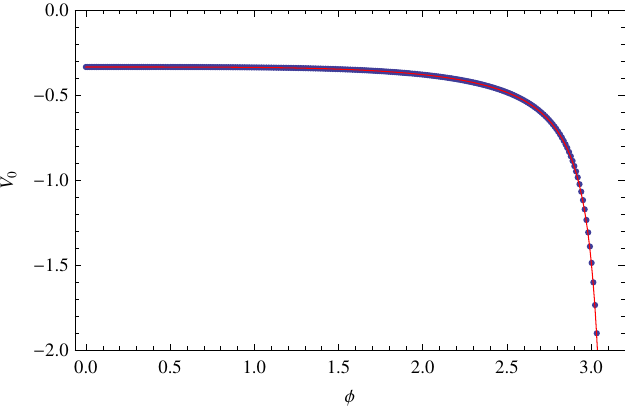}
 \includegraphics[width=8cm]{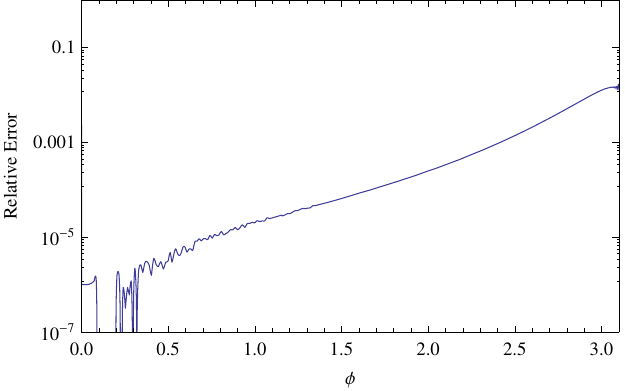}
\end{center}
\caption{\emph{Comparison of numerical and exact analytic calculations of $V_0$ for a massless, minimally coupled scalar field as a function of the angle, $\phi$, along an orbiting null geodesic in Nariai spacetime.} Left: The numerical calculation (blue dots) is a close match with the analytic expression (red line). The coincidence value is $V_0(x,x) = \frac12 (\xi-\frac16)R = -\frac13$, as expected. Right: With parameters so that the code completes in $1$ minute, the relative error in the numerical calculation is within $1\%$ up to the boundary of the normal neighborhood (at $\phi=\pi$).}
\label{fig:nariai-transport-V0-comparison}
\end{figure}

The accuracy of our numerical code may be verified by comparing with the results of Refs.~\cite{Nolan:2009} and \cite{Wardell:PhD}, which give analytic expressions for many of the bi-tensors used in this paper in Nariai spacetime. In Figs.~\ref{fig:nariai-transport-VV-comparison} and \ref{fig:nariai-transport-V0-comparison}, we compare analytic and numerical expressions for $\Delta^{1/2}$ and $V_0$, respectively. We consider the null geodesic which starts at $\rho=0.5$ and moves inwards to $\rho=0.25$ before turning around and going out to $\rho=0.789$, where it reaches  the edge of the normal neighborhood. The affine parameter, $s'$, has been scaled so that it is equal to the angle coordinate, $\phi$. We find that the numerical results faithfully match the analytic solution up to the boundary of the normal neighborhood. For the case of $\Delta^{1/2}$, Fig.~\ref{fig:nariai-transport-VV-comparison}, the error remains less than one part in $10^{-6}$ to within a short distance of the normal neighborhood boundary. The results for $V_0(x,x')$ are less accurate, but nonetheless the relative error remains less than $1\%$.
 
\begin{figure}
 \begin{center}
\begin{tabular}{m{8cm}m{8cm}m{1cm}}
 \includegraphics[width=8cm]{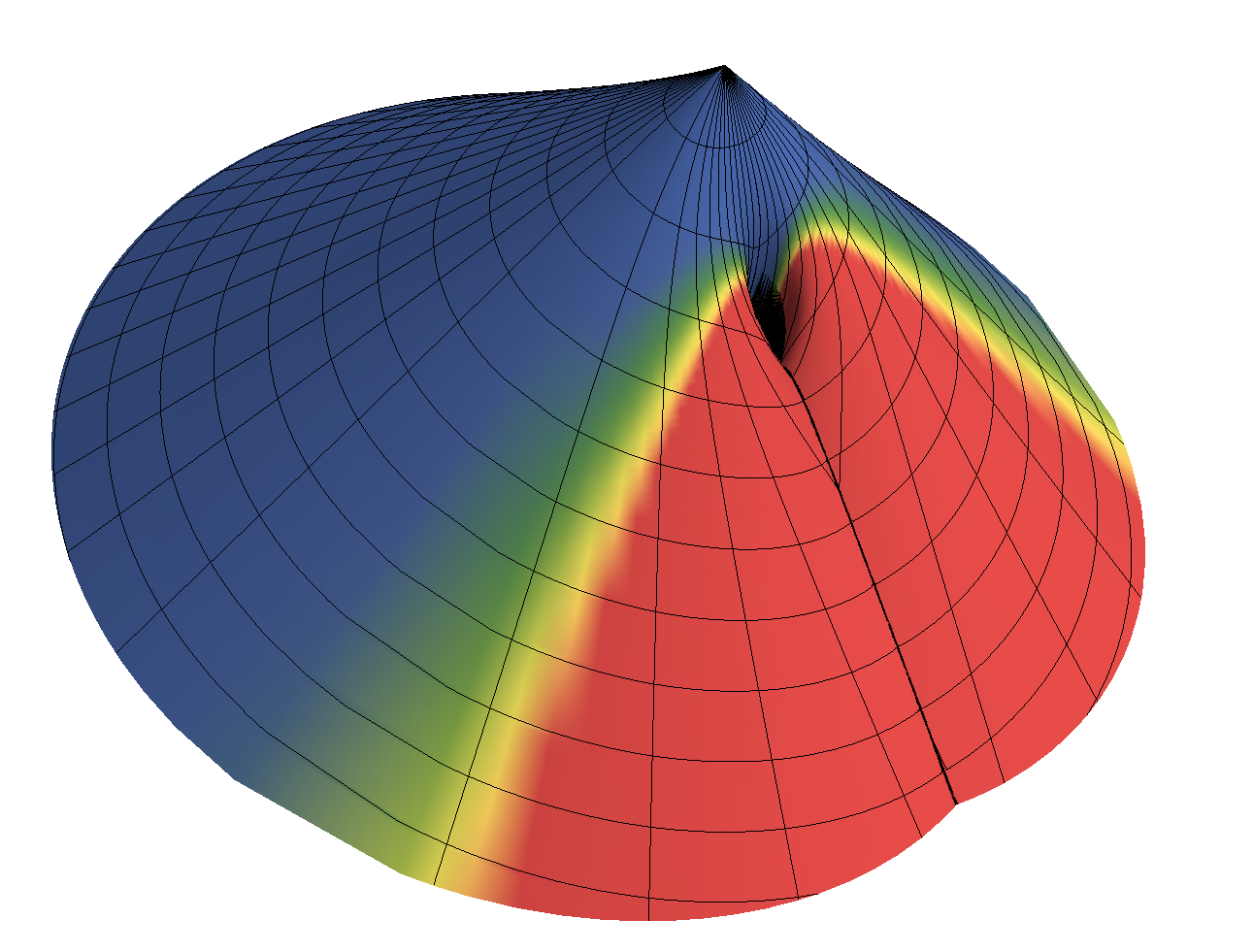} &
 \includegraphics[width=8cm]{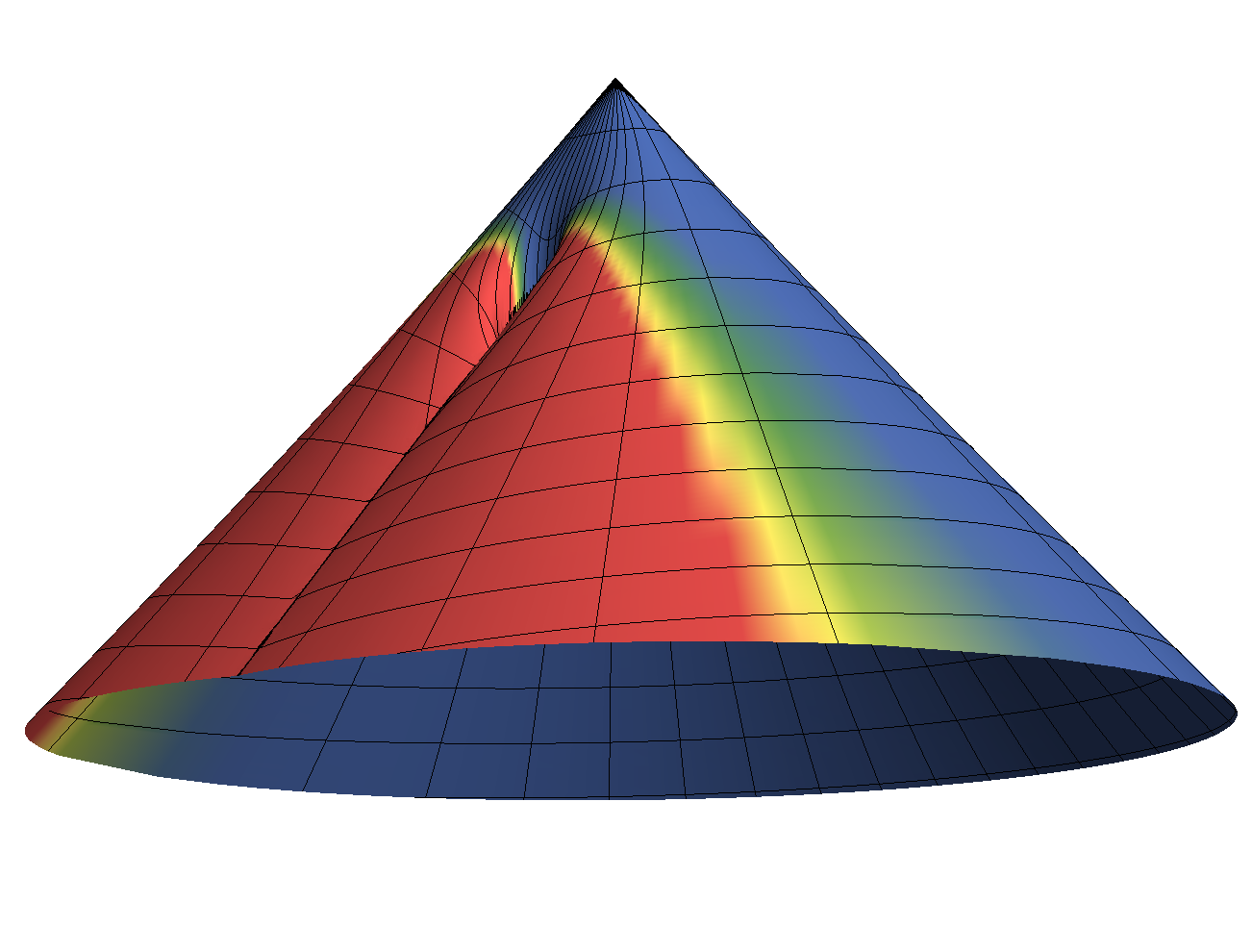} &
 \includegraphics[width=1cm]{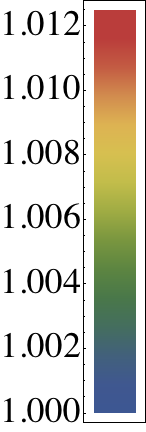}
\end{tabular}
\end{center}
\caption{(Colour online) \emph{$\Delta^{1/2}$ along the light-cone in Schwarzschild spacetime.} The point $x$ at the vertex of the cone is fixed at $r=10M$. $\Delta^{1/2}$ increases along a geodesic up to the caustic where it is singular.}
\label{fig:schw-VV-cone}
\end{figure}

In Fig.~\ref{fig:schw-VV-cone}, we use our numerical code to illustrate how $\Delta^{1/2}$ varies over the whole light-cone in Schwarzschild
spacetime. We find that it remains close to its initial value of $1$ far away from the caustic. As geodesics get close to the caustic, $\Delta^{1/2}$ grows and is eventually singular at the caustic. This is exactly as expected: $\Delta^{1/2}$ is a measure of the strength of focusing of geodesics, where values greater than $1$ correspond to focusing and values less than $1$ correspond to de-focusing. At the caustic, geodesics are focused to a point and correspondingly $\Delta^{1/2}$ is singular there.

\begin{figure}
 \begin{center}
\begin{tabular}{m{7.8cm}m{7.8cm}m{1.4cm}}
 \includegraphics[width=7.8cm]{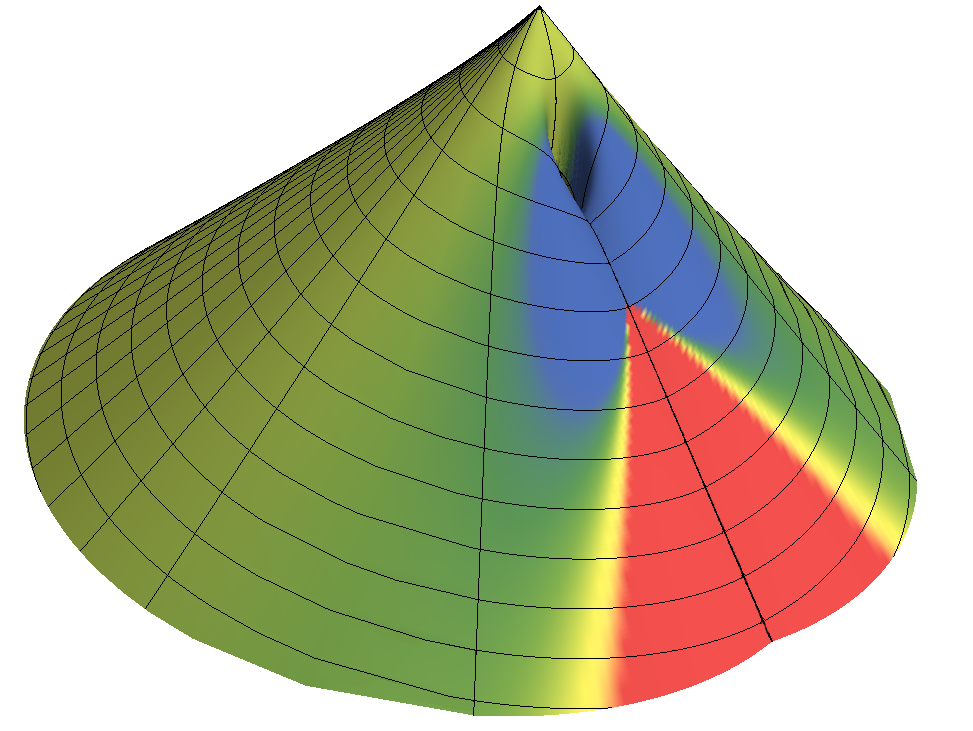} &
 \includegraphics[width=7.8cm]{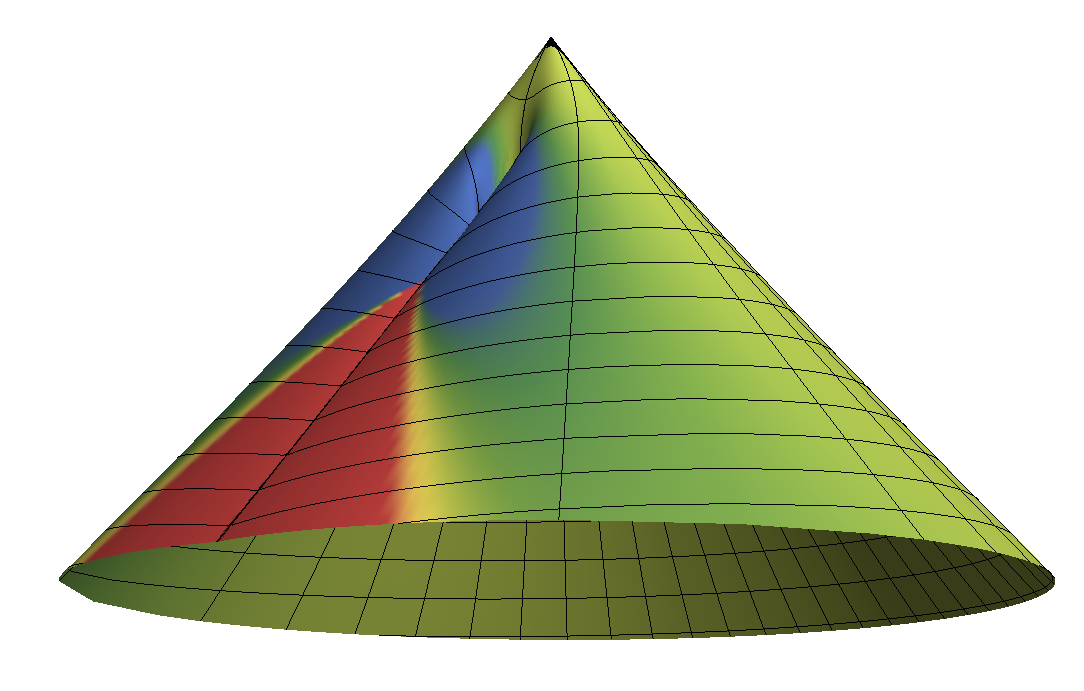} &
 \includegraphics[width=1.4cm]{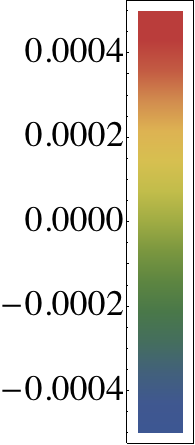}
\end{tabular}
\end{center}
\caption{(Colour online) \emph{$V(x,x')$ for a massless, scalar field along the light-cone in Schwarzschild spacetime.} The point $x$ (the vertex of the cone) is fixed at $r=6M$. $V(x,x')$ is $0$ initially, then, travelling along a geodesic, it goes negative for a period before turning positive and eventually becoming singular at the caustic. (Note that $V$ coincides with $V_0$ on the light cone.)}
\label{fig:schw-V0-cone}
\end{figure}

In Fig.~\ref{fig:schw-V0-cone}, we give a similar plot (again calculated from our numerical code) which indicates how $V(x,x')$ varies over the light-cone in Schwarzschild spacetime. In this case there is considerably more structure than was previously the case with $\Delta^{1/2}$. There is the expected singularity at the caustic. However, travelling along a geodesic, $V(x,x')$ also becomes negative for a period before turning positive and eventually becoming singular at the caustic.

\begin{figure}
 \begin{center}
 \includegraphics[width=8cm]{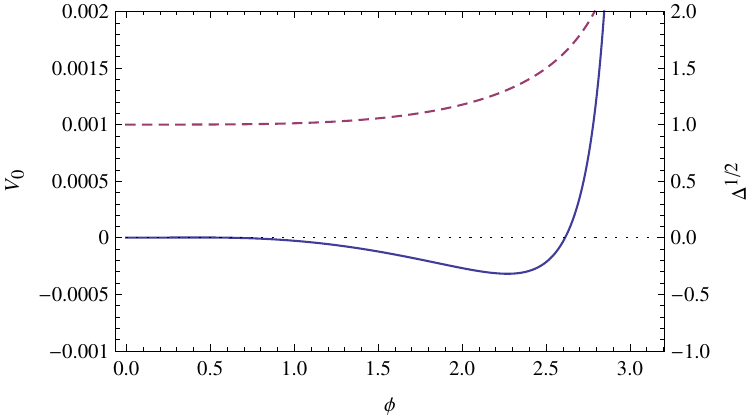}
 \includegraphics[width=8cm]{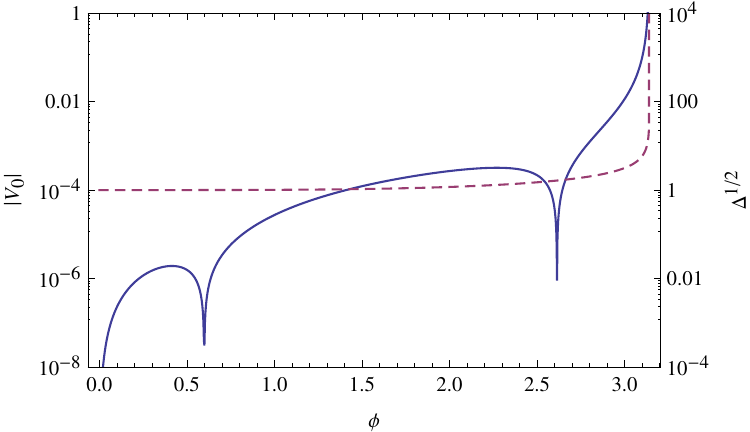}
\end{center}
\caption{(Colour online) \emph{$V_0(x,x')$ (solid blue line) and $\Delta^{1/2}$ (dashed purple line) for a massless, scalar field along the timelike circular orbit at $r=10M$ in Schwarzschild spacetime as a function of the angle, $\phi$ through which the geodesic has passed.} In the logarithmic plot, the absolute value of $V_0(x,x')$ is plotted to illustrate the divergence close to the caustic. (Since $V_0(x,x')$ passes through $0$ at $\phi\sim0.6$ and $\phi \sim 2.6$ there are corresponding features in the logarithmic plot.)}
\label{fig:schw-V0-timelike}
\end{figure}

The transport equations may also be applied to calculate $V_r(x,x')$ along a timelike geodesic. In Fig.~\ref{fig:schw-V0-timelike}, we apply our numerical code to the calculation of $V_0(x,x')$ along the timelike circular orbit at $r=10M$ in Schwarzschild.

\section{Discussion}

Several of the covariant expansion expressions computed by our code using the Avramidi method have been previously given in Ref.~\cite{Decanini:Folacci:2005a}, albeit to considerably lower order (for example, in their paper D\'ecanini and Folacci give $V(x,x')$ to order $\left(\sigma^a\right)^4$ compared to order $\left(\sigma^a\right)^{20}$ here). Comparison between the two results gives exact agreement, providing a reassuring confirmation of the accuracy of both our expressions and those of Ref.~\cite{Decanini:Folacci:2005a} (and confirming the error in Ref.~\cite{Phillips:Hu:2003} found by D\'ecanini and Folacci). Furthermore, several of the expansions not given by D\'ecanini and Folacci may be compared with those found by Christensen \cite{Christensen:1976vb,Christensen:1978yd}. Again, we have found that our code is in exact agreement with Christensen's results.

Our \emph{Mathematica} implementation of the semi-recursive approach (Sec.~\ref{sec:symbolic}) is given as a practical tool for computing high order covariant expansions. While it already exhibits a high level of efficiency, we believe that further improvement could be achieved, particularly in the limiting area of memory requirements. The initial expressions for the DeWitt coefficients as computed by our code are very general. However, they are not necessarily given as a minimal set. For example, with $P=0$ the DeWitt coefficient $a_3$ may be written as a sum of four terms, yet our code produces a sum of seven equivalent terms. It is possible, however, to use a set of transformation rules to reduce our expression to a canonical basis such as that of Ref.~\cite{Fulling:1992}. As our code is already written in \textsl{Mathematica} \cite{Mathematica} and has the ability to output into the \textsl{xTensor} \cite{xTensor} notation used by \textsl{Invar} \cite{Invar1,Invar2}, we were able to quickly canonicalize the scalar invariants appearing in our coincidence limit expressions. An extension of the Invar package to allow for the canonicalization of tensor invariants would allow our non-diagonal coefficients to also be immediately canonicalized with no further effort.

In Sec.~\ref{sec:numerical}, we discussed a numerical implementation of the transport equation approach to the calculation of $V_0$. This implementation is capable of computing $V_0$ for any spacetime, although we have chosen Nariai and Schwarzschild spacetimes as examples. The choice of Nariai spacetime has the benefit that an expression for $V_0$ is known exactly \cite{Nolan:2009}. This makes it possible to compare our numerical results with the analytic expressions to determine both the validity of the approach and the accuracy of the numerical calculation. Given parameters allowing the code to run in under a minute, we find that the numerical implementation is accurate to less than $1\%$ out as far as the location of the singularity of $V_0$ at the edge of the normal neighborhood.

In integrating the transport equations along a specific geodesic, we are not limited to the normal neighborhood. The only difficulty arises at \emph{caustics}, where some bi-tensors such as $\Delta^{1/2}$ and $V_0$ become singular. However, this is not an insurmountable problem. The singular components may be separated out and methods of complex analysis employed to integrate through the caustics, beyond which the bi-tensors once more become regular (but not necessarily real-valued) \cite{Casals:Dolan:Ottewill:Wardell:2009}. 
This is highlighted in Fig.~\ref{fig:schw-V0-timelike}, where our plot of $\Delta^{1/2}$ and $V_0$ extends outside the normal neighborhood, the boundary of which is at $\phi\approx1.25$, where the first null geodesic re-intersects the orbit. It does not necessarily follow, however, that the Green function outside the normal neighborhood is given by this value for $V(x,x')$. Instead, one might expect to obtain the Green function by considering the sum of the contributions obtained by integrating along \emph{all} geodesics connecting $x$ and $x'$ (there will be a discrete number of such geodesics except at caustics).

\section{Acknowledgements}
We would like to thank Antoine Folacci for much helpful correspondence. We also thank Jose M. Mart\'in-Garc\'ia for all of his help and advice on using \textsl{xTensor}. We would also like to thank Marc Casals, Sam Dolan, and Brien Nolan for many interesting and helpful discussions. We thank Justin Vines, Jia Shouqing and Eoin Murphy for identifying a sign error in Eq.~\eqref{eq:A-transport} in an earlier draft of this paper. Some computations were performed on the Damiana cluster at the Albert-Einstein-Institut. BW was supported by the Irish Research Council for Science, Engineering and Technology, funded by the National Development Plan. 

\appendix

\section{Canonical form of Hadamard and DeWitt coefficients} \label{sec:Vn}
In this Appendix, we present expressions for the diagonal DeWitt coefficients  $a_{0\,(0)}$, $a_{1\,(0)}$, $a_{2\,(0)}$, $a_{3\,(0)}$, $a_{4\,(0)}$ and $a_{5\,(0)}$ (where $a_r(x,x)=a_{r(0)}(x)$) in the canonical form produced by \textsl{Invar} \cite{Invar1,Invar2}. These have previously been given in various forms in the literature: $a_1$ and $a_2$ by DeWitt \cite{DeWitt:1965}, $a_3$ by Sakai \cite{Sakai:1971} and by Gilkey \cite{Gilkey:1975}, $a_4$ by Amsterdamski, Berkin and O'Connor \cite{Amsterdamski:Berkin:OConnor:1989} and by Avramidi \cite{Avramidi:1991,Avramidi:1998} and $a_5$ by Van den Ven \cite{Ven:1998}. However, to our knowledge, this is the first time that they have all been given in a simplified canonical form. We also note that our code is capable of producing expressions for $a_6$ and $a_7$ and for the off-diagonal coefficients (in non-canonical form) in a matter of minutes on a laptop computer.

During the canonicalization process, we have allowed Invar to use identities which are valid only in four spacetime dimensions, as our primary motivation is to study black hole spacetimes such as Schwarzschild and Kerr. This additional simplification is not essential, but does lead to more compact expressions.

Although our code is also capable of producing expressions for the off-diagonal coefficients, support for canonicalization of such expressions involving free indices is not yet available in \textsl{Invar}. For this reason, we restrict ourselves here to only the diagonal coefficients. We have also made these expressions, along with the corresponding (non-canonical) off-diagonal coefficients available online as \emph{Mathematica} code \cite{AvramidiCode}.

We also note that Eq.~\eqref{eq:v-a-relation} allows us to directly relate the Hadamard coefficients $V_{0\,(0)}$, $V_{1\,(0)}$, $V_{2\,(0)}$, $V_{3\,(0)}$ and $V_{4\,(0)}$ to these DeWitt coefficients:
\begin{gather}
 V_{0\,(0)} = \frac{1}{2}\left(m^2 a_{0\,(0)}- a_{1\,(0)}\right), \qquad 
 V_{1\,(0)} = \frac{1}{8}\left(m^4 a_{0\,(0)}- 2 m^2 a_{1\,(0)} + 2 a_{2\,(0)} \right),\nonumber \\
 V_{2\,(0)} = \frac{1}{96} \left(m^6 a_{0\,(0)}-3 m^4 a_{1\,(0)} + 6 m^2 a_{2\,(0)} - 6 a_{3\,(0)} \right), \nonumber \\
 V_{3\,(0)} = \frac{1}{2304} \left(m^8 a_{0\,(0)}-4 m^6 a_{1\,(0)} + 12 m^4 a_{2\,(0)} - 24 m^2 a_{3\,(0)} + 24 a_{4\,(0)} \right),\nonumber \\
 V_{4\,(0)} = \frac{1}{92160} \left(m^{10} a_{0\,(0)}-5 m^8 a_{1\,(0)} + 20 m^6 a_{2\,(0)} - 60 m^4 a_{3\,(0)} + 120 m^2 a_{4\,(0)} - 120 a_{5\,(0)} \right).
\end{gather}

Finally, we note that our expressions for the coefficients $V_{0\,(0)}$, $V_{1\,(0)}$ and $V_{2\,(0)}$ are in agreement with Ref.~\cite{Decanini:Folacci:2005a}, after canonicalization. In addition our expressions for the coefficients $a_{3\,(0)}$ and $a_{4\,(0)}$ are in agreement with Ref.~\cite{Matyjasek:Tryniecki:Zwierzchowska:2010}.

In the following we group the expressions in powers of $\xi$ and denote by $a_{r}^{(k)}$ the term involving the $k$-th power of $\xi$ in the diagonal DeWitt coefficient, $a_{r\,(0)}$, so that
\begin{equation}
a_{r\,(0)} = \sum_{k=0}^{r+1} a_{r}^{(k)} \xi^k.
\end{equation}
In this notation, the diagonal DeWitt coefficients are:
\begin{IEEEeqnarray}{rCl}
a_{0}^{(0)} &=& 1,
\end{IEEEeqnarray}
\begin{IEEEeqnarray}{rClrCl}
a_{1}^{(0)} &=& \frac{1}{6} R, \qquad &
a_{1}^{(1)} &=& - R,
\end{IEEEeqnarray}
\begin{IEEEeqnarray}{rClrClrCl}
a_{2}^{(0)} &=& \frac{1}{360} (-2 R{}_{\alpha }{}_{\beta } R{}^{\alpha }{}^{\beta } + 5 R^{2} + 2 R{}_{\alpha }{}_{\beta }{}_{\gamma }{}_{\delta } R{}^{\alpha }{}^{\beta }{}^{\gamma }{}^{\delta } + 12 R{}^{;\alpha }{}_{\alpha }), \qquad &a_{2}^{(1)} &=& \frac{1}{6} (- R^{2} -  R{}^{;\alpha }{}_{\alpha }), \qquad &
a_{2}^{(2)} &=& \frac{1}{2} R^{2},
\end{IEEEeqnarray}
\begin{subequations}
\begin{IEEEeqnarray}{rCl}
a_{3}^{(0)} &=& \frac{1}{15120} (584 R{}_{\alpha }{}^{\gamma } R{}^{\alpha }{}^{\beta } R{}_{\beta }{}_{\gamma } - 654 R{}_{\alpha }{}_{\beta } R{}^{\alpha }{}^{\beta } R + 99 R^{3} + 456 R{}^{\alpha }{}^{\beta } R{}^{\gamma }{}^{\delta } R{}_{\alpha }{}_{\gamma }{}_{\beta }{}_{\delta } + 72 R R{}_{\alpha }{}_{\beta }{}_{\gamma }{}_{\delta } R{}^{\alpha }{}^{\beta }{}^{\gamma }{}^{\delta } \nonumber \\ 
&& - 80 R{}_{\alpha }{}_{\beta }{}^{\epsilon }{}^{\rho } R{}^{\alpha }{}^{\beta }{}^{\gamma }{}^{\delta } R{}_{\gamma }{}_{\delta }{}_{\epsilon }{}_{\rho } + 51 R{}_{;\alpha } R{}^{;\alpha } - 12 R{}_{\alpha }{}_{\gamma }{}_{;\beta } R{}^{\alpha }{}^{\beta }{}^{;\gamma } - 6 R{}_{\alpha }{}_{\beta }{}_{;\gamma } R{}^{\alpha }{}^{\beta }{}^{;\gamma } + 27 R{}_{\alpha }{}_{\beta }{}_{\gamma }{}_{\delta }{}_{;\epsilon } R{}^{\alpha }{}^{\beta }{}^{\gamma }{}^{\delta }{}^{;\epsilon } \nonumber \\ 
&& + 84 R R{}^{;\alpha }{}_{\alpha } + 36 R{}_{\alpha }{}_{\beta } R{}^{;\alpha }{}^{\beta } - 24 R{}^{\alpha }{}^{\beta } R{}_{\alpha }{}_{\beta }{}^{;\gamma }{}_{\gamma } + 144 R{}_{\alpha }{}_{\gamma }{}_{\beta }{}_{\delta } R{}^{\alpha }{}^{\beta }{}^{;\gamma }{}^{\delta } + 54 R{}^{;\alpha }{}_{\alpha }{}^{\beta }{}_{\beta }),\\
a_{3}^{(1)} &=& \frac{1}{360} (2 R{}_{\alpha }{}_{\beta } R{}^{\alpha }{}^{\beta } R - 5 R^{3} - 2 R R{}_{\alpha }{}_{\beta }{}_{\gamma }{}_{\delta } R{}^{\alpha }{}^{\beta }{}^{\gamma }{}^{\delta } - 12 R{}_{;\alpha } R{}^{;\alpha } - 22 R R{}^{;\alpha }{}_{\alpha } - 4 R{}_{\alpha }{}_{\beta } R{}^{;\alpha }{}^{\beta } - 6 R{}^{;\alpha }{}_{\alpha }{}^{\beta }{}_{\beta }),
\end{IEEEeqnarray}
\begin{IEEEeqnarray}{rClrCl}
a_{3}^{(2)} &=& \frac{1}{12} (R^{3} + R{}_{;\alpha } R{}^{;\alpha } + 2 R R{}^{;\alpha }{}_{\alpha }),\qquad
a_{3}^{(3)} &=& -\frac{1}{6} R^{3},
\end{IEEEeqnarray}
\end{subequations}

\begin{subequations}
\begin{IEEEeqnarray}{rCl}
&a_{4}^{(0)} &= \frac{1}{1814400} (-32736 R{}_{\alpha }{}^{\gamma } R{}^{\alpha }{}^{\beta } R{}_{\beta }{}^{\delta } R{}_{\gamma }{}_{\delta } + 8436 R{}_{\alpha }{}_{\beta } R{}^{\alpha }{}^{\beta } R{}_{\gamma }{}_{\delta } R{}^{\gamma }{}^{\delta } + 59136 R{}_{\alpha }{}^{\gamma } R{}^{\alpha }{}^{\beta } R{}_{\beta }{}_{\gamma } R - 43518 R{}_{\alpha }{}_{\beta } R{}^{\alpha }{}^{\beta } R^{2} \nonumber \\ 
&& + 5743 R^{4} + 13944 R{}^{\alpha }{}^{\beta } R{}^{\gamma }{}^{\delta } R R{}_{\alpha }{}_{\gamma }{}_{\beta }{}_{\delta } + 3618 R^{2} R{}_{\alpha }{}_{\beta }{}_{\gamma }{}_{\delta } R{}^{\alpha }{}^{\beta }{}^{\gamma }{}^{\delta } + 168 R{}^{\alpha }{}^{\beta } R{}^{\gamma }{}^{\delta } R{}_{\alpha }{}_{\gamma }{}^{\epsilon }{}^{\rho } R{}_{\beta }{}_{\delta }{}_{\epsilon }{}_{\rho } - 4480 R R{}_{\alpha }{}_{\beta }{}^{\epsilon }{}^{\rho } R{}^{\alpha }{}^{\beta }{}^{\gamma }{}^{\delta } R{}_{\gamma }{}_{\delta }{}_{\epsilon }{}_{\rho } \nonumber \\ 
&& + 14832 R{}^{\alpha }{}^{\beta } R{}^{\gamma }{}^{\delta } R{}_{\alpha }{}^{\epsilon }{}_{\beta }{}^{\rho } R{}_{\gamma }{}_{\epsilon }{}_{\delta }{}_{\rho } - 3282 R{}_{\alpha }{}_{\beta } R{}^{\alpha }{}^{\beta } R{}_{\gamma }{}_{\delta }{}_{\epsilon }{}_{\rho } R{}^{\gamma }{}^{\delta }{}^{\epsilon }{}^{\rho } - 2496 R{}_{\alpha }{}_{\beta }{}^{\epsilon }{}^{\rho } R{}^{\alpha }{}^{\beta }{}^{\gamma }{}^{\delta } R{}_{\gamma }{}_{\epsilon }{}^{\sigma }{}^{\tau } R{}_{\delta }{}_{\rho }{}_{\sigma }{}_{\tau } \nonumber \\ 
&& + 1248 R{}_{\alpha }{}_{\beta }{}^{\epsilon }{}^{\rho } R{}^{\alpha }{}^{\beta }{}^{\gamma }{}^{\delta } R{}_{\gamma }{}_{\delta }{}^{\sigma }{}^{\tau } R{}_{\epsilon }{}_{\rho }{}_{\sigma }{}_{\tau } + 696 R{}_{\alpha }{}_{\beta }{}_{\gamma }{}_{\delta } R{}^{\alpha }{}^{\beta }{}^{\gamma }{}^{\delta } R{}_{\epsilon }{}_{\rho }{}_{\sigma }{}_{\tau } R{}^{\epsilon }{}^{\rho }{}^{\sigma }{}^{\tau } - 65040 R{}^{\beta }{}^{\gamma } R{}_{\beta }{}_{\gamma }{}_{;\alpha } R{}^{;\alpha } + 13740 R R{}_{;\alpha } R{}^{;\alpha } \nonumber \\ 
&& + 6960 R{}^{\beta }{}^{\gamma }{}^{\delta }{}^{\epsilon } R{}_{\beta }{}_{\gamma }{}_{\delta }{}_{\epsilon }{}_{;\alpha } R{}^{;\alpha } + 1440 R{}^{\alpha }{}^{\beta } R{}^{\gamma }{}^{\delta }{}_{;\alpha } R{}_{\gamma }{}_{\delta }{}_{;\beta } - 1560 R{}_{\alpha }{}_{\beta } R{}^{;\alpha } R{}^{;\beta } + 2880 R{}^{\beta }{}^{\gamma } R{}^{;\alpha } R{}_{\alpha }{}_{\beta }{}_{;\gamma } - 2160 R R{}_{\alpha }{}_{\gamma }{}_{;\beta } R{}^{\alpha }{}^{\beta }{}^{;\gamma } \nonumber \\ 
&& - 5760 R{}_{\alpha }{}^{\delta }{}^{\epsilon }{}^{\rho } R{}_{\gamma }{}_{\delta }{}_{\epsilon }{}_{\rho }{}_{;\beta } R{}^{\alpha }{}^{\beta }{}^{;\gamma } - 28920 R R{}_{\alpha }{}_{\beta }{}_{;\gamma } R{}^{\alpha }{}^{\beta }{}^{;\gamma } + 27840 R{}_{\alpha }{}_{\delta }{}_{\beta }{}_{\epsilon } R{}^{\delta }{}^{\epsilon }{}_{;\gamma } R{}^{\alpha }{}^{\beta }{}^{;\gamma } - 7680 R{}^{\alpha }{}^{\beta } R{}_{\gamma }{}_{\delta }{}_{;\beta } R{}_{\alpha }{}^{\gamma }{}^{;\delta } \nonumber \\ 
&& + 4800 R{}^{\alpha }{}^{\beta } R{}_{\beta }{}_{\delta }{}_{;\gamma } R{}_{\alpha }{}^{\gamma }{}^{;\delta } + 85440 R{}^{\alpha }{}^{\beta } R{}_{\beta }{}_{\gamma }{}_{;\delta } R{}_{\alpha }{}^{\gamma }{}^{;\delta } - 1920 R{}_{\alpha }{}_{\beta }{}_{\gamma }{}_{\delta } R{}^{;\alpha } R{}^{\beta }{}^{\gamma }{}^{;\delta } - 1920 R{}_{\beta }{}_{\gamma }{}_{\delta }{}_{\epsilon } R{}^{\alpha }{}^{\beta }{}^{;\gamma } R{}_{\alpha }{}^{\delta }{}^{;\epsilon } \nonumber \\ 
&& - 7680 R{}_{\alpha }{}_{\delta }{}_{\beta }{}_{\epsilon } R{}^{\alpha }{}^{\beta }{}^{;\gamma } R{}_{\gamma }{}^{\delta }{}^{;\epsilon } + 14400 R{}^{\alpha }{}^{\beta } R{}_{\alpha }{}_{\gamma }{}_{\beta }{}_{\epsilon }{}_{;\delta } R{}^{\gamma }{}^{\delta }{}^{;\epsilon } + 34080 R{}^{\alpha }{}^{\beta } R{}_{\alpha }{}_{\gamma }{}_{\beta }{}_{\delta }{}_{;\epsilon } R{}^{\gamma }{}^{\delta }{}^{;\epsilon } + 2700 R R{}_{\alpha }{}_{\beta }{}_{\gamma }{}_{\delta }{}_{;\epsilon } R{}^{\alpha }{}^{\beta }{}^{\gamma }{}^{\delta }{}^{;\epsilon } \nonumber \\ 
&& + 12960 R{}^{\alpha }{}^{\beta } R{}_{\beta }{}_{\delta }{}_{\gamma }{}_{\rho }{}_{;\epsilon } R{}_{\alpha }{}^{\gamma }{}^{\delta }{}^{\epsilon }{}^{;\rho } - 10800 R{}^{\alpha }{}^{\beta }{}^{\gamma }{}^{\delta } R{}_{\gamma }{}_{\delta }{}_{\epsilon }{}_{\rho }{}_{;\sigma } R{}_{\alpha }{}_{\beta }{}^{\epsilon }{}^{\rho }{}^{;\sigma } - 9792 R{}_{\beta }{}_{\gamma } R{}^{\beta }{}^{\gamma } R{}^{;\alpha }{}_{\alpha } + 4608 R^{2} R{}^{;\alpha }{}_{\alpha } \nonumber \\ 
&& + 1296 R{}_{\beta }{}_{\gamma }{}_{\delta }{}_{\epsilon } R{}^{\beta }{}^{\gamma }{}^{\delta }{}^{\epsilon } R{}^{;\alpha }{}_{\alpha } + 432 R{}_{\alpha }{}_{\beta } R R{}^{;\alpha }{}^{\beta } + 1632 R{}^{\gamma }{}^{\delta } R{}_{\alpha }{}_{\gamma }{}_{\beta }{}_{\delta } R{}^{;\alpha }{}^{\beta } + 936 R{}_{;\alpha }{}_{\beta } R{}^{;\alpha }{}^{\beta } - 11136 R{}^{\alpha }{}^{\beta } R{}^{\gamma }{}^{\delta } R{}_{\alpha }{}_{\gamma }{}_{;\beta }{}_{\delta } \nonumber \\ 
&& + 1008 R{}^{;\alpha }{}_{\alpha } R{}^{;\beta }{}_{\beta } + 10464 R{}^{\alpha }{}^{\beta } R{}^{\gamma }{}^{\delta } R{}_{\alpha }{}_{\beta }{}_{;\gamma }{}_{\delta } - 18000 R{}^{\alpha }{}^{\beta } R R{}_{\alpha }{}_{\beta }{}^{;\gamma }{}_{\gamma } + 624 R{}^{;\alpha }{}^{\beta } R{}_{\alpha }{}_{\beta }{}^{;\gamma }{}_{\gamma } + 8352 R R{}_{\alpha }{}_{\gamma }{}_{\beta }{}_{\delta } R{}^{\alpha }{}^{\beta }{}^{;\gamma }{}^{\delta } \nonumber \\ 
&& - 14016 R{}_{\alpha }{}_{\gamma }{}^{\epsilon }{}^{\rho } R{}_{\beta }{}_{\delta }{}_{\epsilon }{}_{\rho } R{}^{\alpha }{}^{\beta }{}^{;\gamma }{}^{\delta } + 384 R{}_{\alpha }{}^{\epsilon }{}_{\beta }{}^{\rho } R{}_{\gamma }{}_{\epsilon }{}_{\delta }{}_{\rho } R{}^{\alpha }{}^{\beta }{}^{;\gamma }{}^{\delta } + 1872 R{}_{\gamma }{}_{\delta }{}_{;\alpha }{}_{\beta } R{}^{\alpha }{}^{\beta }{}^{;\gamma }{}^{\delta } - 4032 R{}_{\alpha }{}_{\gamma }{}_{;\beta }{}_{\delta } R{}^{\alpha }{}^{\beta }{}^{;\gamma }{}^{\delta } \nonumber \\ 
&& + 1872 R{}_{\alpha }{}_{\beta }{}_{;\gamma }{}_{\delta } R{}^{\alpha }{}^{\beta }{}^{;\gamma }{}^{\delta } + 2304 R{}^{\alpha }{}^{\beta } R{}^{\gamma }{}^{\delta }{}^{\epsilon }{}^{\rho } R{}_{\alpha }{}_{\gamma }{}_{\beta }{}_{\epsilon }{}_{;\delta }{}_{\rho } - 216 R{}^{\alpha }{}^{\beta }{}^{;\gamma }{}_{\gamma } R{}_{\alpha }{}_{\beta }{}^{;\delta }{}_{\delta } + 23904 R{}_{\alpha }{}^{\gamma } R{}^{\alpha }{}^{\beta } R{}_{\beta }{}_{\gamma }{}^{;\delta }{}_{\delta } \nonumber \\ 
&& + 8448 R{}^{\alpha }{}^{\beta } R{}_{\beta }{}_{\delta }{}_{\gamma }{}_{\epsilon } R{}_{\alpha }{}^{\gamma }{}^{;\delta }{}^{\epsilon } + 12288 R{}^{\alpha }{}^{\beta } R{}_{\alpha }{}_{\gamma }{}_{\beta }{}_{\delta } R{}^{\gamma }{}^{\delta }{}^{;\epsilon }{}_{\epsilon } + 576 R{}_{\alpha }{}_{\beta }{}_{\gamma }{}_{\delta }{}_{;\epsilon }{}_{\rho } R{}^{\alpha }{}^{\beta }{}^{\gamma }{}^{\delta }{}^{;\epsilon }{}^{\rho } + 2640 R{}^{;\alpha } R{}_{;\alpha }{}^{\beta }{}_{\beta } \nonumber \\ 
&& + 960 R{}_{\alpha }{}_{\beta }{}_{;\gamma } R{}^{;\alpha }{}^{\beta }{}^{\gamma } - 480 R{}^{\alpha }{}^{\beta }{}^{;\gamma } R{}_{\alpha }{}_{\gamma }{}_{;\beta }{}^{\delta }{}_{\delta } - 240 R{}^{\alpha }{}^{\beta }{}^{;\gamma } R{}_{\alpha }{}_{\beta }{}_{;\gamma }{}^{\delta }{}_{\delta } + 5760 R{}_{\alpha }{}_{\gamma }{}_{\beta }{}_{\delta }{}_{;\epsilon } R{}^{\alpha }{}^{\beta }{}^{;\gamma }{}^{\delta }{}^{\epsilon } \nonumber \\ 
&& + 1080 R R{}^{;\alpha }{}_{\alpha }{}^{\beta }{}_{\beta } + 960 R{}_{\alpha }{}_{\beta } R{}^{;\alpha }{}^{\beta }{}^{\gamma }{}_{\gamma } - 240 R{}^{\alpha }{}^{\beta } R{}_{\alpha }{}_{\beta }{}^{;\gamma }{}_{\gamma }{}^{\delta }{}_{\delta } + 1920 R{}_{\alpha }{}_{\gamma }{}_{\beta }{}_{\delta } R{}^{\alpha }{}^{\beta }{}^{;\gamma }{}^{\delta }{}^{\epsilon }{}_{\epsilon } + 480 R{}^{;\alpha }{}_{\alpha }{}^{\beta }{}_{\beta }{}^{\gamma }{}_{\gamma }),\\
&a_{4}^{(1)} &= \frac{1}{15120} (-584 R{}_{\alpha }{}^{\gamma } R{}^{\alpha }{}^{\beta } R{}_{\beta }{}_{\gamma } R + 654 R{}_{\alpha }{}_{\beta } R{}^{\alpha }{}^{\beta } R^{2} - 99 R^{4} - 456 R{}^{\alpha }{}^{\beta } R{}^{\gamma }{}^{\delta } R R{}_{\alpha }{}_{\gamma }{}_{\beta }{}_{\delta } - 72 R^{2} R{}_{\alpha }{}_{\beta }{}_{\gamma }{}_{\delta } R{}^{\alpha }{}^{\beta }{}^{\gamma }{}^{\delta } \nonumber \\ 
&& + 80 R R{}_{\alpha }{}_{\beta }{}^{\epsilon }{}^{\rho } R{}^{\alpha }{}^{\beta }{}^{\gamma }{}^{\delta } R{}_{\gamma }{}_{\delta }{}_{\epsilon }{}_{\rho } + 12 R{}^{\beta }{}^{\gamma } R{}_{\beta }{}_{\gamma }{}_{;\alpha } R{}^{;\alpha } - 135 R R{}_{;\alpha } R{}^{;\alpha } - 36 R{}^{\beta }{}^{\gamma }{}^{\delta }{}^{\epsilon } R{}_{\beta }{}_{\gamma }{}_{\delta }{}_{\epsilon }{}_{;\alpha } R{}^{;\alpha } + 102 R{}_{\alpha }{}_{\beta } R{}^{;\alpha } R{}^{;\beta } \nonumber \\ 
&& + 24 R{}^{\beta }{}^{\gamma } R{}^{;\alpha } R{}_{\alpha }{}_{\beta }{}_{;\gamma } + 12 R R{}_{\alpha }{}_{\gamma }{}_{;\beta } R{}^{\alpha }{}^{\beta }{}^{;\gamma } + 6 R R{}_{\alpha }{}_{\beta }{}_{;\gamma } R{}^{\alpha }{}^{\beta }{}^{;\gamma } + 24 R{}_{\alpha }{}_{\beta }{}_{\gamma }{}_{\delta } R{}^{;\alpha } R{}^{\beta }{}^{\gamma }{}^{;\delta } - 27 R R{}_{\alpha }{}_{\beta }{}_{\gamma }{}_{\delta }{}_{;\epsilon } R{}^{\alpha }{}^{\beta }{}^{\gamma }{}^{\delta }{}^{;\epsilon } \nonumber \\ 
&& + 30 R{}_{\beta }{}_{\gamma } R{}^{\beta }{}^{\gamma } R{}^{;\alpha }{}_{\alpha } - 123 R^{2} R{}^{;\alpha }{}_{\alpha } - 18 R{}_{\beta }{}_{\gamma }{}_{\delta }{}_{\epsilon } R{}^{\beta }{}^{\gamma }{}^{\delta }{}^{\epsilon } R{}^{;\alpha }{}_{\alpha } - 48 R{}_{\alpha }{}_{\beta } R R{}^{;\alpha }{}^{\beta } - 72 R{}^{\gamma }{}^{\delta } R{}_{\alpha }{}_{\gamma }{}_{\beta }{}_{\delta } R{}^{;\alpha }{}^{\beta } \nonumber \\ 
&& - 72 R{}_{;\alpha }{}_{\beta } R{}^{;\alpha }{}^{\beta } - 84 R{}^{;\alpha }{}_{\alpha } R{}^{;\beta }{}_{\beta } + 24 R{}^{\alpha }{}^{\beta } R R{}_{\alpha }{}_{\beta }{}^{;\gamma }{}_{\gamma } - 24 R{}^{;\alpha }{}^{\beta } R{}_{\alpha }{}_{\beta }{}^{;\gamma }{}_{\gamma } - 144 R R{}_{\alpha }{}_{\gamma }{}_{\beta }{}_{\delta } R{}^{\alpha }{}^{\beta }{}^{;\gamma }{}^{\delta } \nonumber \\ 
&& - 210 R{}^{;\alpha } R{}_{;\alpha }{}^{\beta }{}_{\beta } - 36 R{}_{\alpha }{}_{\beta }{}_{;\gamma } R{}^{;\alpha }{}^{\beta }{}^{\gamma } - 96 R R{}^{;\alpha }{}_{\alpha }{}^{\beta }{}_{\beta } - 36 R{}_{\alpha }{}_{\beta } R{}^{;\alpha }{}^{\beta }{}^{\gamma }{}_{\gamma } - 18 R{}^{;\alpha }{}_{\alpha }{}^{\beta }{}_{\beta }{}^{\gamma }{}_{\gamma }),\\
&a_{4}^{(2)} &= \frac{1}{720} (-2 R{}_{\alpha }{}_{\beta } R{}^{\alpha }{}^{\beta } R^{2} + 5 R^{4} + 2 R^{2} R{}_{\alpha }{}_{\beta }{}_{\gamma }{}_{\delta } R{}^{\alpha }{}^{\beta }{}^{\gamma }{}^{\delta } + 34 R R{}_{;\alpha } R{}^{;\alpha } - 12 R{}_{\alpha }{}_{\beta } R{}^{;\alpha } R{}^{;\beta } + 32 R^{2} R{}^{;\alpha }{}_{\alpha } \nonumber \\ 
&& + 8 R{}_{\alpha }{}_{\beta } R R{}^{;\alpha }{}^{\beta } + 8 R{}_{;\alpha }{}_{\beta } R{}^{;\alpha }{}^{\beta } + 10 R{}^{;\alpha }{}_{\alpha } R{}^{;\beta }{}_{\beta } + 24 R{}^{;\alpha } R{}_{;\alpha }{}^{\beta }{}_{\beta } + 12 R R{}^{;\alpha }{}_{\alpha }{}^{\beta }{}_{\beta }),
\end{IEEEeqnarray}
\begin{IEEEeqnarray}{rClrCl}
&a_{4}^{(3)} &= \frac{1}{36} (- R^{4} - 3 R R{}_{;\alpha } R{}^{;\alpha } - 3 R^{2} R{}^{;\alpha }{}_{\alpha }),\qquad
&a_{4}^{(4)} &= \frac{1}{24} R^{4},
\end{IEEEeqnarray}
\end{subequations}

\begin{subequations}

&a_{5}^{(2)} &= \frac{1}{30240} (584 R{}_{\alpha }{}^{\gamma } R{}^{\alpha }{}^{\beta } R{}_{\beta }{}_{\gamma } R^{2} - 654 R{}_{\alpha }{}_{\beta } R{}^{\alpha }{}^{\beta } R^{3} + 99 R^{5} + 456 R{}^{\alpha }{}^{\beta } R{}^{\gamma }{}^{\delta } R^{2} R{}_{\alpha }{}_{\gamma }{}_{\beta }{}_{\delta } + 72 R^{3} R{}_{\alpha }{}_{\beta }{}_{\gamma }{}_{\delta } R{}^{\alpha }{}^{\beta }{}^{\gamma }{}^{\delta } \nonumber \\ 
&& - 80 R^{2} R{}_{\alpha }{}_{\beta }{}^{\epsilon }{}^{\rho } R{}^{\alpha }{}^{\beta }{}^{\gamma }{}^{\delta } R{}_{\gamma }{}_{\delta }{}_{\epsilon }{}_{\rho } - 24 R{}^{\beta }{}^{\gamma } R R{}_{\beta }{}_{\gamma }{}_{;\alpha } R{}^{;\alpha } - 26 R{}_{\beta }{}_{\gamma } R{}^{\beta }{}^{\gamma } R{}_{;\alpha } R{}^{;\alpha } + 257 R^{2} R{}_{;\alpha } R{}^{;\alpha } \nonumber \\ 
&& + 17 R{}_{\beta }{}_{\gamma }{}_{\delta }{}_{\epsilon } R{}^{\beta }{}^{\gamma }{}^{\delta }{}^{\epsilon } R{}_{;\alpha } R{}^{;\alpha } + 72 R R{}^{\beta }{}^{\gamma }{}^{\delta }{}^{\epsilon } R{}_{\beta }{}_{\gamma }{}_{\delta }{}_{\epsilon }{}_{;\alpha } R{}^{;\alpha } + 90 R{}_{\alpha }{}^{\gamma } R{}_{\beta }{}_{\gamma } R{}^{;\alpha } R{}^{;\beta } - 300 R{}_{\alpha }{}_{\beta } R R{}^{;\alpha } R{}^{;\beta } \nonumber \\ 
&& + 48 R{}^{\gamma }{}^{\delta } R{}_{\alpha }{}_{\gamma }{}_{\beta }{}_{\delta } R{}^{;\alpha } R{}^{;\beta } - 48 R{}^{\beta }{}^{\gamma } R R{}^{;\alpha } R{}_{\alpha }{}_{\beta }{}_{;\gamma } - 12 R^{2} R{}_{\alpha }{}_{\gamma }{}_{;\beta } R{}^{\alpha }{}^{\beta }{}^{;\gamma } - 6 R^{2} R{}_{\alpha }{}_{\beta }{}_{;\gamma } R{}^{\alpha }{}^{\beta }{}^{;\gamma } \nonumber \\ 
&& - 48 R R{}_{\alpha }{}_{\beta }{}_{\gamma }{}_{\delta } R{}^{;\alpha } R{}^{\beta }{}^{\gamma }{}^{;\delta } + 27 R^{2} R{}_{\alpha }{}_{\beta }{}_{\gamma }{}_{\delta }{}_{;\epsilon } R{}^{\alpha }{}^{\beta }{}^{\gamma }{}^{\delta }{}^{;\epsilon } + 312 R{}^{;\alpha } R{}^{;\beta } R{}_{;\alpha }{}_{\beta } - 60 R{}_{\beta }{}_{\gamma } R{}^{\beta }{}^{\gamma } R R{}^{;\alpha }{}_{\alpha } + 162 R^{3} R{}^{;\alpha }{}_{\alpha } \nonumber \\ 
&& + 36 R R{}_{\beta }{}_{\gamma }{}_{\delta }{}_{\epsilon } R{}^{\beta }{}^{\gamma }{}^{\delta }{}^{\epsilon } R{}^{;\alpha }{}_{\alpha } + 60 R{}_{\alpha }{}_{\beta } R^{2} R{}^{;\alpha }{}^{\beta } + 144 R{}^{\gamma }{}^{\delta } R R{}_{\alpha }{}_{\gamma }{}_{\beta }{}_{\delta } R{}^{;\alpha }{}^{\beta } + 200 R R{}_{;\alpha }{}_{\beta } R{}^{;\alpha }{}^{\beta } + 252 R{}_{;\alpha } R{}^{;\alpha } R{}^{;\beta }{}_{\beta } \nonumber \\ 
&& + 238 R R{}^{;\alpha }{}_{\alpha } R{}^{;\beta }{}_{\beta } + 168 R{}_{\beta }{}_{\gamma }{}_{;\alpha } R{}^{;\alpha } R{}^{;\beta }{}^{\gamma } - 312 R{}^{;\alpha } R{}_{\alpha }{}_{\beta }{}_{;\gamma } R{}^{;\beta }{}^{\gamma } + 56 R{}_{\beta }{}_{\gamma } R{}^{;\alpha }{}_{\alpha } R{}^{;\beta }{}^{\gamma } - 128 R{}_{\alpha }{}_{\gamma } R{}^{;\alpha }{}^{\beta } R{}^{;\gamma }{}_{\beta } \nonumber \\ 
&& - 24 R{}^{\alpha }{}^{\beta } R^{2} R{}_{\alpha }{}_{\beta }{}^{;\gamma }{}_{\gamma } - 72 R{}^{;\alpha } R{}^{;\beta } R{}_{\alpha }{}_{\beta }{}^{;\gamma }{}_{\gamma } + 48 R R{}^{;\alpha }{}^{\beta } R{}_{\alpha }{}_{\beta }{}^{;\gamma }{}_{\gamma } + 144 R^{2} R{}_{\alpha }{}_{\gamma }{}_{\beta }{}_{\delta } R{}^{\alpha }{}^{\beta }{}^{;\gamma }{}^{\delta } + 224 R{}_{\alpha }{}_{\gamma }{}_{\beta }{}_{\delta } R{}^{;\alpha }{}^{\beta } R{}^{;\gamma }{}^{\delta } \nonumber \\ 
&& + 588 R R{}^{;\alpha } R{}_{;\alpha }{}^{\beta }{}_{\beta } + 72 R R{}_{\alpha }{}_{\beta }{}_{;\gamma } R{}^{;\alpha }{}^{\beta }{}^{\gamma } + 36 R{}_{;\alpha }{}_{\beta }{}_{\gamma } R{}^{;\alpha }{}^{\beta }{}^{\gamma } + 102 R{}^{;\alpha }{}_{\alpha }{}^{\beta } R{}_{;\beta }{}^{\gamma }{}_{\gamma } + 72 R{}_{\beta }{}_{\gamma } R{}^{;\alpha } R{}^{;\beta }{}^{\gamma }{}_{\alpha } \nonumber \\ 
&& - 102 R{}_{\alpha }{}_{\beta } R{}^{;\alpha } R{}^{;\beta }{}^{\gamma }{}_{\gamma } + 144 R{}^{;\alpha }{}^{\beta } R{}_{;\alpha }{}_{\beta }{}^{\gamma }{}_{\gamma } + 138 R^{2} R{}^{;\alpha }{}_{\alpha }{}^{\beta }{}_{\beta } + 72 R{}_{\alpha }{}_{\beta } R R{}^{;\alpha }{}^{\beta }{}^{\gamma }{}_{\gamma } + 84 R{}^{;\alpha }{}_{\alpha } R{}^{;\beta }{}_{\beta }{}^{\gamma }{}_{\gamma } \nonumber \\ 
&& + 108 R{}^{;\alpha } R{}_{;\alpha }{}^{\beta }{}_{\beta }{}^{\gamma }{}_{\gamma } + 36 R R{}^{;\alpha }{}_{\alpha }{}^{\beta }{}_{\beta }{}^{\gamma }{}_{\gamma }),\\
&a_{5}^{(3)} &= \frac{1}{2160} (2 R{}_{\alpha }{}_{\beta } R{}^{\alpha }{}^{\beta } R^{3} - 5 R^{5} - 2 R^{3} R{}_{\alpha }{}_{\beta }{}_{\gamma }{}_{\delta } R{}^{\alpha }{}^{\beta }{}^{\gamma }{}^{\delta } - 66 R^{2} R{}_{;\alpha } R{}^{;\alpha } + 36 R{}_{\alpha }{}_{\beta } R R{}^{;\alpha } R{}^{;\beta } - 36 R{}^{;\alpha } R{}^{;\beta } R{}_{;\alpha }{}_{\beta } \nonumber \\ 
&& - 42 R^{3} R{}^{;\alpha }{}_{\alpha } - 12 R{}_{\alpha }{}_{\beta } R^{2} R{}^{;\alpha }{}^{\beta } - 24 R R{}_{;\alpha }{}_{\beta } R{}^{;\alpha }{}^{\beta } - 30 R{}_{;\alpha } R{}^{;\alpha } R{}^{;\beta }{}_{\beta } - 30 R R{}^{;\alpha }{}_{\alpha } R{}^{;\beta }{}_{\beta } - 72 R R{}^{;\alpha } R{}_{;\alpha }{}^{\beta }{}_{\beta } \nonumber \\ 
&& - 18 R^{2} R{}^{;\alpha }{}_{\alpha }{}^{\beta }{}_{\beta }),
\end{IEEEeqnarray}
\begin{IEEEeqnarray}{rClrCl}
&a_{5}^{(4)} &= \frac{1}{144} (R^{5} + 6 R^{2} R{}_{;\alpha } R{}^{;\alpha } + 4 R^{3} R{}^{;\alpha }{}_{\alpha }),\qquad
&a_{5}^{(5)} &= -\frac{1}{120} R^{5},
\end{IEEEeqnarray}
\end{subequations}

\bibliography{references}{}
\bibliographystyle{apsrev4-1}

\end{document}